\documentstyle[amssymb]{elsart}
\newcommand{\rij}{r_{ij}} 

\newcommand{\p}{\partial} 

\newcommand{\rd}{\,{\rm d}} 
\newcommand{\re}{\,{\rm e}} 
\newcommand{\dx}{\,{\rm d}x} 
\newcommand{\dz}{\,{\rm d}z} 

\newcommand{\dom}{\,{\rm d}\omega} 
\newcommand{\Dom}{\,{\rm d}\Omega} 
\newcommand{\dph}{\,{\rm d}\phi} 
\newcommand{\hh}{\frac{\hbar^2}{m}}
\newcommand{\om}{\omega}
\newcommand{\Om}{\Omega}

\newcommand{\ma}{{m_\alpha}}
\newcommand{\la}{{\ell_\alpha}}
\newcommand{\mb}{{m_\beta}}
\newcommand{\lb}{{\ell_\beta}}
\newcommand{\ja}{{j_\alpha}}
\newcommand{\jb}{{j_\beta}}

\newcommand{\Laa}{{\Lambda_\alpha}}

\newcommand{\rr}{r_{ij}}
\newcommand{\Ll}{{[L]}}

\newcommand{\Lm}{{[L_m]}}
\newcommand{\Lz}{{[0]}}

\newcommand{\Lc}{{\cal L}}
\newcommand{\LLr}{\frac{\Lc(\Lc+1)}{r^2}}

\begin{document}
\begin{frontmatter}
%=============================================
%                 secT.tex:     TITLE                                 
%=============================================
 \title{Method for Solving  the Many-Body Bound State Nuclear Problem}
 \author{ M. Fabre de la Ripelle, S. A. Sofianos$^*$, 
and R. M. Adam}\footnote{Permanent address: Department of 
Science and Technology, Private Bag X894, Pretoria 0001, South Africa.}
\address
{ Physics Department, University of South Africa, P.O. 
 Box 392, Pretoria 0003, South Africa.}
\address { E-mail:$^*$sofiasa@science.unisa.ac.za}
\date{\today}
\begin{abstract}
 We present  a method based on hyperspherical harmonics
to solve the nuclear many-body problem. It is an extension of 
accurate methods used for studying  few-body systems
to many bodies and is  based on the  assumption that  nucleons 
in nuclei interact mainly via pairwise forces. 
This leads to a two-variable  integro-differential equation  
which is easy to solve. Unlike  methods that utilize 
effective interactions, the present one employs directly  
nucleon-nucleon potentials and therefore nuclear correlations 
are included in an unambiguous way. Three body forces can also 
be included in the formalism. Details on how to obtain
the various ingredients entering into the equation for the A-body system 
are given. Employing our formalism we calculated the binding
energies  for  closed  and open shell nuclei with central forces 
where the bound states are defined by a single hyperspherical harmonic.
The results found are  in agreement with those obtained by other methods.
\end{abstract}
%\maketitle
\end{frontmatter}
%cccccccccccccccccccccccccccccccccccccccccccccccccccccccccc
%               secI.tex : Introduction
%cccccccccccccccccccccccccccccccccccccccccccccccccccccccccc
 \section{Introduction}
 %%%%%%%%%%%%%%%%%%%%%%
 The discovery by Jensen and Goeppert-Mayer of the shell model 
structure of nuclei \cite{JM}, was a great step forward in  
understanding 
 the organization of nucleons in nuclei.  Since the magic numbers 
occurring in the shell model can be explained in terms of  harmonic 
oscillator quantum numbers, with the occurrence of a large spin-orbit 
force, it was taken for granted that the nucleons move inside an 
average  one-body potential.  Indeed, Goeppert-Mayer 
and Jensen \cite{JM} assumed that {\it ``~$\cdots$  each nucleon moves in an 
average field of force $V(r)$, of spherical symmetry, and independent of the 
exact instantaneous position of all the other nucleons.''} In order to explain 
the existence  of  magic numbers, this one body field was completed by the 
introduction of a strong spin-orbit force. Some interesting remarks,
however, have  been put forward by de Shalit and Talmi  about the validity of 
this interpretation \cite{deShalit}, namely, that if such  a  self-consistent 
central field is found, then one should  consider the residual interaction, 
which is not accounted for, as a perturbation.\par
The shell model was introduced at the beginning of the fifties at the time
where the nucleon-nucleon potential was not yet well known because the
collision   energy was  not large enough to detect the strong repulsive 
core needed to reproduce the $S$-phase shifts beyond 300\,MeV.
This strong repulsive core generates important two-body correlations in 
nuclei in such a way that Hartree-Fock methods applied with actual realistic 
nucleon-nucleon potentials produce a very small binding energy.
Thus, the residual interaction responsible for the largest part of the binding
energy cannot be considered as a perturbation.  Of course if one considers
that the free nucleon-nucleon potential is not the true interaction between
nucleons in nuclei, the previous remarks do not apply but then one has to 
state it explicitly and indicate what interaction should be used instead.
For instance in the self-consistent mean-field models \cite{BH} a
Hartree-Fock method is used with Skyrme interactions.\par
The shell model was first established for atoms but as pointed out by 
de Shalit and Talmi, there is an important difference between nuclei 
and atoms, namely,  in atoms electrons  have as a natural reference point
the nucleus and  {\it \dots  no such reference  point exists in the 
nucleus, and   at a first look it seems hard  to understand why a central 
potential may form a  good starting point for a nuclear model''}
\cite{deShalit}.\par
Fifty years after the introduction of the shell model, the situation was 
still the same.  After a lengthy discussion about the validity of 
the shell model, Talmi \cite{Talmi} pointed out that {\it ``\dots 
 Today, as in 1949, the best proof for the validity of the shell model is 
the good agreement of predictions with experiments"}.
The only well established property is that the quantum numbers 
associated with the  magic numbers are those of the harmonic 
oscillator in a ground state corrected by a strong spin-orbit force.\par
In order to find an alternative  explanation, to that given by the 
 Independent Particle Model (IPM), to interpret the magic numbers, let us 
start from the  harmonic oscillator (HO) model. For an $A$ 
identical-particle system, the HO potential, neglecting 
the  parameter defining the strength, is proportional to the 
sum of the squares of the linear coordinates of the particles, 
$V_{HO}(r)\sim \sum_1^A x_i^2$. 
This sum,  when expressed in the  center of mass frame in polar coordinates 
in the $D=3(A-1)$ dimensional space spanned by the particle coordinates 
${\vec x}_i$ ($i=1,\cdots, A$), is nothing else but the radial coordinate, 
called hyperradius. The other space variables are the angular coordinates 
$\Omega$. Let $\Omega$ and $r$ be the polar coordinates fixing the position of
the particles. For any potential $V(r)$ which depends only on the radial 
coordinate  $r$,  such as for the  HO potential where $V(r)\sim r^2$,  the 
rotational and vibrational motion described by  a function of  the radial 
coordinate are  independent. The rotational motion in the physical 
three-dimensional space is described by spherical harmonics which 
are harmonic polynomials (HP) when the radial coordinate is 
equal to unity. In the $D$--dimensional space, the rotations 
are similarly described by hyperspherical harmonics which are HPs 
when the hyperradius is equal to one.  These harmonics
are  independent of the shape of the potential $V(r)$ since 
the rotational and vibrational motions are independent.  
Nevertheless, the speed of the 
rotation enters into  the radial equation through the repulsive centrifugal 
barrier  $\Lc(\Lc+1)/r^2$, $\Lc=L+(D-3)/2$, where  $L$ is the 
degree of the HP defining the state in the $D$--dimensional space.\par
Whatever the potential $V(r)$ is, the ground state is obtained when the 
repulsive centrifugal barrier, {\em i.e}, the degree of the HP defining the 
state, is the smallest. For bosons where all particles can be in the same
$S$-state, the degree is $L_m=0$. For identical fermions where  two 
particles cannot be in the same state, the wave function, which 
must be antisymmetric in the exchange  of two particles, 
is  the product  of an antisymmetric HP and a function
of the hyperradius $r$. This HP is a Slater determinant constructed
from  the individual polynomials 
     $s^j_i t^j_i x_i^{2n_j+\ell_j} Y^{m_j}_{\ell_j}(\om_i)$
where $s$ and $t$ are the spin and isospin  states and $i$  refers to the  
rows and $j$ to the columns for $i$ and $j$  running from $1,\cdots,A$.
 The degree of this HP is the smallest when all the $m_j$  quantum
 numbers are used for the smallest possible values of $2n_j+\ell_j$.
 It corresponds, as we shall show in the next section, to a HO Slater 
 determinant in ground state and, therefore,
 to the filling of the states giving
 the minimum energy for the HO potential leading 
 to the shell model. From this analysis it turns out that 
 the ``{\it  average field of force $V(r)$ of spherical
 symmetry}''  Goeppert-Mayer and Jensen speak about \cite{JM}, which
 ``{\it represents the action of the other nucleons}''
 is that part of the nuclear potential, called hypercentral, which
 is invariant by rotation in the $D$-space and thus a function of the
 hyperradius only.\par
 For the spin-orbit force one uses the property that a sum over 
 all pairs of pairwise spin-orbit operators is equivalent to a sum 
 over all particles of one--body spin-orbit operators \cite{deShalit}.
 When the spin-orbit interaction is the product of a hypercentral
 potential and a sum of pairwise spin-orbit operators, then this
is equivalent 
to the product of a hypercentral potential and a sum of one-body spin-orbit 
operators  \cite{JM}, but this time the potential is not the one of an 
independent particle model but a function of the collective  radial 
coordinate, {\em i.e}, the hyperradius. The solution obtained from the 
wave equation where the potential  is purely hypercentral does not contain 
any correlations since $r$ is  a collective variable that describes 
only monopolar  excitations when the radial wave function has nodes.
The wave function has therefore to be improved  when we have to deal 
with  two-body potentials generating correlations.\par
In this paper a method for introducing the correlations generated by the
nuclear potential is proposed.  It leads to a system of coupled 
integro-differential equations taking the various operators of 
the potential into account.  It is an extension to many bodies
of accurate methods used for studying the few-body systems.\par
In Sect. II we  briefly describe the   antisymmetric harmonic polynomials
and in Sect. III the handling of the nuclear problem that neglects 
two-body correlations namely via the hypercentral approximation and 
the spin-orbit force. In Sect. IV  we discuss the Integro-differential 
equation formalism while in Sect. V  we describe  in details  the  
construction of the so-called weight and pseudo-weight functions 
associated with exchange operators followed by a  
description of the projection function in 
Sect. VI. The use of nuclear potentials is described in Sect. VII while 
a brief discussion concerning  spin-isospin exchange generated elements 
is given in Sect. VIII. Our results are given in Sect. IX and our 
conclusions in Sect. X. Finally, details on the coordinates, and on the
various coefficients needed in the formalism are given in  appendices A, 
and B.
%ccccccccccccccccccccccccccccccccccccccccccccccccccccccccccc  
%               secII.tex 
%cccccccccccccccccccccccccccccccccccccccccccccccccccccccccc
 \section{Antisymmetric Harmonic Polynomials}
 %%%%%%%%%%%%%%%%%%%%%%%%%%%%%%%%%%%%%%%%%%%%
 Any analytical function, including the wave function of a many-body system,
 can be expanded in terms of  HPs or equivalently
 in terms of Hyperspherical Harmonics (HHs) multiplied by coefficients 
 depending only on the hyperradius.  Any such expansion starts from a 
 HP of minimal degree $L_m$  called ground polynomial.  When 
 we have to deal with a hypercentral potential, {\em i.e},
 with  a potential invariant by  rotation in the $D$--dimensional 
space spanned by the 
 coordinates of the particles of an $A$--body system in the center of mass 
 frame, this polynomial describes
 a ground state when the repulsive centrifugal barrier is minimal.  When
 the potential is not hypercentral, the product of the potential with
a ground  polynomial generates new harmonic polynomials of higher degree.\par
 For the ground state and low excited states we expect, as was  
 found for medium light nuclei, that the ground polynomial brings the largest
 contribution to the HH expansion of the wave function of the 
 state under investigation.
 Since the other polynomials are obtained by starting from the ground 
 polynomial, we define the state by the quantum numbers defining the
 ground polynomial.  Ground polynomials for identical
 fermions must be antisymmetric in any exchange of two particles.
  In order to construct a ground polynomial suitable for
 describing a system of $A$ fermions, we introduce homogeneous polynomials
 \begin{equation}
	 \phi_i(n,\ell,m)=x^{2n+\ell}_i Y_{\ell\, m} (\omega_i)
\label{phiex} 
 \end{equation}
 of degree $2n+\ell$ in terms of the polar coordinates 
 $(x_i,\omega_i)$ of the $i$th particle  where $i=1,\cdots,A$.\par
 In nuclear physics each nucleon $i$ has a spin $s$ and an isospin $t$ 
and thus to each $\phi_i(n,\ell,m)$ polynomial  a spin-isospin state
 $s\,t$ is associated. 
 Let us construct  a Slater determinant with the individual states
 $\phi_i(n_j,\ell_j,m_j) s_i^j t^j_i$ where $i$ defines the
 rows and $j$ the columns for $i$ and $j$ running from 1 to $A$
\begin{equation}
	 D=\parallel s^j_i t^j_i \phi_i (n_j, \ell_j, m_j)\parallel
%	   \qquad i,j=1,\cdots,A
\label{sla1}
 \end{equation}
 If for each  independent state ($s,t,\ell,m$) the determinant is 
 constructed by using all the $n$ from $n=0$ to $n_{\rm max}(s,t,\ell,m)$,
 where $n_{\rm max}$ is defined independently for each $s,t,\ell,m$, then 
 this determinant is a harmonic polynomial.  Indeed, if the Laplace operator 
 $$
	 \Delta=\sum^A_{i=1}\Delta_i
 $$ 
 is applied to any column it decrease by 2 the degree of the 
 homogeneous polynomial $D$ without changing the quantum numbers 
($s,t,\ell,m$).  
 Since $D$ has been constructed in such a way as to  generate
 the lowest degree  polynomial for the selected ($s,t,\ell,m$)
quantum numbers, such a polynomial does not exist. Thus 
	 $\Delta D=0$
 and the antisymmetric homogeneous  polynomial $D$ is harmonic.\par
 Substituting  the Laguerre polynomial 
	 $L^{\ell+1/2}_n(x_i^2)$ for $x_i^{2n}$ in $\phi_i$ 
 one gets a new determinant 
 $$
	 D=\parallel s^j_i t^j_i x_i^{\ell_j}Y_{\ell_j}^{m_j}(\om_j)
	      L^{\ell+1/2}_n(x_i^2)\parallel\,,
	    \qquad i,j=1,\cdots,A     
 $$
 If the same procedure of selecting the quantum numbers is applied,
then   according to the rule in which a determinant with
 proportional columns disappears, the new determinant, 
except for a normalization constant, is identical to the 
 previous one.  This new determinant multiplied by an
 exponential
 $$
	 D\, {\rm exp}(-\sum^A_1 x^2_i/2)
 $$
 is a harmonic oscillator Slater determinant \cite{FN79a}. 
 This property will enable  us to calculate two-body matrix elements 
by using HO Slater determinants.\par
 Instead of using spin-isospin and HO quantum numbers 
 $n,\ell,m_\ell$ one can  as well combine the spin $s$ and 
 angular momentum $\ell$ to generate
 a total angular momentum $j=\ell\pm\frac{1}{2}$ defined for each  particle.
 Then the state of each nucleon is defined by the isospin $t$ and 
the quantum numbers $n,\ell,j,m_j$ where $m_j$ is the 
projection of the total angular momentum $j$.
 This  description will be used for open shell nuclei where the 
 spin-orbit interaction contributes significantly
 to the binding energy.\par
The removal of excitations of the center of mass motion occurring in the
product of individual wave functions, is a difficult problem in the case of
an IPM. As discussed by Talmi (\cite{Talmi}, p. 51) {\em
$``\cdots $ it is only in the case of the HO potential that
eigenstates can be written as product of an intrinsic function and a
function of center of mass coordinates $\cdots$. Removing the 
contribution of the center of mass motion can be done only 
approximately $\cdots$.  It is still important to realize that such 
corrections are the price we pay for the great convenience
of using shell model wave functions of nucleons moving independently
in a potential well.''} In the hyperspherical model the expansion
of wave function starts from a ground HP translationally
invariant. Indeed, by substitution of $\vec x_i-\vec X$ 
($\vec X$ for the
center of mass) for $x_i$ in Eq. (\ref{phiex})
and according to the properties of  determinants, Eq. (\ref{sla1}),
constructed according to the rule explained above, the $\vec X$ 
dependence disappears. Since the next terms to be included in the
wave function are correlations described by
functions of the relative coordinates $\vec \rij=\vec x_i-\vec x_j$
the overall wave function is translationally invariant and the
center of mass problem is solved.\par
Ground states are defined by ground HP where all available nucleons
eigenstates of the shells are occupied  up to the last one which, 
except for closed shell nuclei, are not fully filled. Excited 
states where one or several nucleons of the last shell are raised 
to the next shell are still described by a ground HP. But deep hole
states where a nucleon is extracted from the core and
raised to a higher shell, does not generate a ground HP and
therefore such states cannot be described by our scheme.  
This results agree with the remark of Talmi (\cite{Talmi}, p. 7) that 
{\em ``$\cdots$ simple shell model states, like single nucleon or single hole
states, if they lie at sufficient high energy, are not pure single hole
states of deep-lying orbits have high excitations and may no
longer be accurately  described in terms of a single  hole wave 
functions, the deeper the state of the missing nucleon, the higher 
the excitation energy of the hole state. The fragmentation  of single 
nucleon or single hole states  at higher excitations impose practical 
limitations to the simple shell model picture.''}
%
%ccccccccccccccccccccccccccccccccccccccccccccccccc
%                  secIII.tex
%ccccccccccccccccccccccccccccccccccccccccccccccccc
 \section{The Hypercentral  Approximation}
 %%%%%%%%%%%%%%%%%%%%%%%%%%%%%%%%%%
\label{SHCA}
Let us start the description of our formalism by
considering the case where correlations are ignored in the 
 study of the $A$-body system, {\em i.e}, by discussing the 
 hypercentral model first. The HO potential, neglecting the HO parameter,  is 
 \begin{equation}
	 V_{HO}(\vec x)=\sum^A_{i=1}x_i^2=\sum^A_{i=1}(\vec x_i-\vec X)^2+AX^2
 \label{VHO} 
  \end{equation}
 where $\vec X$ is the center of mass. 
 The last term in the right-hand-side is irrelevant for isolated systems
and therefore can be omitted. 
The other term is the sum of the square of the radial coordinates
of the  $D=3(A-1)$ dimensional space spanned by the coordinates 
$\vec x_i$, $i=1,\cdots,A$,  of the A-particles in the center 
of mass frame expressed  in polar coordinates $(r,\Omega)$. 
This sum defines the hyperradius $r$,
\begin{equation}
       r^2=2\sum^A_{i=1}(\vec x_i-\vec X)^2
           =\frac{2}{A} \sum_{i,j=i}^A(\vec x_i-\vec x_j)^2\,,
\label{hypr}
 \end{equation}
where the factor 2 is introduced so that for two particles 
 $r^2=(\vec x_1-\vec x_2)^2$. \par
 The hyperradius is invariant by rotation in the $D$-dimensional space.  
Here `hyper' refers to a  $D$-dimensional space with $D>3$.  Since
 the HO potential is invariant by rotation in the $D$-space one 
may write the  Schr\"odinger equation in polar coordinates 
$(r,\Omega)$ where $\Omega$ is the set of $D-1$ angular coordinates 
on the unit hypersphere ($r=1$) in the  $D$-space which together 
with $r$ fixes the coordinates $\vec x(\vec x_1, \cdots, \vec x_A)$ 
of the particles. Let $V(r)$ be a potential  invariant by rotation 
in the $D$-space. The Schr\"odinger equation,
 \begin{equation}
	 \left\{-\frac{\hbar^2}{2m}
	      \sum^A_{i=1}\Delta_i+V(r)-E\right\}\Psi=0 \,,
 \label{SchrE} 
\end{equation}
 in polar coordinate becomes 
 \begin{equation}
	 \left\{-\hh\left[\frac{1}{r^{D-1}}\frac{\p}{\p r}
	   r^{D-1}\frac{\p}{\p r}+\frac{L^2(\Omega)}
	 {r^2}\right]+V(r)-E\right\}\Psi=0
 \label{SchrP} 
 \end{equation}
 where $L^2(\Omega)$ is the square of the grand orbital operator
 which is a  generalization    of the orbital momentum $\ell(\omega)$ in the 
3-dimensional space  to the D-dimensional space.
It should be noted that in Eq. (\ref{SchrP}), 
the vibrational and  rotational motions are independent.\par
Let  $H_L(\vec x)$ 
 be a HP of degree $L$.  Since it is a  homogeneous polynomial,
 the $r$-dependence can be factorized out, $H_L(\vec x)=r^L H_L(\Omega)$.
 The $H_L(\Omega)$ is a HH, {\em i.e.}, the value of $H_L(\vec x)$ 
 on the unit hypersphere $r=1$.
 From the assumption that $H_L(\vec x)$ is a HP, 
 with the definition of $\Delta$ in
 terms of $r$ and $\Omega$ given in Eq. (\ref{SchrP}), one obtains
 \begin{equation}
	 [L^2(\Omega)+L(L+D-2)]H_L(\Omega)=0\,.
 \label{Leq} 
  \end{equation}
 The wave function $\Psi$ can be written as a product of a HH and a 
hyper-radial function, 
 \begin{equation}
	 \Psi=\frac{1}{r^{(D-1)/2}}\,u_L(r) H_\Ll (\Omega)
 \label{Psi}
  \end{equation}
  where the hyper-radial function $u_L(r)$ is a solution of
 \begin{equation}
	 \left\{\hh\left[-\frac{\rd^2}{\rd r^2}+\LLr
	 \right]+V(r)-E\right\} u_L(r)=0
 \label{ueq} 
  \end{equation}
 with $\Lc=L+(D-3)/2$ and where $\Ll$ stands for the set of quantum
 numbers including spin and isospin defining the state of grand 
orbital $L$.\par
 Whatever  the potential $V(r)$ is, the ground 
 state is obtained when the repulsive
 central barrier is the smallest.  It corresponds to a HP of minimum degree 
 $L_m$.  For a system of bosons where all particles can
 be in the $S$-state this degree is $L_m=0$. For fermions, $L$ is 
 the smallest when  $H_{L_m}(\vec x)$
 is a Slater determinant constructed from the individual polynomials
 $ x_i^{2n_j+\ell_j}Y^{m_j}_{\ell_j}(\omega_i)$, where $i$ refers to the 
 rows and $j$ to the 
 columns, when all the $m_j$ are used for the smallest possible values of
 $2n_j+\ell_j$.  As we explained in the previous section, 
this corresponds  to a Slater determinant for a HO in a  ground state.  
 The minimum degree is the sum over the degree of the individual 
 polynomials in $H_\Lm(\vec x)$ which is, as we have 
 seen in the previous section, the sum over the radial and 
 orbital quantum numbers of all the HO occupied states,
 \begin{equation}
	 L_m=\sum^A_{j=1}(2n_j+\ell_j)\,.
 \label{Lm}
  \end{equation}
 The $L_m$  is related only to the rotation in the $D$-space and
is independent of the shape of the hypercentral potential
 $V(r)$.
% Therefore, it is not surprising that it is precisely the one given by
% a harmonic oscillator where $V(r)=c r^2$.
  It is thus clear that in the hyperspherical scheme the
 {\em ``average field of force $V(r)$ of spherical symmetry''}
  Mayer and Jensen speak about, which  {\em "represents the action of the
 other nucleons"} , is the hypercentral potential generated by the two body
 nuclear potential $V(\vec r_{ij})$.  This average potential is given by 
 \begin{equation}
	 V(r)=\frac{A(A-1)}{2}\int H^*_\Lm(\Omega)V(\vec r_{ij})H_\Lm
	 (\Omega)\rd\Omega
 \label{Vr}
  \end{equation}
 where $ \vec r_{ij}=\vec x_i-\vec x_j$ and $V(\vec r_{ij})$ 
 contains the spin and isospin exchange operators and
 in general all the components of the nuclear interaction $V(\vec r_{ij})$
 and where 
 \begin{equation}
	 \int| H_\Lm (\Omega)|^2\rd\Omega=1\,.
 \label{HHn}
  \end{equation}
 It depends on the collective variable $r$ and does not generate 
any correlations  since when $r$ is fixed, a variation in 
the distance $\vec r_{ij}$ between two
 particles does not affect the hypercentral potential and the
 corresponding  wave function.\par
 To investigate the effect of a spin-orbit force we recall that 
 it operates  only on the last  open shells.  Following the derivation
 of de Shalit and Talmi we  write for the  spin-orbit operator for the pair 
 $(i,j)$, $(\vec\sigma_i+\vec\sigma_j)\cdot \vec\ell_{ij}$ where
 $$
   \vec\ell_{ij}=\frac{1}{2}
	 (\vec x_i-\vec x_j)\times(\vec p_i-\vec p_j)
 =
 \frac{1}{2}
	 (\vec \ell_i+\vec \ell_j)-\frac{1}{2}\left[
	       \vec x_j\times\vec p_i+
	 \vec x_i\times\vec p_j\right]
 $$
 After elimination of the center of mass, the sum over all pairs of the 
 spin--orbit operator gives 
 \begin{equation}
	 \sum_{i,j>i}(\vec\sigma_i+\vec\sigma_j)\cdot\vec\ell_{ij}
	  =A\sum^A_{i=1}
	 \vec\ell_i\cdot\vec s_i+\vec S\cdot\vec L
 \label{lso}
 \end{equation}
 where
 $$
  \vec s_i= \vec \sigma_i/2,\qquad 	\vec S= \sum^A_{i=1}\vec s_i,
\qquad \vec L=\sum^A_{i=1}\vec \ell_i\,.
 $$
 The first term in the right hand side of Eq. (\ref{lso}) 
 is the sum of one body spin-orbit operator occurring in the
 standard shell model and the last one, a collective spin-orbit 
 operator, is a small perturbation.\par
 In the $jj$-coupling scheme, the spin-orbit operator does not modify the
 structure of the Slater determinant defining the state and thus one has
 $$
	 \sum_{i,j>i}(\sigma_i+\sigma_j)\cdot\vec\ell_{ij}
	 \,H_\Ll (\Omega)=C_{LS}H_\Ll (\Omega)\,.
 $$
 The coefficient $C_{LS}$ is determined by the filling of the shell of
 the determinant in the $jj$-coupling.  \par
 The hypercentral potential, including the spin-orbit force 
 $V_{\ell s}(r_{ij})$\ $(\vec \sigma_i+~\vec \sigma_j)\cdot~\vec\ell_{ij}$
 operating on a Slater determinant in the $jj$-scheme, is $
	 V(r)+C_{LS}V_{LS}(r)
 $.
 The ground state results from a competition between the repulsive centrifugal
 barrier  $\hbar^2/m\, \Lc(\Lc+1)/r^2$ and the potential including 
 the effect of the spin-orbit force
 through the coefficient $C_{LS}$ fixed by the filling of the individual
 states in the last shell.\par
 In summary the quantum numbers defining the ground state are those of the
 HO in ground state for which the repulsive central barrier is the smallest
 corrected by the effect of the spin--orbit force.  It is the model proposed
 by Goeppert-Mayer and Jensen but this time the potential is not a one body
 potential in which all nucleons move but a potential of the collective 
 symmetrical hyper-radial coordinate whose shape should be deduced from the 
 realistic nucleon-nucleon potential.
 More information about the Hypercentral Approximation (HCA) where only the
 HH of minimal order $L_m$ is taken into account in the wave function can be 
 found in Refs. \cite{FN79a,F83,FFS89}.
We only note here that  for a two-body potential $V(\rij)$, the
HCA provides a ground state binding energy for nuclei a little
 stronger  but nearly identical to the one given by a variational 
calculation using a HO Slater determinant where the strength parameter
is adjusted to give the lowest eigen-energy \cite{FN79a}. 
The HCA  variationally
 provides an upper bound to the eigen-energy. A further improvement
in the solution can be achieved with the inclusion of  the two-body
correlations generated by the two-body potential.

%
%ccccccccccccccccccccccccccccccccccccccccccccccccccc
%                 secIV
%ccccccccccccccccccccccccccccccccccccccccccccccccccc
 \section{Integro-differential Equations Approach}
 %%%%%%%%%%%%%%%%%%%%%%%%%%%%%%%%%%%%%%%%%%%%%%%%%
%
 For nuclear potentials with a strong repulsive core, the 
 Hartree-Fock method, which reduces the interaction to  a sum of
 individual potentials leading to an IPM, gives very poor results 
 because the residual interaction responsible for the correlations 
 is large and  is not taken into account in the procedure. A Jastrow 
 function might be added to take care of the correlations.
 In the method developed in this paper, one operates 
 in the $D$-dimensional space spanned by the Jacobi coordinates 
 and uses a radial coordinate system ($r,\Omega$) where $r$ is
 the hyperradius   and $\Omega$ stands for the angular coordinates at 
the surface of the unit  hypersphere $r=1$.\par
 When  we have to deal with the one-body problem  in
 polar coordinates and  the potential in the physical
 three-dimensional space is not  central, the wave function
 can be expanded in terms of spherical harmonics and the 
 Schr\"odinger equation, projected on the spherical harmonics basis,
 generates a system of radial, coupled, second order, 
 differential equations that have to be integrated to solve the problem.
 When we have to deal with many-body systems, treated
 in polar coordinates in the $D$-dimensional space spanned by the the
 coordinates of the particles, we have the same situation 
 as for the one-body problem, except that in this
 case the space is larger.
 The  interactions occurring in many body systems, for instance a 
 sum of two-body  potentials,  are generally 
 not invariant by rotation in  the $D$-dimensional space, 
 and thus  we have  to deal
 in this space with a deformed potential. The expansion method
 can be applied in which HH  must be substituted for Spherical
 Harmonics.  Like for one  particle in a deformed potential, the 
 Schr\"odinger equation is transformed into a system of
 coupled, second order, differential equations in the hyperradius $r$.
 In this procedure we are facing a different problem namely 
 that of degeneracy.
 While   the ($ 2\ell+1 $)-degeneracy of the spherical harmonics
  for each orbital $\ell$ is moderate,
  leading to the use of a rather small number of significant 
 terms in the expansion in the  three-dimensional space, in contrast 
 the degeneracy of the HH basis  for a  grand orbital $L$ increases 
rapidly with the dimension $D$ of the space leading to an 
intractable large number of  significant coupled equations.\par
 A partial solution to this problem has been obtained by selecting 
 the HH describing only two-body correlations, namely, the
 Potential Harmonics (PH) \cite{F83,F87}. With this restricted basis,
 a good  approximation  can be achieved  for the solution of the  
 few-body problem. The rate of convergence can be improved by 
 introducing functions limiting the number of significant 
 coupled equations \cite{Viviani}. But for more particles, the  number
 of coupled  equations to be solved to obtain a good 
 accuracy  becomes again  too large \cite{HHE}. \par
To overcome this  difficulty the Integro-Differential Equation 
Approach (IDEA)  has been proposed 
\cite{F84,FFS87,FFS88,FFS89,OSFF91a,OSFF91b} in which
the  two-body correlations are taken into account. 
In the IDEA method, the Schr\"odinger equation is  transformed 
into two-variable integro-differential equations,
 whatever the number of particles is.
The approximation is justified in saturated systems, like nuclei,
 by the rather low kinetic energy occurring between two particles
 because the volume occupied by the system increases proportionally to
the  number of particles. At this low kinetic energy, the pairs  
are mainly in  $S$-state  and many-body correlations are expected 
to be rather small. 
 It should be noted that the  IDEA equation is identical
 to the Faddeev equation for three-particles when the interaction 
 operates on pairs  in an $S$-state only.\par
The capability of the IDEA to solve the Schr\"odinger equations 
for nuclei  arises from the large component of the potential 
which is invariant by rotation in the D-space,  {\em i.e}, 
from the hypercentral potential. The contribution of  
this potential to the binding energy balances the one of  the kinetic 
energy in such a way that the binding energy in nuclei is mainly provided 
by correlations when a strong repulsive core (that give rise to 
large correlations) exists in the potential. 
In other words, we agree with Talmi \cite{Talmi} that {\em ``$\cdots$
the relevance to nuclear many-body theory is in  realizing that 
two-nucleon effective interactions are all that is necessary 
to calculate nuclear energies. 
 The two-nucleon matrix elements 
obtained by this procedure determines the structure of nuclei 
 in their ground state and at low excitations.''}
In what follows we shall describe the IDEA formalism that 
takes into account  two-body correlations and discuss 
various aspects of it. 
\subsection{Two-Body Correlations}
%%%%%%%%%%%%%%%%%%%%%%%%%%%%%%%%%%%%%%%%
We assume that we are dealing  with a system of $A$ 
identical fermions of mass $m$, in particular nucleons. 
Our aim is to solve the Schr\"odinger equation
 \begin{equation}
		   (T+V(\vec x)-E)\Psi(\alpha,\vec x)=0
 \label{Schr}
 \end{equation}
 where $\vec x=(\vec x_1,\cdots,\vec x_A)$ are the particle 
coordinates of an $A$-body system, $\alpha$ denotes the space 
independent  degrees of freedom like spin and isospin,   $T$ is 
the kinetic  energy operator, $V(\vec x)$ is the interaction 
potential, and   $E$  is the energy state of the system.
 The wave function can be expanded in terms of harmonic polynomials
 and the expansion begins with a ground polynomial  $H_\Lm(\vec x)$
 of degree $L_m$ for the state under consideration,
 \begin{equation}
\hspace*{-6mm}	   \Psi(\alpha,\vec x)=\sum_{L=L_m}^\infty  
		       \sum_{\Ll} H_\Ll(\vec x)\, u_\Ll(r)
	 =\sum_{L=L_m}^\infty  
		       \sum_{\Ll} H_\Ll(\Omega)\, r^L u_\Ll(r) \
 \label{psiax}
 \end{equation}
 where $\Ll$ is the set of quantum numbers, including the space independent 
 degrees of freedom $\alpha$, defining $H_\Ll(\vec x)$. The notation 
 $H_\Ll(\Omega)=H_\Ll(\vec x)/r^L$
 is used for the associated Hyperspherical Harmonics (HH) which
 is the value of $H_\Ll(\vec x)$ on the unit hypersphere $r=1$.
 Antisymmetric Harmonic polynomials are denoted by
 $D_\Ll(\vec x)$ (to remind us that they are constructed as Slater 
 determinants) and the associated HH by $D_\Ll(\Omega)$.\par
 To solve the Schr\"odinger Eq. (\ref{Schr})
 one needs to specify the properties of the interaction. When the
 interaction is a sum of one-body potentials 
	$	V(\vec x)=\sum_{i=1}^A\,V(\vec x_i)$
 we have to deal with an Independent Particle Model (IPM) and
 the wave   function is a product of individual states eigenfunctions
 of the one-body problem. When the potential is a many-body one,
 various methods have been proposed to obtain the solution of 
 Eq. (\ref{Schr}). In 
 the most popular one,  the state is described by the product of a
 Slater determinant constructed with individual states adequately
 chosen and  a Jastrow function  which is the product of two-body 
 functions $\prod_{i,j>i}f(r_{ij})$ for describing two-body correlations
 originating from a two-body potential.
 For most  systems of identical particles like nuclei, the dominant 
 part of the interaction is a sum of two-body potentials. The three-body 
 potentials occurring next, can be reduced, except for a small 
 contribution, to a sum of two-body potentials as well \cite{v3p}. Thus in the 
 present work we concentrate  to the case where the interaction
 is a sum of two-body potentials only.\par
In order to understand the structure of the solution,  one   starts 
from the HCA (discussed in Sect. \ref{SHCA})
where only the ground harmonic, {\em i.e}, only the first 
term in the HH expansion of the wave function is taken into account.
In this case, the  wave function,  in the center of mass 
coordinate, is the product of a HH and a radial function
 \begin{equation}
	 \Psi(\alpha,\vec x)=\Psi_0(r,\Omega)=D_\Lm (\Omega)
	   \frac{1}{r^{(D-1)/2)}}u_0(r)
 \label{psi0}
 \end{equation}
 where $D_\Lm (\Omega)$ is the ground
 antisymmetric HH for the investigated state.\par
A projection of   Eq. (\ref{Schr}) on $D_\Lm(\Om)$ for the
wave function given by Eq. (\ref{psi0}), results in
the radial equation 
 \begin{equation}
		 \left[\hh \left\{ -\frac{{\rm d}^2}{{\rm d}r^2}
	+\LLr\right\}  +V_\Lm(r)-E\right]u_0(r)=0
 \label{u0h}
 \end{equation}
where $\Lc=L_m+(D-3)/2$.
 The hypercentral potential $V_\Lm(r)$ is given by the integral
 \begin{equation}
		 V_\Lm(r)=\int D^*_\Lm(\Om)V(r,\Om)D_\Lm(\Om)\,{\rm d}\Om
 \label{VLm}
 \end{equation}
 in terms of ($r,\Om$), the polar coordinates of $\vec x$, taken over
 the surface of the unit hypersphere $r=1$.  In the three-dimensional 
 space it would be the central part of the potential.
When $ V(r,\Om)$ is a sum of two-body potentials over all pairs and 
$D_\Lm(\Om)\equiv H_\Lm(\Om)$ is normalized according to Eq. (\ref{HHn})
then Eq. (\ref{Vr}) holds.
 For the infinite hyperspherical well, the Coulomb,
 and the Harmonic Oscillator potentials the radial solutions
 $u_0(r)$ are  known analytically \cite{FN79a}. 
 In the last case the Schr\"odinger  equation  can be solved 
 either as a collective or as an independent particle  model.\par
 To proceed beyond the HCA and obtain  a  solution of Eq. (\ref{Schr})
 with a better accuracy, one writes
 \begin{equation}
	 \Psi(\alpha,\vec x)=\Psi(\vec x)= \Psi_0(r,\Om)+\Psi_1(r,\Om) 
 \label{psi01}
 \end{equation}
 where $\Psi_1(r,\Om)$ is the next improvement in the wave function 
 which  depends on the structure of the potential. Here we 
 are interested  for potentials which can be written as
 a a sum of two-body interactions
 \begin{equation}
	 V(\vec x)=\sum_{i<j}V(\vec \rr)\,, \qquad 
	  \vec \rr=\vec x_i-\vec x_j
 \label{vsum}
 \end{equation}
 Using  Eq. (\ref{psi01}), Eq. (\ref{Schr}) becomes
 \begin{equation}
	 (T-E)(\Psi_0+\Psi_1)=\sum_{i<j}V(\vec \rr)(\Psi_0+\Psi_1)
 \label{Schr1}
 \end{equation}
 Omission of the correction term on the right hand side, implies 
 a structure of $\Psi_1$ of the form
 \begin{equation}
		 \Psi_1=\Psi_0\sum_{i<j}F(\vec \rr,r)
 \label{psi1}
 \end{equation}
 where $F(\vec \rr,r)$ is a two body  amplitude. Thus, for
 nuclei with central  potentials we may write
 \begin{equation}
		 \Psi(\vec x)=D_\Lm(\Om)\,\sum_{i<j}F(\rr,r)
 \label{Psic}
 \end{equation}
 in terms of the two-body amplitude $F(\rr,r)$ where 
 the pair is assumed to be in an $S$-state.\par
 In order to find the amplitude $F(\rr,r)$ we have two options: 
 Either to solve the amplitude equation 
 \begin{equation}
   \hspace*{-.5cm}   (T-E)D_\Lm(\Om)F(\rr,r)=-V(\rr)D_\Lm(\Om)
             \sum_{\ell,\, k<\ell}F(r_{kl},r)
 \label{Feq}
 \end{equation}
 or to solve the Schr\"odinger equation 
 \begin{equation}
	 (T-E)\Psi(\vec x)=-\sum_{j,\,i<j}V(\rr)\Psi(\vec x)
 \label{Schr2}
 \end{equation}
 where now $\Psi$ is given by Eq (\ref{Psic}). The first option, 
 Eq. (\ref{Feq}), where the sum over all pairs reproduces the 
 Schr\"odinger equation, leads to the IDEA 
 \cite{F84,FFS87,FFS88,FFS89,OSFF91a,OSFF91b},  while the second 
 leads to the Variational Integro-Differential Equation (VIDE)
 \cite{vide0,vide1}. It should be noted that for three-body the 
 IDEA becomes a  Faddeev-type equation while the VIDE
 does not have any counterpart. 
 The solutions are not the same, as the VIDE is variational 
 while the IDEA is not, but the latter is easier to solve.
 When  the potential $V(\rr)$ operates only on pairs
 in $S$-state the Faddeev equation provides the exact 
 solution but when the potential is local and operates on all orbitals,
 the VIDE gives far more accurate solutions. In  order to obtain 
 the same accuracy, a few coupled Faddeev equations have to be solved. 
 In both cases the wave function has the structure of a sum of 
 two-body amplitudes.\par
 The correlations are described by a  product when  Jastrow functions 
are used. However, the  product structure of the  Jastrow function 
commonly   used, does not have the correct structure of the solution for three
or more particles with an $S$-projected interaction.  In contrast, in 
the IDEA formalism  we only assume that  a sum of  two-body amplitudes 
solution of  Eq. (\ref{Feq}) is sufficient to  provide  us with a good 
solution although   many-body correlations  are not taken into account.  
\subsection{Reduction of the  IDEA Equation}
 %%%%%%%%%%%%%%%%%%%%%%%%%%%%%%%%%%%%%%%%%%%%%%
 We assume that the wave function is given by Eq. (\ref{Psic}),
{\em i.e}, by  a product of an antisymmetric
 HH and a symmetrized two-body amplitudes.
 Our aim is to solve Eq. (\ref{Feq}) to obtain the two-body 
 amplitude $F(\rr,r)$ associated with the ground
 polynomial $D_\Lm(\Om)$ defined by the occupied states in a Slater
 determinant. In order to have an equation for
 $\rr$ only, for instance for a reference pair $\vec \rij=\vec \xi_N$
with $N=A-1$, $i=1$, and $j=2$,  we have to eliminate the part which in 
$F(r_{k\ell},r)$ depends on  the other Jacobi variables 
$\vec \xi_i$, $i<N$, which are related to many-body correlations 
(For the A-particle coordinates in the center of mass frame,
see Appendix A). For this purpose, one 
 projects $F(\rr,r)$ for $(i,j)\neq(1,2)$ on the pair $(1,2)$ in
 an $S$-state. Let us call ${\cal P}_c$  the operator projecting
 the connected pairs like $(1,i)$ and $(2,i)$, $i=3,\cdots, A$ and
 ${\cal P}_d$  the  disconnected pairs like for $i$ and $j>2$.
 Then the total projection operator is
 \begin{equation}
	 {\cal P}^0=2(A-2){\cal P}^0_c+\frac{(A-2)(A-3)}{2}{\cal P}^0_d
 \label{P0}
 \end{equation}
 where ${\cal P}^0_c$ operates on one connected pair, e.g. the pair 
$(2,3)$,  and ${\cal P}^0_d$
 on one disconnected  pair, e.g. the pair $(3,4)$.
 Since our aim is to calculate correlations, we isolate on the 
 left hand side the hypercentral part of the potential and in 
 the right hand side only that part of the potential generating 
 correlations.  Then  Eq. (\ref{Feq}) becomes
 \begin{eqnarray}
 \nonumber
&&\hspace*{-7mm}
	 \bigg(T\bigg.+\bigg.\frac{A(A-1)}{2}V_\Lm(r)-E\bigg)
               D_\Lm(\Om)F(r_{12},r)
        = - \bigg (V(r_{12})-V_\Lm(r)\bigg)
\\  &&\hspace*{-2mm}
\times D_\Lm(\Om)\bigg\{ F(r_{12},r)\bigg.
       +
       \left. (A-2)\left[
		2 {\cal P}^0_cF(r_{23},r)+\frac{A-3}{2}
		{\cal P}^0_dF(r_{34},r)\right]\right\}
\label{Feqp}
 \end{eqnarray}
 where the functions within the bracket $\{\ \}$ depend only  
 on $r_{12}$ and $r$. Eq. (\ref{Feqp}) is written for the pair
 $(i,j)=(1,2)$. It is noted that by summing over all pairs 
$(i,j)$ one recovers the  Schr\"odinger equation (\ref{Schr2}).\par
%s
 The two variables equation  in $r$ and $z$    we are  
 looking for, which determine the two-body amplitudes, is obtained 
 by multiplying  Eq. (\ref{Feqp}) at left by $D^*_\Lm(\Om)$
 and by integrating over the surface element  ${\rm d}\Om_{N-1}$ 
(see Appendix A).  First we require the integral
 \begin{equation}
	\int D^*_\Lm(\Om)D_\Lm(\Om)\rd\Om_{N-1}= W_\Lm^{(D)}(z,\om)\,,
 \label{WDL}
 \end{equation}
taken over all angular coordinates $\Om_{N-1}$ excluding 
$ z=\cos2\phi$ with $\rij/r=\cos\phi$ and $\om$,
the angular coordinate of $\vec \rij$,  and thus this integral
 is a function of $z$ and $\om$.
 We  may define the weight function $W_\Lm(z,\om)$ for the state $\Lm$,
 \begin{equation}
	 W_\Lm(z,\om)=W_0(z)W_\Lm^{(D)}(z,\om)\,,
 \label{wlm}
 \end{equation}
where 
\begin{equation}
        W_0(z)=2^{-D/2}(1-z)^{(D-5)/2}(1+z)^{1/2}\,,
\label{W0}
\end{equation}
 in such a way that  for normalized $D_\Lm(\Om)$  (see Eq. \ref{dOm})
 \begin{equation}
	 \int D^*_\Lm(\Om)D_\Lm(\Om)\rd\Om=\int_{-1}^{+1}
		 \,W_\Lm(z,\om)\rd z\rd\om=1
 \label{idd}
 \end{equation}
 where ${\rm d}\Om$ is  the surface element on the
 unit hypersphere  given by ${\rm d}\Om=W_0(z)\dz {\rm d}
 \om{\rm d}\Om_{N-1}$.
 The construction of $W_\Lm(z,\om)$  is discussed in the next section.
 We only note here that for bosons in ground state $\Lm=0$,
 $$
	 D_\Lz(\Om)=Y_{[0]}=\frac{\Gamma(D/2)}{2\pi^{D/2}}\,,
 $$ 
 and $W_{[0]}(z,\om)$ reduces to $W_0(z)$ except for a normalization 
 constant.\par
 To proceed we must calculate the kinetic energy and the projection terms.
For the former term we use the relation  (\ref{lapla}) with the
center of mass at rest omitted,
 $$
	 T=-\hh\Delta_\xi=-\hh\sum_{i=1}^{A-1}\nabla^2_{\xi_i}\,, 
	      \qquad i=1,\cdots,A-1
 $$
 and set
 \begin{equation}
	 F(\xi_N,r)=\frac{1}{r^{(D-1)/2}}P(z,r)\,.
 \label{deff}
 \end{equation}
Therefore, we have to calculate 
 $$
	 \Delta \left[D_{L_m}(\Om) \frac{1}{r^{(D-1)/2}}P(z,r)\right]
 $$
 where 
 $$
	 \Delta\equiv \frac{1}{r^{D-1}}\frac{\p}{\p r}r^{D-1}\frac{\p}{\p r}
	 +\frac{L^2(\Om)}{r^2}
 $$
 with the grand orbital operator being written as
 \begin{eqnarray}
 \nonumber
	 L^2(\Om)&=&\frac{4}{W_\Lz(z)}\frac{\p}{\p z}(1-z^2)W_\Lz(z)
		 \frac{\p}{\p z}+ \frac{2\ell^2(\om)}{1+z}\\
	 &+&\ \ {\rm derivatives\  in}\ \Om_{N-1} \ {\rm coordinates}
 \label{L2Om}
 \end{eqnarray}
 Thus, multiplying Eq. (\ref{Feqp}) from left by $D^*_\Lm(\Om)$, 
integrating  over the surface element ${\rm d}\Om_{N-1}$, and 
taking into account the relation
 $$
	 2D^*_\Lm(\Om)\frac{\p}{\p z}D_{\Lm}(\Om)=\frac{\p}{\p z}
		  \left [D^*_{\Lm}(\Om)D_{\Lm}(\Om)\right]
 $$
 one gets the basic equation \cite{F84,FFS87,FFS88}
 %
% \begin{widetext}
 \begin{eqnarray}
 \nonumber
\left\{\hh\left[\right.\right. &-&\frac{\p^2}{\p r^2}
          +\LLr	 -  \left.\frac{4}{r^2}
	  \frac{1}{W_\Lm(z)}\frac{\p}{\p z}(1-z^2)W_\Lm(z)
	  \frac{\p}{\p z}\right]
\\ \nonumber
& +&\left. \frac{A(A-1)}{2}V_\Lm(r)-E\right\}P(z,r)
\\
 & =&-\left(V(r\sqrt{(1+z)/2})-V_\Lm(r)\right)
	    \left[ P(z,r)+{\cal P}^0P(z,r)\right]
\label{IDEA1} 
 \end{eqnarray}
% \end{widetext}
 %
 where ${\cal P}^0P(z,r)$ is the projection of the two-body 
 amplitudes for all the pairs $(i,j)$ for $(i,j)\ne (1,2)$ 
 on the reference pair $(1,2)$ in $S$-state \cite{FFS88} 
(see forth Sect. \ref{Sproj}).\par
Eq. (\ref{IDEA1}) can be easily solved directly as a two-variable
integro-differential equation to obtain the binding energy
of the nucleus under consideration. However, it is desirable
to use also the adiabatic and the more accurate uncoupled
adiabatic approximations which can provide us not only 
reliable solutions but also the eigen-potentials 
from which further physical information, such as
low energy scattering states, can be extracted. 
\subsection{The Extreme Adiabatic  Approximation}
 %%%%%%%%%%%%%%%%%%%%%%%%%%%%%%%%%%%%%%%%%%%%%%%%
 In most  systems the energy  contained in the rotation in 
 the $D$-dimensional space is very much larger than the one
 in the  radial motion.  Indeed, in nuclei the radial energy of 
the ground state is of the order of half the monopolar
 excitation energy (breathing mode), {\em i.e}, about 10\,MeV,
 compared to the several
 hundreds or thousands of  MeV generated by the rotation
 which can be estimated from the Fermi gas model giving
 the average kinetic energy per particle 
 $E_{\rm kin}/A=28.7 r_0^{-2}$\,fm\,MeV, {\em i.e}, $\sim23.7$\,MeV
 for $r_0=1.1$\,fm .
 This means that  the rotation and the vibration are nearly decoupled
 and thus one can use an adiabatic approximation  
 in which one freezes the $r$-motion and solve the rotational motion
 equation to obtain for each $r$ an eigenpotential that is subsequently
 used in the radial equation \cite{F72}. More specifically, one writes
 $P(z,r)=P_\lambda(z,r)u_\lambda(r)$ and solves
 %
% \begin{widetext}
 \begin{eqnarray}
\nonumber
	&& \left[\frac{4}{r^2}\hh\left\{ 
	  \frac{1}{W_\Lm(z)}\frac{\p}{\p z}(1-z^2)W_\Lm(z)
	  \frac{\p}{\p z}\right\}\right.+
		 \left.U_\lambda(r)\right]P_\lambda(z,r)
\\ && \ \ \ =
	 [V(r\sqrt{(1+z)/2}-V_\Lm(r)]
       \left[ P_\lambda(z,r)+{\cal P}^0P_\lambda(z,r)\right]
 \label{EAA1}
 \end{eqnarray}
% \end{widetext}
 %
 for fixed $r$ to obtain the eigenpotential $U_\lambda(r)$
 which is then used in the radial equation
 \begin{equation}
 \label{urS}
\hspace{-5mm} \left\{\hh\left[-\frac{\rd ^2}{\rd r^2} +\LLr\right]
     + \frac{A(A-1)}{2}V_\Lm(r)+
	 U_\lambda(r)-E\right\}u_\lambda(r)=0
 \nonumber
 \end{equation}
 to obtain the total energy $E$ and the radial function $u_\lambda(r)$
\cite{F84}. The Eq. (\ref{EAA1}) and (\ref{urS}) constitute the
so-called Extreme Adiabatic Approximation (EAA).
\par
The wave function, within this approximation, is obtained
from  $P_\lambda(z,r)$ and $u_\lambda(r)$,
\begin{equation}
\label{wfeaa}
        \Psi_{\rm EAA}(\vec x)=\frac{1}{r^{3A/2-2}}u_\lambda(r)
             \sum_{i<j\le A}P_\lambda(2r^2_{ij}/r^2-1,r)\,.
\end{equation}
 When the hypercentral potential $V_\Lm(r)$ is ignored, in Eqs. 
 (\ref{IDEA1}-\ref{urS})  for $A=3$, one finds the Faddeev equation for
 $S$-state projected potentials \cite{F86,FF86}. For $A>3$ it will be called 
 $S$-State Integro-Differential Equation (SIDE) and should be applied
 when we are dealing with $S$-state projected  potentials since ${\cal P}^0$
 in  Eqs. (\ref{IDEA1}-\ref{urS}) projects any other pair$(i,j)$ on the 
 pair $(1,2)$ in $S$-state.  In the IDEA only the residual interaction 
generating  correlations operates on pairs in $S$-state.
\subsection{The Uncoupled Adiabatic Approximation}
 %%%%%%%%%%%%%%%%%%%%%%%%%%%%%%%%%%%%%%%%%%%%%%%%%%
In order to separate the two-variable integrodifferential
equation (\ref{IDEA1}) into two, one-variable, equations (\ref{EAA1} 
and \ref{urS}) we assumed that the amplitude can be written as a product
 $P(z,r)=P_\lambda(z,r)u_\lambda(r)$ and the EAA is obtained when 
the variation of  $P_\lambda(z,r)$ with respect to $r$ is neglected.
It provides a lower bound to the eigen-energy \cite{BFL82,FF86}.
A further improvement in the binding can be achieved if such a 
variation is taken into account. \par
The new equation is obtained by substitution of  
$P_\lambda(z,r)u_\lambda(r)$  for $u(r)$ in (\ref{urS}) and by taking
the derivatives of $P_\lambda(z,r)$ with respect  to $r$ into account.
One first  normalizes the $P_\lambda(z,r)$ according to
\begin{equation}
\label{Pnorm}
       \langle P_\lambda | P_\lambda\rangle\equiv \int_{-1}^{+1}
           P_\lambda^2(z,r)W_\Lm(z) \rd z=1
\end{equation}
which leads to the derivatives
\begin{equation}
\label{der}
      \langle P_\lambda|\frac{\rd P_\lambda}{\rd r}\rangle=0\,,\qquad
      \langle P_\lambda|\frac{\rd^2 P_\lambda}{\rd r^2}\rangle=
          - \langle\frac{\rd P_\lambda}{\rd r}
               |\frac{\rd P_\lambda}{\rd r}\rangle
\end{equation}
Multiplying Eq. (\ref{urS}) with $P_\lambda(z,r)W_\Lm(z)$ from  right 
and integrating over $z$, we obtain the new equation for $u_\lambda$
 \begin{eqnarray}
 \nonumber
  	\left\{\hh\left[-\frac{\rd ^2}{\rd r^2}\right.\right.
       &+&\left. \LLr\right]
    + \frac{A(A-1)}{2}V_\Lm(r)
\\
     &+&\left.
      U_\lambda(r)+\frac{\hbar^2}{m}
         \langle\frac{\rd P_\lambda}{\rd r} 
          |\frac{\rd P_\lambda}{\rd r}\rangle	 
         -E\right\}u_\lambda(r)=0\,.
 \label{uruaa}
 \end{eqnarray}
The solution thus obtained is called the Uncoupled Adiabatic 
Approximation (UAA). We note that the extra term  introduced 
in (\ref{uruaa})  is always negative and  the new effective potential
\begin{equation}
   V_{\rm eff}(r)\equiv \frac{A(A-1)}{2}V_\Lm(r)+U_\lambda(r)+
    \hh\langle\frac{{\rm d}P_\lambda}{{\rm d}r}
               |\frac{{\rm d}P_\lambda}{{\rm d}r}\rangle
\label{vuaeff}
  \end{equation}
always provides an upper bound to the eigen-energy.\par
\subsection{A Variational Equation for the IDEA}
 %%%%%%%%%%%%%%%%%%%%%%%%%%%%%%%%%%%%%%%%%%%%%%%%%%
 The IDEA Equation (\ref{IDEA1}) is a two-variable integro-differential 
equation  where the integral part comes from the projection ${\cal P}^0$ 
of all  pairs $(i,j)$   on the reference pair $(1,2)$. The two-body 
amplitude $P(z,r)$ is a solution of an  integro-differential 
equation which must fulfill certain  asymptotic  conditions for 
$z=\pm1$  and  $r=0$  and $r=\infty$. 
A variational   equation can also be obtained in the adiabatic 
approximation by multiplying Eq. (\ref{EAA1}) by the weight function,
normalized to unity,  and by integrating  over $z$ in the range 
$[-1,+1]$.  The normalization is
 \begin{equation}
   \int_{-1}^{+1}P_\lambda(z,r)W_\Lm(z)\rd z=1
 \label{PW}
  \end{equation}
 implying that the hypercentral part of $P_\lambda(z,r)$ is taken
 to be 1.  It is the dominant contribution in the HH expansion of 
 $P_\lambda(z,r)$. The kinetic  energy term in the left hand side 
disappears and the equation for $U_\lambda(r)$ becomes
 \begin{eqnarray}
\nonumber
       U_\lambda(r)&=& \int_{-1}^{+1}\left[V(r\sqrt{(1+z)/2}- V_\Lm(r)\right]
     \\  &&\left[ P(z,r)+{\cal P}^0P(z,r)\right]\,W_\Lm(z)\dz
 \label{IDEAV}
 \end{eqnarray}
 Thus one may search for the solution $P_\lambda(z,r)$  for which 
$U_\lambda(r)$ is a minimum  which is then  introduced in the 
radial equation  (\ref{urS}) to calculate the ground state energy. 
One can also control the quality of $P_\lambda(z,r)$  solution of 
Eq. (\ref{EAA1}) by calculating  the eigen-potential $ U_\lambda(r)$ 
from  (\ref{IDEAV}). This should be identical (within numerics) to the 
one obtained from   Eq. (\ref{EAA1}).
%cccccccccccccccccccccccccccccccccccccccccccccccc
%             secV.tex
%cccccccccccccccccccccccccccccccccccccccccccccccc
 %%%%%%%%%%%%%%%%%   WEIGHT  %%%%%%%%%%%%%%%%%%%%%%%
%
 \section{Construction of the Weight Function}
 %%%%%%%%%%%%%%%%%%%%%%%%%%%%%%%%%%%%%%%%%%%%%%%
 The HO wave function can be written either in the $D$-space
 as a product of a HP $D_\Lm(\vec x)$ and an exponential
  in the hyperradius  or as a Slater determinant constructed 
 from individual HO eigenfunctions. The weight function can then 
 be obtained by identification of the Fourier transform
in the relative coordinate $\vec r_{ij}$ of
 $|D_\Ll(\vec x)|^2$ 
 calculated  first from the HP 
 representation in the $D$-space and then from the
 standard HO Slater determinant.\par
 Let $D^{HO}_\Ll(\vec x)$ be the normalized Slater determinant
 describing the states of an $A$-nucleon system
 \begin{equation}
	 D^{HO}_\Ll(\vec x)=\frac{1}{\sqrt A!} || s^j_i\,t^j_i
	 \psi_{n_j,\ell_j,m_j}(\vec x_i)||
 \label{slater}
  \end{equation}
 where $i$ refers to the rows, $j$ to the columns ($i,j=1,\cdots,A$) and 
 where $s$ and $t$ denote the spin and isospin states.
 The normalized HO eigenfunction for one particle in 
polar coordinates ($x,\om$) is
 \begin{equation}
  \hspace{-3mm} \psi_{n,\ell,m}(\vec x)=\left[\frac{2n!}{b^3
        \Gamma(n+\ell+\frac {3}{2})}\right]^{1/2}
       \,Y_\ell^m(\omega)(x/b)^\ell
     L_n^{\ell+1/2}\left((x/b)^2\right)
	      \,{\rm e}^{-(x/b)^2/2}
 \label{psilnm}
 \end{equation}

 where  $b$ is a parameter related to the size of the system.
 The determinant describing the HO ground  state of the $A$-nucleon
 system is constructed in such a way that for a
 given set of quantum numbers $s^j,t^j,\ell_j,m_j$ all 
 the quantum numbers $n_j$, for $n_j$ running from 0 to a maximum 
 value $n_j(s^j,t^j,\ell_j,m_j)$ chosen
 independently for each set,  are filled.  According to the 
 properties of determinants it can also be written as
 \begin{equation}
	 D^{HO}_\Ll(\vec x)=C_L\parallel s^j_i\,t^j_i\,x_i^{2n_j+\ell_j}
	 Y^{m_j}_{\ell_j}(\omega_i)\parallel\,{\rm e}^{-(\sum^A_1x^2_i)/
	 \varrho_0^2}
 \label{det1}
  \end{equation}
 where $\varrho_0^2=2b^2$, $\sum^A_1x^2_i=r^2/2+AX^2$, $\vec X$
 being the center of mass coordinate, and $C_L$ 
 is a normalization constant. This determinant is a 
 translationally invariant  ground HP of degree 
 $L=\sum^A_{j=1}(2n_j+\ell_j)$.  \par
 One may factorize $r^L$ and rewrite Eq. (\ref{det1}) as
 \begin{equation}
    D^{HO}_\Ll(\vec x)=D_\Ll(\Omega)\left[\frac{2r^{2L}\,{\rm e}^
	 {-r^2/\varrho_0^2}}{\varrho^{2L+D}_0\Gamma(L+D/2)}
 \right]^{1/2}
	 \left[
 \ \frac{2{\rm e}^{-R^2/b^2}}{b^3\Gamma(3/2)}\right]^{1/2}
 \label{DHO}
 \end{equation}
 where $R^2=AX^2$ and $D_\Ll(\Omega)$ is a normalized HH 
of grand orbital $L$.  
 Indeed, by integrating $|D^{HO}_\Ll(\vec x)|^2$ over the 
 $3A$--dimensional space with 
 \begin{equation}
	 \rd^{3A}x=\rd^3x_1\,\cdots\,\rd^3\,x_A=\rd\Omega\, 
		 r^{D-1}\,\rd r\, \rd^3R
 \label{d3a}
  \end{equation}
 one finds
 \begin{equation}
	 \int|D^{HO}_\Ll(\vec x)|^2\,\rd^{3A}x=\int|D_\Ll(\Omega)|^2
	 \,\rd\Omega=1\,.
 \label{normD}
   \end{equation}
 The integral
 $
	 \int|D_\Ll(\Omega)|^2\,\rd\Omega_{N-1}
 $
 (occurring in the weight function)  can be obtained from the Fourier 
 transform
 \begin{equation}
	 {\cal D}(y,\om_k)\equiv\langle D^{HO}_\Ll(\vec x)|
	 {\rm e}^{i\vec k\cdot\vec \xi_N}|
		D^{HO}_\Ll(\vec x)\rangle,
	 \qquad \vec\xi_N= \vec r_{12}
 \label{DeD}
  \end{equation}
calculated  by two different methods.  Since $r^LD_\Ll(\Omega)$ is 
a polynomial of degree $L$  homogeneous in the Jacobi coordinates
 $\vec \xi_i$, $i=1,\cdots,N(=A-1)$,  then $|r^LD_\Ll(\Omega)|^2$ 
is an even  homogeneous polynomial and the integral
 over $\Om_{N-1}$  in Eq. (\ref{WDL}) is a homogeneous polynomial 
in the  $\vec\xi_N$ and $ \rho^2=r^2-\xi_N^2 $ variables
 \begin{equation}
%\hspace*{-7mm}
 	\int|r^L\,D_\Ll(\Omega)|^2\,\rd\Omega_{N-1}=
  \sum_{n,\ell\atop{\ell\ {\rm even}}}\langle\Ll |n,\ell\rangle  
          Y^0_\ell(\omega) 
\xi^{2n+\ell}_N\rho^{2L_m-(2n+\ell)}
 \label{coef}
 \end{equation}
 where $\langle\Ll | n,\ell\rangle$ are coefficients that have
 to be defined. The volume 
 element for the coordinate system $(\Omega_{N-1},\rho,\vec\xi_N)$ in the
 $D=3N=3(A-1)$--dimensional space is 
 \begin{equation}
	 \rd^{3N}\xi=\rd\Omega_{N-1}\,\rho^{D-4}\,\rd \rho\,\xi^2_N\,
	   \rd\xi_N\,	\rd\omega
 \label{xi3n}
  \end{equation}
 where $\omega$ is for the angular coordinates of $\vec\xi_N$.  Since
 the operator $\exp[i\vec k\cdot \vec r_{12}]$ is independent of 
the center of  mass, the integral  over $\vec R$ in (\ref{DeD})
gives 1 and  the Fourier transform becomes
 %
 %\begin{widetext}
 \begin{eqnarray}
 \nonumber
 {\cal D}(y,\om_k)&=&\frac{2}{\Gamma(L+D/2)} 
       \sum_{{n,\ell}\atop{\ell \ {\rm even}}}\,
	   \langle\Ll|n,\ell\rangle 
	\int\,\re^{i\vec k\cdot\vec\xi_N}\, Y^0_\ell(\omega)
\\\nonumber
&\times&\,\rd\omega \,
	  {\rm e}^{-(\xi_N^2+\rho^2) /\varrho_0^2}
	 \left(\frac{\xi_N}{\varrho_0}\right)^{2(n+1)+\ell}
\rd\left(\frac{\xi_N}{\varrho_0}\right)
\\&\times&
	     \left(\frac{\rho}{\varrho_0}\right)^{2L+D-4-(2n+\ell)}
	  \,\rd \left(\frac{\rho}{\varrho_0}\right)
 \label{DeD1}
 \end{eqnarray}
 %\end{widetext}
 %
 One substitutes for the plane wave ${\rm e}^{i\vec k\cdot\vec x}$,
 $(\vec x=\xi_N$),  the expansion in the polar coordinates $(\omega, x)$
 of $\vec x$, 
 \begin{equation}
	 {\rm e}^{i\vec k\cdot\vec x}=(2\pi)^{3/2}
	 \sum_{\ell,m}
	  Y^m_\ell (\omega_k)
	 Y^{m*}_\ell(\omega)\frac{1}{\sqrt{ kx}}J_{\ell+1/2}(kx)\,.
 \label{eexp}
  \end{equation}
 where $(k,\omega_k)$ are the polar coordinates of $\vec k$.
 After integration over the $(\rho,\vec\xi_N)$ variables, the Fourier
 transform becomes
 \begin{eqnarray}
 \nonumber
	 {\cal D}(y,\om_k)
	   &=&\frac{\pi^{3/2}}{\Gamma(L+D/2)}\sum_{n,\ell}(-1)^{\ell/2}
	 \langle\Ll|n,\ell\rangle n! 
\\& \times&
	\Gamma(L+(D-3)/2-n-\ell/2)
	 Y^0_\ell(\omega_k)y^\ell L_m^{\ell+1/2}(y^2)\,{\rm e}^{-y^2}
 \label{Fourier1}
 \end{eqnarray}
 where $	y=kb/\sqrt{2}=k\varrho_0/2$.\par
 The Laguerre polynomials $L^\alpha_n(x)$, $x=y^2$,  constitute a 
 complete orthogonal polynomial basis associated with the weight function
 $x^\alpha \,{\rm e}^{-x}$. Thus, multiplying Eq. (\ref{Fourier1})
  by 	$Y^0_\ell(\omega_k)y^\ell L_n^{\ell+1/2}(y^2)$ 
 and integrating  over $y$ in the range $0\le y\le\infty$ and 
over $\omega_k$ we  obtain the  expansion coefficient occurring in 
Eq. (\ref{coef})
 \begin{eqnarray}
 \nonumber
	 \langle\Ll |n,\ell\rangle&=&(-1)^{\ell/2}\frac{2}{\pi^{3/2}}
% \\&&\hspace{-1.3cm}
 %\nonumber
%	 \times
\frac{\Gamma(L+D/2)}{\Gamma[L+(D-3)/2-(n+\ell/2)]
	\Gamma(n+\ell+3/2)}
 \\&\times&
	  \int{\cal D}(y,\om_k)
	 Y^0_\ell(\omega_k)y^\ell \,L_n^{\ell+1/2}(y^2)
	y^2\,\rd y \rd\om_k
 \label{cof1}
 \end{eqnarray}
 According to Eq. (\ref{Fourier1}), the Fourier transform is a polynomial
 in $Y^0_\ell(\omega_k)y^{2\nu+\ell}$ which can also be written as 
 \begin{equation}
 {\cal D}(y,\om_k)
 =\sqrt {4\pi}
	 \sum_{\nu,\lambda}\, (-1)^{\lambda/2}
	 (\Ll|\nu,\lambda)\, Y^0_\lambda(\omega_k)
	 y^{2\nu+\lambda}\,\re^{-y^2}
 \label{ddn}
  \end{equation}
 where $(\Ll|\nu,\lambda)$ are new coefficients  to be determined.
 Introducing (\ref{ddn})
 in (\ref{cof1}) and using the analytical expression
 \begin{equation}
    \int_0^\infty\,y^{2(\nu+\ell)}\,L^{\ell+1/2}_n(y^2)
	  \re^{-y^2}y^2\,{\rm d}y       =\frac{(-1)^n}{2}
	 { \nu\choose n} \Gamma(\nu+\ell+3/2)
\label{ddnn}
 \end{equation}
 we find the relation between the two kinds of coefficients 
 $       \langle\Ll | n, \ell\rangle$ and $(\Ll|\nu,\ell)$
 \begin{eqnarray}
 \nonumber
	\langle\Ll | n, \ell\rangle&=&
	       \frac{2}{\pi}(-1)^n \frac{\Gamma(L+D/2)}
		{\Gamma(L+(D-3)/2-n-\ell/2)}\\
 & \times&
		\sum_{\nu\ge n}{\nu\choose n}
		\Gamma(\nu+\ell+3/2)(\Ll|\nu,\ell)\,.
 \label{cofL}
 \end{eqnarray}
  Introducing  (\ref{cofL}) in (\ref{coef}) with
 $\xi_N=r\cos\phi$, $\rho=r\sin \phi$, and $z=\cos2\phi$,
dividing by $r^{2L}$,
 and using the definition of the Jacobi polynomials 
 $P_n^{\alpha,\beta}(x)$ \cite{Erdelyi} one obtains 
 \begin{eqnarray}
 \nonumber
&& \int|\,D_\Ll(\Omega)|^2\,\rd\Omega_{N-1}=
 \frac{2}{\pi}\Gamma(L+D/2)
     \sum_{n,\ell\atop{\ell\ {\rm even}}}  \frac{(-1)^n n!}
         {\Gamma(L+(D-3-\ell)/2)}
\\&& \ \ \ \ \times (\Ll|n,\ell) Y^0_\ell(\omega)
       (\cos\phi)^\ell(\sin\phi)^{2L-2n-\ell}
	 {\sl  P}_n^{\alpha-n-\ell/2,\beta}(\cos2\phi)
 \label{coefn}
 \end{eqnarray}
 where $\alpha=L+(D-5)/2$ and $\beta=\ell+1/2$.\par
 Since $\ell$ is even  while $\sin^2\phi=(1-z)/2$ and $\cos^2\phi=(1+z)/2$
 the sum over $\ell$ and $n$ generates a polynomial in $z$.
 We have now two expressions at our disposal for $W^{(D)}_\Lm$
 occurring in the weight function (\ref{wlm}) according to whether
 one uses $ \langle\Lm | n, \ell\rangle$ or $ (\Lm | n, \ell)$ coefficients.
 In the first case 
 \begin{equation}
	W^{(D)}_\Lm(z,\om)=\frac{1}{2^{L_m}}
	    \,\sum_{n,\ell}\langle\Lm | n, \ell\rangle\,Y^0_\ell(\om)
	  (1+z)^{n+\ell/2}\,(1-z)^{L_m-n-\ell/2}
 \label{WD1}
 \end{equation}
 and in the second
 \begin{eqnarray}
 \nonumber
    W^{(D)}_\Lm(z,\om)&=&\frac{1}{2^{L_m}}\frac{2}{\pi}
	    \,\sum_{n,\ell}\,\frac{ (-2)^n n!\Gamma(L_m+D/2)}
	   {\Gamma(L_m+(D-3-\ell)/2)}
	   (\Lm | n, \ell)
\\ 
&\times&(1-z)^{L_m-n-\ell/2}(1+z)^{\ell/2} 
   Y^0_\ell(\om)\,P^{\alpha-n-\ell/2,\ell+1/2}_n(z)
 \label{WD2}
 \end{eqnarray}
In order to find explicit expressions for the coefficients occurring in 
either (\ref{WD1}) or (\ref{WD2}) we have to calculate the Fourier 
transform, Eq. (\ref{DeD}), where $D^{HO}_\Ll(\vec x)$ is  the IPM 
representation of the Harmonic Polynomial $D_\Ll(\vec x)$, 
Eq. (\ref{slater}). For this purpose we first write
 $$
	 \re^{ i\vec k\cdot\vec \xi_N}=\re^{ i\vec k\cdot\vec r_{12}}=
	    \re^{ i\vec k\cdot\vec x_1}\,
	 \re^{ -i\vec k\cdot\vec x_2}
 $$
 and then develop the normalized  Slater determinant $D^{HO}_\Lm(\vec x)$
in Eq. (\ref{slater}) with respect to the first two rows for
 $\vec x_1$, and $\vec x_2$ 
\begin{equation}
	 D^{HO}_\Ll (\vec x)=\frac{1}{A(A-1)}
         \sum_{i,i<j}\, d_{ij}(\vec x_1,\vec x_2)
	\,D^{HO}_{ij}(\vec x_3,	\cdots,\vec x_A)
 \label{dhoex}
 \end{equation}
 where 
 $$
	  d_{ij}(\vec x_1,\vec x_2)= \left |
  \matrix{
      s_1^i\,t_1^i\,\psi_i(\vec x_1),  & s_1^j\,t_1^j\,\psi_j(\vec x_1)   \cr
      s_2^i\,t_2^i\,\psi_i(\vec x_2),  & s_2^j\,t_2^j\,\psi_j(\vec x_2)
	  }
 \right |.
 $$
 Since $D^{HO}_{ij}$ and $D^{HO}_{k\ell}$ do not contain the same HO 
 individual state except for $i=k$ and $j=\ell$ we have
 \begin{equation}
	 \langle D^{HO}_{ij} | D^{HO}_{k\ell} \rangle 
	    =\delta_{ik}\delta_{j\ell}
  \end{equation}
 Integration over all  $\vec x_i$ for $i>2$ leads to
\begin{equation}
	 \int | D^{HO}_\Ll(x) |^2 \rd^3x_3\cdots\rd^3x_A
=
	\frac{1}{A(A-1)}\sum_{i,j>i}	|d_{ij}(\vec x_1,\vec x_2)|^2
 \label{dho2}
 \end{equation}
 where
$$
	 d_{ij}(\vec x_1,\vec x_2)=\left|\matrix {| i_1 \rangle& |j_1\rangle\cr
				|i_2 \rangle& |j_2 \rangle\cr}
	 \right| 
 $$
 and 
 $$
	 \int |d_{ij}(\vec x_1,\vec x_2)|^2\,\rd^3x_1\,\rd^3x_2=2\,.
 $$
 Therefore,
 $$
	 \int | D^{HO}_\Ll(\vec x) |^2 \rd^{3}x_1\cdots\rd^{3}x_A=1\,.
 $$
 The $d_{ii}$ contains  identical columns and thus it 
 vanishes. Further, since $d^*_{ij}\,d_{ij}$ is invariant 
 by exchange of  $i$ and $j$, the sum over $i$ and $j>i$ can be 
 transformed into a sum over all states independently of their 
 position in the determinant $D^{HO}_\Ll(\vec x)$
 \begin{equation}
	 \sum_{i,j>i}\,|\,d_{ij}(\vec x_1,\vec x_2)\,|^2=\frac{1}{2}
	 \sum_{i,j}\,|\,d_{ij}(\vec x_1,\vec x_2)\,|^2
 \label{det2}
  \end{equation}
 For simplicity, we introduce a single set of quantum numbers $\zeta$ for
 spin and isospin states, labeled according to
 $$
 \matrix{
 \zeta=
	 +\frac{3}{2}& {\rm for\ proton\ spin\ up}\cr
 \zeta=
	 +\frac{1}{2}&\ \ \ {\rm for\ proton\ spin\ down}\cr
 \zeta=
	 -\frac{1}{2}& {\rm for\ neutron\, spin\, up}\cr
 \zeta=
	 -\frac{3}{2}&\ \ \ \ {\rm for\ neutron\ spin\ down}\cr
 }
 $$
 where we count 1 or -1 for proton and  neutron respectively
  and $\frac{1}{2}$,  $-\frac{1}{2}$ for  spin up or down and we add 
 the values to obtain $\zeta$.  In order to calculate the Fourier 
 transform we do not have anymore to refer
 to the columns but only to the spin-isospin and space occupied states.
 Then a sum over all pairs of occupied states is substituted for the 
 sum over $i$ and $j>i$ in Eq. (\ref{dho2}). \par
 Let us consider the pair of states $\zeta_\alpha\,|\alpha>$ 
 and $\zeta_\beta\,|\beta>$
 where $|\alpha>$  and $|\beta>$ stand for the HO eigenfunctions
 (\ref{psilnm}) with quantum numbers ($n_\alpha,\ell_\alpha,m_\alpha$)
 and  ($ n_\beta,\ell_\beta,m_\beta$).  Then one may distinguish
 two cases in the calculation of the Fourier transform
 \begin{equation}
	 F_{\alpha\beta}=\langle d^*_{\alpha\beta}(12)|\re^{i\vec k
	    \cdot(\vec x_1-\vec x_2)}|d_{\alpha\beta}(12)\rangle\,.
 \label{fab}
\end{equation}
 In the first case one  defines the direct term
 \begin{equation}
      D_{\alpha\beta}(\vec k)=
          \langle\alpha|\re^{i\vec k\cdot\vec x}|\alpha \rangle
         \langle\beta|\re^{-i\vec k\cdot \vec x}|\beta\rangle
         +\{\vec k\rightarrow\, -\vec k\}
 \label{dab}
\end{equation}
 and the exchange term
 \begin{equation}
	 E_{\alpha\beta}(\vec k)=\langle\alpha|\re^{i\vec k\cdot\vec x}|
	   \beta\rangle\langle \beta|\re^{-i\vec k\cdot
	    \cdot\vec x}|\alpha\rangle+  \{\vec k\rightarrow-\vec k\}
 \label{eab}
\end{equation}
 where $\vec k\rightarrow-\vec k$ means the first term in which $\vec k$ is 
 changed to  $-\vec k$.  These Fourier transforms
are real and  even functions of $\vec k$ and
 thus they contain  $Y^0_\ell(\om)$ for even $\ell$ only, where
$\ell>0$ is related to the deformation of nuclei.\par
In Nuclear interactions spin and isospin can be exchanged between
two nucleons $i$ and $j$. The exchange operators are
$P^{(\epsilon)}_{ij}$ where $\epsilon=0$ without exchange while
$\epsilon=\sigma,\ \tau,\ \sigma\tau$ where spin, isospin or spin and 
isospin are exchanged respectively. These operators are
traditionally known as Wigner ($\epsilon=0$, Bartlett ($\epsilon=\sigma$),
Heisenberg ($P^H_{ij}=-P^\tau_{ij}$,
and Majorana ($P^M_{ij}=-P^{\sigma\tau}$), exchanged operators.
In the above,  the Fourier transform has been calculated without
exchanged operators i.e for $\epsilon=0$ only. This case 
corresponds  to the Wigner force
for which the  weight function is given by Eq. (\ref{wlm}) with 
(\ref{WD1}) or (\ref{WD2}). Therefore, we have still to calculate
 the so-called {\em pseudo-weight functions} associated with the 
 operators $P^\epsilon_{ij}$,
where $\epsilon=\sigma,\ \tau$, or $\sigma\tau$.\par
To evaluate the  contribution brought by the direct
and exchange terms $D_{\alpha\beta}$ and $E_{\alpha\beta}$,
to the Fourier transform 
\begin{equation}
       F_{\alpha\beta}^{(\epsilon)}= \langle d^*_{\alpha\beta}(12)|
      \re^{i\vec k\cdot(\vec x_1-\vec x_2)}  P^{(\epsilon)}_{12}|
            d_{\alpha\beta}(12)\rangle\,,
\label{faben}
\end{equation}
including  the exchange operator $P^{(\epsilon)}_{12}$ for
$\epsilon=0,\sigma,\tau,\sigma\tau$, we have to distinguish
four cases:
\begin{itemize}
\item[1)] 
    The spin-isospin states $\zeta_\alpha$ and $\zeta_\beta$
    are the same, {\em i.e}, $\zeta_\alpha-\zeta_\beta=0$.
\item[2)] 
    The spin-isospin states are different, {\em i.e}, 
    $\zeta_\alpha\ne \zeta_\beta$ but the  $P^{(\epsilon)}_{ij}$
    operator does not modify the spin-isospin states.
\item[3)] 
    The operator $ P^{(\epsilon)}_{ij}$ exchanges the two 
    spin-isospin states.
\item[4)] 
    The operator $  P^{(\epsilon)}_{ij}$
    generates new  spin-isospin states orthogonal to the 
    original $\zeta_\alpha$ and $\zeta_\beta$ states.
\end{itemize}
In terms of the Kronecker symbol $\delta_{ab}=1$ for $a=b$ and zero
otherwise, the contribution in the first case is 
$$
      \delta_{\zeta_\alpha \zeta_\beta}(D_{\alpha\beta} -E_{\alpha\beta}),
$$
in the second 
$$
        (1-\delta_{\zeta_\alpha \zeta_\beta})D_{\alpha\beta}\,,
$$
in the third
$$
          -(1-\delta_{\zeta_\alpha \zeta_\beta})E_{\alpha\beta}\,,
$$
while in the last is zero. Applying these rules  one finds that
for the Wigner operator $ P^{(0)}_{12}=1$ we have the Fourier transform
\begin{equation}
 F_{\alpha\beta}^{(0)}=D_{\alpha\beta}-
         \delta_{\zeta_\alpha \zeta_\beta}E_{\alpha\beta}.
\label{fab0}
\end{equation}
Instead of considering  the $P^\sigma_{12}$ and $P^\tau_{12}$
operators separately, it is more convenient to use the
combinations 1/2($P^\sigma_{12}\pm P^\tau_{12}$) leading to the 
expression
 \begin{eqnarray}
 \nonumber
%\hspace{-0.8cm}
 F^{(\epsilon)}_{\alpha\beta}&=& \frac{1}{2}
	(1-\delta_{ (\zeta_\alpha+\zeta_\beta)\, 0})
 \bigg\{ (1+\delta_{\zeta_\alpha\zeta_\beta})
    \frac{1}{2}(\delta_{\epsilon\sigma}+\delta_{\epsilon\tau})
      \bigg[D_{\alpha,\beta}  -E_{\alpha,\beta}\bigg]
\\
   & -&(-1)^{\zeta_\alpha+\zeta_\beta}
   (1-\delta_{\zeta_\alpha \zeta_\beta})
\frac{1}{2}
    (\delta_{\epsilon\sigma}-\delta_{\epsilon\tau})
    \bigg [ D_{\alpha,\beta} + E_{\alpha,\beta}\bigg]\bigg\}
\label{fab12}
 \end{eqnarray}
where $\epsilon=\sigma$ or $\epsilon=\tau$.
Finally, for the spin-isospin exchange operator $\sigma\tau$ we have
\begin{equation}
    F_{\alpha\beta}^{(\sigma\tau)}=
         \delta_{\zeta_\alpha \zeta_\beta}D_{\alpha\beta}
    - E_{\alpha\beta}\,.
\label{fab3}
\end{equation}
From the above analysis we see that  in order to calculate the weight
and  pseudo-weight  functions associated with the operators occurring 
in realistic nuclear  potentials one needs to know the Fourier transform 
of the product of two individual HO eigenfunctions. Indeed, since we 
have to deal with a two-body weight function, which depends on
 $z=2\rr^2/r^2-1$ (here $(i,j)=(1,2)$), the Fourier transform of 
the product of two
 individual HO eigenfunctions occur twice for the states $|\alpha>$
 and  $|\beta>$ in the calculation of the direct and exchange terms 
 in Eqs. (\ref{dab}) and (\ref{eab}).
 \subsection{ Weight and Pseudo-Weight  Functions: General Case}
\label{wgc}
 %%%%%%%%%%%%%%%%%%%%%%%%%%%%%%%%%%%%%%%%%%%%%%%%%%%%%%%%%%%%%%
%
 Let us define the Fourier transform
 \begin{equation}
    \int \psi^*_{n_\alpha, \ell_\alpha,m_\alpha}
	  \,\re^{i\vec k\cdot\vec x}
	  \psi_{n_\beta, \ell_\beta,m_\beta}\rd^3x
	  =\langle \alpha| \,\re^{i\vec k\cdot\vec x}|\beta\rangle
\label{ftpsi}
 \end{equation}
 where $|\alpha>$ and $|\beta>$  refer, respectively, 
  to the HO eigenstates with quantum numbers 
 $n_\alpha, \ell_\alpha,m_\alpha$ and $n_\beta, \ell_\beta,m_\beta$ 
in the HO wave function (\ref{psilnm}). The  Fourier transform is the 
sum over the quantum numbers $n$ and $\ell$
 \begin{equation}
       \langle \alpha| \,\re^{i\vec k\cdot\vec x}|\beta\rangle
       =\sqrt {4\pi}
	 \sum_{n,\ell}\, (i)^\ell
	 (\alpha|n,\ell|\beta)\, Y^m_\ell(\omega_k)
	 y^{2n+\ell}\,\re^{-y^2/2}
 \label{Fakb}
 \end{equation}
 where $m=m_\beta-m_\alpha$. The coefficients $(\alpha|n,\ell|\beta)$,
  normalized according to $(\alpha|0,0|\beta)=\delta_{\alpha\beta}$,
are given  in the appendix. For $\beta=\alpha$ the simplified notation 
 $(\alpha|n,\ell|\alpha)=(\alpha|n,\ell)$ is also used with 
$(\alpha|00)=1$.\par
The calculation of the Fourier transform (\ref{ddn}) depends
on the  direct and exchange terms $D_{\alpha\beta}(\vec k)$ and
$E_{\alpha\beta}(\vec k)$.
 However, the question of  normalization must be addressed  first.
 In Ref. \cite{FFW82} it was shown that for
 closed shell nuclei  the coefficient
 $(\Ll|\nu,\lambda)$ occurring in (\ref{ddn}) and in the weight 
 function written as in Eq. (\ref{WD2}) are simple and most of them
 integers and  the  normalization in Eq. (\ref{normD}) is $A(A-1)/2$ 
instead of 1. To obtain afterwards the normalization to 1 obviously one 
divides by  $A(A-1)/2$.  To obtain these coefficients
 one must calculate the Fourier transform for the expression
 \begin{equation}
     {\cal F}=  \frac{1}{4} \sum_{i,j}\langle d_{ij} |
           \re^{i\vec k\cdot(\vec x_1
	   -\vec x_2)}|d_{ij}\rangle\,.
 \label{dedn}
 \end{equation}
where the factor 4  stems from the normalization to $A(A-1)/2$
of this expression for $k=0$ and from the fact that  $d_{ii}=0$
$d_{ij}=-d_{ji}$ with $\langle  d_{ij}|d_{ij}\rangle =2$.
 Using  Eq. (\ref{Fakb}) and the expansion of the product of
 two spherical harmonics
\begin{equation}
	Y^{m_1}_{\ell_1}(\om_k) Y^{m_2}_{\ell_2}(\om_k)=
       \sum_\lambda\langle Y^\mu_\lambda |Y^{m_1}_{\ell_1}
	    | Y^{m_2}_{\ell_2}\rangle \,Y^\mu_\lambda(\om_k),
 \label{y1y2}
 \end{equation}
 with $\mu=m_1+m_2$, the direct term, Eq. (\ref{dab}), for the couple of
 states $|\alpha>$ and  $|\beta>$ with $m_1=m_2=0$ becomes
\begin{equation}
	 D_{\alpha\beta}(\vec k)=\sqrt{4\pi}\,\sum_{\lambda, \nu}
	  (-1)^{\lambda/2}\,d(\alpha,\beta|\nu,\lambda)
	 Y^0_\lambda(\om_k)y^{2\nu+\lambda}\,\re^{-y^2}
 \label{dab1}
 \end{equation}
 where the direct coefficient
\begin{equation}
	d(\alpha,\beta|\nu,\lambda)=\sqrt{4\pi}\,
	 \sum_{n1\,\ell_1\atop{n_2\,\ell_2}}
	 (-1)^{(\ell_1+\ell_2-\lambda)/2}
(\alpha|n_1,\ell_1)(\beta|n_2,\ell_2)
	\langle Y^0_\lambda |Y^0_{\ell_1}
	    | Y^0_{\ell_2}\rangle 
 \label{dcoe1}
 \end{equation}
for $ \nu=n_1+n_2+(\ell_1+\ell_2-\lambda)/2$ where $\ell_1$, $\ell_2$, and 
 $\lambda$ are even and $|\ell_1-\ell_2|\le\lambda\le \ell_1+\ell_2$.
A factor of 2, stemming from the two terms in (\ref{dab}) for $\vec k$
 and  $\vec k\rightarrow-\vec k$, has also been taken into account.  
One notices that for 
 $n_1=n_2=\ell_1=\ell_2=0$ for which $\nu=\lambda=0$
 the normalization $d(\alpha\beta|00)=1$ leads to 
 $D_{\alpha\beta}(\vec k=0)=1$.\par
 A similar procedure is applied to obtain the exchange term
\begin{equation}
	 E_{\alpha\beta}(\vec k)=
      \sqrt{4\pi}\,\sum_{\lambda, \nu}
	  (-1)^{\lambda/2}\,e(\alpha,\beta|\nu,\lambda)
	 Y^0_\lambda(\om_k)y^{2\nu+\lambda}\,\re^{-y^2}
 \label{eab1}
 \end{equation}
 where the exchange coefficient
 \begin{eqnarray}
 \nonumber
	e(\alpha,\beta|\nu,\lambda)&=&\sqrt{4\pi}\,
	 \sum_{n1\,\ell_1\atop{n_2\,\ell_2}}
	 (-1)^{(\ell_1+\ell_2-\lambda)/2}
 (\alpha|n_1,\ell_1|\beta)(\beta|n_2,\ell_2|\alpha)
\\
   &\times&	\langle Y^0_\lambda |Y^{m_1}_{\ell_1}
	    | Y^{m_2}_{\ell_2}\rangle 
 \label{ecoe1}
 \end{eqnarray}
 for $ \nu=n_1+n_2+(\ell_1+\ell_2-\lambda)/2$,  $\lambda$ even,
 $m_1=m_\beta-m_\alpha$, $m_2=-m_1$ and the normalization 
      $e(\alpha,\beta|0,0)=\delta_{\alpha\beta}$\par
 Let us come back to the Fourier transform, Eqs. (\ref{fab}) and
\begin{equation}
       \int {\cal D}^{HO\,*}_\Ll(\vec x)
	 {\rm e}^{i\vec k\cdot(\vec x_1-\vec x_2)}
		D^{HO}_\Ll(\vec x)\rd^{3A}x=\frac{1}{4}
	 \sum_{{{\rm all\ occupied}\atop{\rm  states}}}
	 F_{\alpha\beta}
 \label{DeDn}
 \end{equation}
for the normalization (\ref{dedn}) where the wave function is
normalized to $A(A-1)/2$ as explained above.  
Using (\ref{dab1}) and (\ref{eab1}) one finds,
by identification with (\ref{ddn}), the coefficient
for $\varepsilon=0$ {\em i.e} for $P^{(0)}_{ij}=1$ in Eq. (\ref{fab0}),
 \begin{equation}
      (\Ll|\nu,\lambda)^{(0)}= \frac{1}{2}
	 \sum_{{{\rm all\ occupied}\atop{\rm  states}}}
      \left\{  d(\alpha,\beta|\nu,\lambda)
     -\delta_{ \zeta_\alpha \zeta_\beta}\, 
       e(\alpha,\beta|\nu,\lambda)\right\}
 \label{LnlW}
 \end{equation}
where the sum is taken  independently over all HO occupied states in the
 Slater determinant $D^{HO}_\Ll(\vec x)$ defining the state
 under consideration.  This coefficient, introduced in 
Eq. (\ref{WD2}), defines the weight function
 $$
     W_\Lm(z,\om)\equiv W^{(0)}_\Lm(z,\om)=W_0(z)W^{(D)}_\Lm(z,\om)
 $$
 according to Eq. (\ref{wlm}). The upper index (0) is to recall that 
 we are dealing with a Wigner potential.\par
 For the pseudo-weight function corresponding to the spin, isospin, and 
 spin-isospin exchange operators $P^{\epsilon}_{ij}$ respectively
 for $\epsilon=\sigma,\ \tau$, $ \sigma\tau$ one uses Eq. (\ref{fab12})
  and  (\ref{fab3}) leading to  the new
 spin and isospin coefficients 
 \begin{eqnarray}
 \nonumber&&
\hspace{-0.5cm}
 (\Ll|\nu,\lambda)^{(\epsilon)}= \frac{1}{2}
	\sum_{{\rm all\ occupied}\atop{{\rm states}}}
 \frac{1}{2}(1-\delta_{ (\zeta_\alpha+\zeta_\beta)\, 0})
\bigg\{ (1+\delta_{\zeta_\alpha\zeta_\beta})
    \frac{1}{2}(\delta_{\epsilon\sigma}+\delta_{\epsilon\tau})
 \\\nonumber &&\ \     \times 
\bigg\{ (1+\delta_{\zeta_\alpha\zeta_\beta})
    \frac{1}{2}(\delta_{\epsilon\sigma}+\delta_{\epsilon\tau})
     \bigg[d(\alpha,\beta|\nu,\lambda)
                -e(\alpha,\beta|\nu,\lambda)\bigg]
\\
&& \ \
    -(-1)^{\zeta_\alpha+\zeta_\beta}
   (1-\delta_{\zeta_\alpha \zeta_\beta})\frac{1}{2}
    (\delta_{\epsilon\sigma}-\delta_{\epsilon\tau})
%\\
%&&\hspace{1.5cm}
%    \times 
\bigg [ d(\alpha,\beta|\nu,\lambda)
 + e(\alpha,\beta|\nu,\lambda)\bigg]
\bigg\}
 \label{LnlB}
 \end{eqnarray}
for the Bartlett ($\epsilon=\sigma$) and Heisenberg ($\epsilon=\tau$)
forces and to
 \begin{equation}
      (\Ll|\nu,\lambda)^{\sigma\tau}= \frac{1}{2}
	 \sum_{{{\rm all\ occupied}\atop{\rm  states}}}     
	\bigg\{\delta_{ \zeta_\alpha\,\zeta_\beta}\, 
      \times d(\alpha,\beta|\nu,\lambda)-e(\alpha,\beta|\nu,\lambda)
       \bigg\}
 \label{LnlM}
 \end{equation}
for the Majorana force ($\epsilon=\sigma\tau$).
 The weight and pseudo-weight functions are thus  defined by 
 \begin{equation}
     W_\Lm^{(\epsilon)}(z,\om)=W_0(z)W^{D,\epsilon}_\Lm(z,\om)\,,
     \qquad \epsilon=0,\sigma,\tau,\sigma\tau
 \label{wwwe}
 \end{equation}
 where in Eq. (\ref{WD2}) one uses $(\Lm|n,\ell)^\epsilon$
 instead of 
 $(\Lm|n,\ell)$ and with normalization
 \begin{equation}
    \int |D^{HO}_\Lm(x)|^2\dx^{3A}=\int W_\Lm^{(0)}\rd z\rd\om=\frac{A(A-1)}{2}
 \label{w0a}
 \end{equation}
 \subsection{Weight Function for Coulomb Potential}
 %%%%%%%%%%%%%%%%%%%%%%%%%%%%%%%%%%%%%%%%%%%%%%%%%%%%%%
%
The Coulomb interaction is of Wigner type {\em i.e},  it does
not contain any  spin-isospin exchange operator,  the interaction 
being considered  only between charged particles, in our case 
the  protons. The projection operator applied to $d_{\alpha\beta}(1,2)$ is
 \begin{equation}
       P^{(c)}_{12}d_{\alpha\beta}(1,2)=\frac{1}{4}(1+{\rm sgn\,}
      {\zeta_\alpha})
        (1+{\rm sgn\,}{\zeta_\beta})d_{\alpha\beta}(1,2)
\label{pc}
 \end{equation}
which cancels the neutron states.\par
In the Fourier transform  (\ref{faben}) with $\varepsilon=c$ (for Coulomb)
only the first two cases contribute. Using the same procedure as for the 
other projection operators $P_{ij}^{(\varepsilon)}$ the two-body
 coefficients are given by
 \begin{eqnarray}
 \nonumber
       (\Ll|\nu,\lambda)^{c}&=& \frac{1}{8}
	 \sum_{{{\rm all\ occupied}\atop{\rm  states}}}
              (1+{\rm sgn}( \zeta_\alpha))
           (1+{\rm sgn}(\zeta_\beta))
\\&\times&\bigg\{
      d(\alpha,\beta|\nu,\lambda)
         -\delta_{\zeta_\alpha\zeta_\beta}e(\alpha,\beta|\nu,\lambda)
\bigg\}
 \label{LnlC}
 \end{eqnarray}
The hypercentral potential for the Coulomb  interaction can be be obtained  
directly from the Fourier transform of the Coulomb weight function 
\cite{FFW82}  with
 \begin{equation}
     v(k)=\frac{e^2}{2\pi^2k^2}
\label{vck}
 \end{equation}
the result being
\begin{equation}
         V^c_\Lm(r)=\frac{2}{\pi}\frac{\Gamma(L_m+D/2)}{\Gamma(L_m+(D-1)/2)}
          \frac{e^2}{r}
   \sum_{n=0}^{2\ell_m}\Gamma(n+1/2)(2I_0(n)-I_{\alpha\beta}(n))
\label{vcr}
\end{equation} 
for closed shell nuclei. The coefficients $I_0(n)$ and  $I_{\alpha\beta}(n)$
\cite{FFW82} are given in appendix B. 
 \subsection{ Weight and Pseudo-Weight  Functions for Closed Shell Nuclei}
 %%%%%%%%%%%%%%%%%%%%%%%%%%%%%%%%%%%%%%%%%%%%%%%%%%%%%%%%%%%%%%%%%%%%%
 The various ingredients occurring in the calculation of the weight function
 $W^{(\epsilon)}(z,\om)$ have been defined in terms of the spin-isospin 
 and the eigenstates occupied in the state determinant $D^{HO}_\Lm(\vec x)$
 describing the ground harmonic defining the state of the nucleus.
 The $(\alpha|n,\ell|\beta)$     coefficients are known analytically. 
 Therefore, it is only a matter  of computer  programming to  get the 
 the needed $W^{(\epsilon)}(z,\om)$ functions. Nevertheless, analytical
 calculations can be pushed further when we have to deal
 with closed shell or sub-shell nuclei by using the summation over 
 the spherical harmonics. The analytical expressions are interesting
 because the nucleus is constituted by a core of nucleons in
 closed shells and by other nucleons outside the core. The core
 is spherical with a total angular momentum  $J=0$. The 
 particular state of the nucleus is defined by the configuration of the
 last open shell. The analytical derivation of the coefficients
 for closed shells are given in  appendix B.\par
Since closed shell nuclei are spherical only terms with $\ell=0$
appear in (\ref{WD1}) and (\ref{WD2}).
 A shell $\Lambda_\alpha$ is defined by the value 
$\Lambda_\alpha=2n_\alpha+\la$ in terms of the HO quantum numbers 
of the occupied state. The contribution to the direct term in 
Eqs. (\ref{LnlW}-\ref{wwwe}) for each spin-isospin state when
 all the shells for which $\Laa=2n_\alpha+\la$ are filled
 from $\Laa=0$ to the last shell $\Laa=\ell_m$ is \cite{FFW82}
 \begin{equation}
	 I_0(\ell_m,n)=\frac{(-1)^n}{n!}\sum_{p=0}^n
	  {n\choose p}{\ell_m+3\choose p+3}{\ell_m+3\choose n+3-p}
 \label{I0a}
 \end{equation}
 By taking the four spin-isospin states and the normalization  (\ref{w0a})
  into account, the contribution of the direct term in Eqs. 
(\ref{LnlW}-\ref{wwwe})  for closed shell $N=Z$ nuclei is  \cite{FFW82}
 \begin{equation}
    (\Lm|\nu,0)_d^\epsilon=2I_0(\ell_m,n)\left[4\delta_{0\epsilon}
       +2(\delta_{\sigma\epsilon}+\delta_{\tau\epsilon})
         +\delta_{(\sigma\tau)\epsilon} 
\right]
 \label{lmn0}
 \end{equation}
 where $\epsilon=0,\sigma,\tau,\sigma\tau$.\par
The contribution coming for the exchange term, denoted by $I_{\alpha\beta}$,
cannot be  obtained analytically. However, the procedure for its computation 
is explained  in the appendix. For closed shells  with $N=Z$ spherical nuclei
one has the relation
 \begin{eqnarray}
 \nonumber
    (\Lm|\nu,0)^\epsilon&=&2\bigg \{ I_0(\ell_m,\nu)\left[4\delta_{0\epsilon}
     +2(\delta_{\sigma\epsilon}+
	 \delta_{\tau\epsilon})
       +\delta_{(\sigma\tau)\epsilon}\right]\bigg. 
 \\& -&
    \bigg. I_{\alpha\beta}(\ell_m,\nu)\left[\delta_{0\epsilon} 
	 + 2(\delta_{\sigma\epsilon}+\delta_{\tau\epsilon})
        +4\delta_{(\sigma\tau)\epsilon}\right]\bigg\}
 \label{lmn1}
 \end{eqnarray}
 for the normalization (\ref{w0a}).\par
For the Coulomb case, where only the proton states are taken 
into account in the 
calculation for $I_0$ and $I_{\alpha\beta}$, we have the relation
 \begin{equation}
        (\Ll|\nu,\lambda)^{c}=2I_0(\nu)-I_{\alpha\beta}(\nu)
 \label{LnlCcs}
 \end{equation}
A more compact expression  for the weight functions $W^{(\epsilon)}(z)$ 
can be obtained by using the Rodrigues' formula (Ref. \cite{Erdelyi},
Vol II p. 169)
\begin{equation}
  (-2)^nn!P_n^{a,b}(x)=(1-x)^{-a}(1+x)^{-b}
   D_x^n  \left[(1-x)^{a+n}(1+x)^{b+n}\right]
\label{ca}
\end{equation}
where $D_x^n$ denotes the operator $D_x^n={\rm d}^n/\dx^n$,
$a=L_m+(D-5)/2-n$, and  $b=1/2$. Then, for $\ell=0$ Eq. (\ref{wwwe})
becomes, after integrating Eq. (\ref{WD2}) over $\om$,
\begin{eqnarray}
\nonumber
   W^{(\epsilon)}_\Lm(z)
         &=&\frac{ 2^{-(L_m+D/2-2)}}{\sqrt{\pi}}
        \frac{\Gamma(L_m+D/2)}{\Gamma(L_m+(D-3)/2}
\\&\times& \sum_{n=0}^{2\ell_m}(\Lm|n,0)^{(\epsilon)}
D_z^n
    \left[(1-z)^{L_m+(D-5)/2}(1+z)^{n+1/2}\right]\,.
\label{cwe}
\end{eqnarray}
Alternatively we may write
\begin{eqnarray}
\nonumber
     W^{(\epsilon)}(z)
     &=&\frac{4}{\sqrt\pi}\left(\frac{1}{2}\right)^{L_m+D/2}
        \frac{\Gamma(L_m+D/2)}{\Gamma(L_m+(D-3)/2)}
    \,(1-z)^\alpha(1+z)^\beta
\\ &\times&
      \sum_{n=0}^{2\ell_m}(-2)^n n!\, (\Lm|n,0)^{(\epsilon)}
     (1-z)^{2\ell_m-n}
 P_n^{L_m+(D-5)/2-n,\beta}(z)
\label{rholme}
\end{eqnarray}
The weight function can thus be written as a product
 \begin{equation}
   W_\Lm^{(\epsilon)}(z)=(1-z)^\alpha
          (1+z)^\beta\rho_\Lm^{(\epsilon)}(z)
\label{wlmrho}
 \end{equation}
where now $\alpha=L_m+(D-5)/2-2\ell_m$, $\beta=1/2$,
 $\ell_m$ refers to the sum
$2n+\ell$ of radial and orbital quantum numbers in the last shell,
and  $\rho_\Lm^{(\epsilon)}(z)$ is a polynomial of degree $2\ell_m$,
\begin{equation}
\hspace{-8mm}     \rho_\Lm^{(\epsilon)}|(z)=C^{(\epsilon)}
         \sum_{n=0}^{2\ell_m}(-2)^n n!\, (\Lm|n,0)^{(\epsilon)}
       (1-z)^{2\ell_m-n} P_n^{L_m+(D-5)/2-n,\beta}(z)
\label{rholma}
\end{equation}
The normalization of (\ref{cwe}) is given by the first term, 
{\em i.e} for $n=0$,
\begin{equation}
    \int_{-1}^{+1} W^{(\epsilon)}_\Lm(z)\dz=(\Lm|0,0)^{(\epsilon)}
\label{cc}
 \end{equation}
where we used the fact that for $n>0$
$$
    \int_{-1}^{+1} D_z^n\left[(1-z)^{\alpha}(1+z)^{n+1/2}\right]\dz=0\,.
$$
The  $\rho_\Lm^{(\epsilon)}(z)$, being a polynomial of degree $2\ell_m$,
can be expressed in terms of its roots
\begin{equation}
    \rho_\Lm^{(\epsilon)}(z)=C^{(\epsilon)}
            \prod_{n=1}^{2\ell_m} (z-z_m)
\label{roots}
 \end{equation}
which can simplify the numerical calculations. \par
A similar expression can be obtained in terms of the variable
$X=(1+z)/2=\rij^2/r^2$, $(i,j)=(1,2)$,  for $0\le X\le 1$. 
By substitution one gets
\begin{eqnarray}
\nonumber
   W_\Lm^{(\epsilon)}(z)\dz&=&\frac{ 2}{\sqrt{\pi}}
 \frac{\Gamma(L_m+D/2)}{\Gamma(L_m+(D-3)/2)}
\\&
      \times&\sum_n(\Lm|n,0)^{(\epsilon)}\,D^n_X((1-X)^{L_m+(D-5)/2}
       X^{n+1/2})\rd X
\label{cddd}
\end{eqnarray}
Analytical formulas for the hyperradial part of the two-body potentials
written in terms of $X$ can be obtained from (\ref{cddd}) since
\begin{equation}
\hspace{-0.7cm} \int_0^1f(X)D_X^n\left[(1-X)^{\alpha}
         X^{n+1/2}\right]\rd X
    =(-1)^n\int_0^1(1-X)^{\alpha} X^{n+1/2}
      \frac{{\rm d}^nf(X)}{{\rm d}X^n}{\rm d}X
\label{ce}
\end{equation}
The integral can be obtained  analytically when $f(X)$
is either a polynomial or an exponential or the product of both.

%ccccccccccccccccccccccccccccccccccccccccccc
%              secVI.tex
%ccccccccccccccccccccccccccccccccccccccccccc
\section{The Projection Function}
\label{Sproj}
%%%%%%%%%%%%%%%%%%%%%%%%%%%%%%%%% 
We have seen that in order to have an equation for the
calculation of the two-body correlations, using Eq. (\ref{Feq})
one must extract from the amplitude $F(r_{k\ell},r)$ where 
the pair particles  $(k,\ell)$ is not the reference pair
$(i,j)$, the part of $F(r_{k\ell},r)$ which, nevertheless, depends on 
$r_{ij}$.  There are two cases: Either one of the $(k,\ell)$
is either $i$ or $j$, in which case  we have a connected 
pairs like $(i,k)$ or $(j,k)$ where $k\ne i$ or $j$; 
or $k$ and $\ell$ are neither $i$ nor $j$ and  we  have
disconnected pairs. The procedure is the following
i) Expand $F(r_{k\ell},r)$ in terms of HPs
which are polynomials in $r_{k\ell}^2$ and $r^2$ (which contains
$\rij^2$). ii)  Write $r_{k\ell}^2$  in terms of the kinematical 
rotation vector  (\ref{krov}) \cite{F83,krv}. iii) Extract
 from each polynomial the part which depends
on  $\rij^2$ and  ignore the residual part which contains many body 
correlations \cite{F87}.\par
Practically since the two-body amplitude  in Eq. (\ref{IDEA1}) 
is expressed in terms of the $z=\cos{2\phi}=2\rij^2/r^2-1$  and $r^2$
(for $(i,j)=(1,2)$), the expansion is done for any pair in terms of 
Potential Harmonics (PH) for pairs in $S$-state. They
have been designed in order to provide a
complete expansion basis for pairs of particles in $S$-state. For 
bosons in ground state the PH are the Jacobi polynomials 
$P^{(D-5)/2,1/2}_K(z)$ associated with the  weight function  
$W_0(z)$ (see Eq. (\ref{dOm})).\par
Let $z=\cos{2\phi}=2\rij^2/r^2-1$  for the reference pair
($i=1$, and $j=2$ in Eq. (\ref{Feqp})) and 
$z_{k\ell}=2r_{k\ell}^2/r^2-1$ for another pair. The two-body 
amplitude can be expanded as
\begin{equation}
 \hspace{-6mm}    P(z_{k\ell},r)= \sum_{K=0}^\infty\bigg\{ \frac{1}
  { h^{\alpha,\beta}_K}\int_{-1}^{+1 } (1-z')^\alpha(1+z')^\beta
    P_K^{\alpha,\beta}(z')
   P_K^{\alpha,\beta}(z_{k\ell})\bigg\} P(z',r)\dz'
\label{Pexp}
\end{equation}
where $P_K^{\alpha,\beta}(z) $ are Jacobi polynomials
associated with the weight function 
$ (1-z)^\alpha(1+z)^\beta$ with $\alpha=(D-5)/2$ and $\beta=1/2$. The 
$h_K ^{\alpha,\beta}$ 
 is the normalization constant for the Jacobi polynomials
 \begin{equation}
    h_K^{\alpha,\beta}=\frac{2^{\alpha+\beta+1}\Gamma(K+\alpha+1)
          \Gamma(K+\beta+1)}{\Gamma(2K+\alpha+\beta+1)
      K! \Gamma(K+\alpha+\beta+1)}
\label{hkab}
 \end{equation}
The expression in the braces in (\ref{Pexp}) is the $\delta$-function
$\delta(z'-z_{k\ell})$.\par
It was shown in Ref. \cite{F83} that for the connected pairs where
  $z_c =z_{ik}$ or $z_c =z_{jk}$, $k\ne i$ or $j$, and for equal 
mass particles
$P_K^{\alpha,\beta}(z_c)$ can be separated into two terms
 \begin{equation}
     P_K^{\alpha,\beta}(z_c)=\frac{P_K^{\alpha,\beta}(-1/2)}
      {P_K^{\alpha,\beta}(+1)}P_K^{\alpha,\beta}(z)
       + {\rm other \ terms}
\label{pcon}
\end{equation}
and for disconnected pairs $z_d=z_{k\ell}$, $k,\ell\ne i,j$, 
 \begin{equation}
     P_K^{\alpha,\beta}(z_d)=\frac{P_K^{\alpha,\beta}(-1)}
      {P_K^{\alpha,\beta}(+1)}P_K^{\alpha,\beta}(z)
       + {\rm other \ terms}
\label{pdis}
\end{equation}
Since $P_K^{\alpha,\beta}(z)$  is a HH we have 
 $\Delta r^{2K}P_K^{\alpha,\beta}(z)=0$ whatever the pair is. 
Therefore, $\Delta [r^{2K}\times\left\{{\rm other \ terms}\right\}]=0$
and the residual part is also a HH
orthogonal to $P_K^{\alpha,\beta}(z)$. Since only $ P_K^{\alpha,\beta}(z)$ 
contains two-body correlations, the other terms are related to many-body 
correlations that  we neglect. Finally, the projection function for 
pairs in $S$-states is the sum of the projection functions of the 
$2(A-2)$ connected pairs and the $(A-2)(A-3)/2$ disconnected pairs.
Thus we may write
 \begin{equation}
     {\cal P}^0P(z,r)=\int_{-1}^{+1 } f_\Lz (z,z')P(z',r)\dz'
\label{proj1}
 \end{equation}
with 
\begin{eqnarray}
\nonumber
      f_\Lz (z,z')&=&(1-z')^\alpha(1+z')^\beta(A-2)
\sum_{K=0}^\infty\frac{2 
      P_K^{\alpha,\beta}(-1/2)+\frac{A-3}{2}
          P_K^{\alpha,\beta}(-1)}{ P_K^{\alpha,\beta}(1)}
\\ &\times&
       P_K^{\alpha,\beta}(z) P_K^{\alpha,\beta}(z')/h_K^{\alpha,\beta}
\label{fex}
\end{eqnarray}
An analytical  expression of the projection function has been derived
  \cite{F87,FFS88}.\par
The projection of a potential harmonic for a pair $(k,\ell)$ 
on a reference pair $(i,j)$,  both in $S$-state, is given by \cite{F83}
$$
     \frac{{\tt P}^{\alpha,\beta}_K(\cos{2\delta})}
           {{\tt P}^{\alpha,\beta}_K(1)}\,,\qquad \alpha=(D-5)/2,\ \beta=1/2
$$ where $\delta=2\pi/3$ for connected pairs and $\delta=\pi/2$ 
for disconnected pairs. The projection function in terms of the angular 
parameter $\delta$ is for one pair
\begin{equation}
\hspace{-2mm}
 f_\Lz(z,z',\delta)=(1-z')^\alpha(1+z')^\beta\sum_{K=0}^\infty
      \frac{{\tt P}_K^{\alpha,\beta}(\cos{2\delta})}
     { {\tt P}_K^{\alpha,\beta}(1)}
   {\tt P}_K^{\alpha,\beta}(z)
     {\tt P}_K^{\alpha,\beta}(z')/h_K^{\alpha,\beta}
\label{projf}
\end{equation}
The sum over the series, Eq. (\ref{projf}), can be carried out
analytically \cite{F87,FFS88} providing the projection function 
in terms of $\cos\delta$  which is related to the choice of the 
pair to be projected
where  $\delta=2\pi/3$ or $\delta=\pi/2$ for connected  or
 disconnected pairs  of equal mass particle. In terms of
 $\cos\phi=r_{12}/r=\sqrt{(1+z)/2}$ for the reference pair
\begin{eqnarray}
\nonumber
   && \int_{-1}^{+1 } f_\Lz(z,z',\cos{2\delta}) P(z',r)\dz'
  = \frac{4}{\sqrt\pi}\frac {\Gamma(\lambda+1/2)}
              {\Gamma(\lambda)}
    \frac{1}{\sin{2\delta}\sin 2\phi}
 \\
&&\ \ \ \times\left[\frac{1}{\sin{\delta}\sin\phi}\right]^{2(\lambda-1)}
   \int_a^b\left[(u-a)(b-u)\right]^{(\lambda-1)}P(2u^2-1)u{\rm d}u
\label{fsc}
\end{eqnarray}
where $a=\cos(\phi+\delta)$, $b=\cos(\phi-\delta)$, and $\lambda=
\alpha+1/2=D/2-2$. For disconnected pairs since $\sin2\delta\to 0$
we have to take the limit of Eq. (\ref{fsc}) for $\delta \to \pi/2$
the result being
\begin{eqnarray}
\nonumber
  \int_{-1}^{+1 } f_\Lz(z,z',-1) P(z',r)\dz'
         &=& \frac{2}{\sqrt\pi}\frac {\Gamma(\lambda+1/2)}
              {\Gamma(\lambda-1)}(1-z)^{1/2-\lambda}\\
&\times&
   \int_{-1}^{-z}\left[-(z+z')\right]^{\lambda-2}
      (1+z')^{1/2}P(z',r)\dz'
\label{fsd}
\end{eqnarray}
For fermions and  in particular for nuclei one follows the same 
procedure as for bosons  in $S$-state. One starts from the weight 
function $W_\Lm(z)$ to find the associated polynomials ${\tt P}_K^\Lm(z)$
 which fulfill the normalization condition
 \begin{equation}
       \int_{-1}^{+1}{\tt P}_K^\Lm(z){\tt P}_{K'}^\Lm(z)W_\Lm(z)\dz
        =\delta_{KK'}
\label{pkn}
 \end{equation}
For bosons in ground states we have seen that they are the 
normalized Jacobi polynomials $P_K^{\alpha,\beta}(z)/\sqrt{
    h_K^{\alpha,\beta}}$. From ${\tt P}_K^\Lm(z)$ one generates
the projection function
\begin{eqnarray}
\nonumber
    f_\Lm (z,z') &=&(A-2)\sum_{K=0}^\infty\frac{2 
      {\tt P}_K^\Lm(-\frac{1}{2})+\frac{A-3}{2}
          {\tt P}_K^\Lm(-1)}{ {\tt P}_K^\Lm(1)}\\
    &\times&
      {\tt P}_K^\Lm(z) {\tt P}_K^\Lm(z')
           W_\Lm(z')
\label{fexf}
\end{eqnarray}
where, whatever the normalization of $ W_\Lm(z)$ is,  
the polynomials ${\tt P}_K^\Lm(z)$ are normalized according to 
(\ref{pkn}).\par
Let be $ z_1\cdots, z_{2\ell_m}$ the zeros of $\rho_\Lm(z)$. 
According to the Christoffel's formula \cite{Erdelyi},
the polynomial $\tt P_K^\Lm(z)$ is given by the determinant
\begin{equation}
{\tt P}^\Lm_K(z)=N_K^\Lm/\rho_\Lm(z)
   \left |\matrix { P_K^{\alpha,\beta}(z), &P_{K+1}^{\alpha,\beta}(z),
    &\cdots,  & P_{K+n}^{\alpha,\beta}(z)\cr
	P_K^{\alpha,\beta}(z_1),&P_{K+1}^{\alpha,\beta}(z_1), &\cdots, 
				 & P_{K+n}^{\alpha,\beta}(z_1)\cr
		       \vdots &\vdots&\vdots&\vdots \cr
      P_K^{\alpha,\beta}(z_n),&P_{K+1}^{\alpha,\beta}(z_n), &\cdots, 
			& P_{K+n}^{\alpha,\beta}(z_n)
	    }\right|
 \label{det}
 \end{equation}
 where $N_K^\Lm$ is a normalization constant fixed by Eq. 
(\ref{pkn}) and $P_K^{\alpha,\beta}(z)$ are Jacobi polynomials.
Obviously for bosons in ground state where $L_m=0$
one recovers the Jacobi polynomials for $\alpha=(D-5)/2$
quoted in Eq. (\ref{Pexp}).\par
The expression of ${\tt P}^\Lm_K(z)$ in terms of a determinant,
Eq. (\ref{det}),  presupposes that the roots of the polynomial
$\rho_\Lm(z)$  ($\epsilon=0$) are known.  An alternative way  to obtain
${\tt P}^\Lm_K(z)$ is to use the moments formula \cite{Erdelyi}
\begin{equation}
{\tt P}^\Lm_K(z)=C_K^\Lm   \left |\begin{array}{cccc}
      1,  &(1+z)/2,&\cdots,  & ((1+z)/2)^K\\
         C_0,   &C_1,   &\cdots,  &C_K\\
         C_1    &C_2,   &\cdots, &C_{K+1}\\
       \vdots &\vdots&\vdots&\vdots \cr
         C_{K-1}    &C_K,   &\cdots, &C_{2K-1}
	\end{array}    
     \right|
 \label{det2}
 \end{equation}
where the elements $C_K$ are given by
 \begin{equation}
        C_K=\int_{-1}^{+1}\left(\frac{1+z}{2}\right)^KW_\Lm(z)\dz
\label{moments}
 \end{equation}
and $C_K^\Lm$ is the normalization constant.

%ccccccccccccccccccccccccccccccccccccccccccc
%              secVII.tex
%ccccccccccccccccccccccccccccccccccccccccccc
\section{The Effective Nuclear Potential}
%%%%%%%%%%%%%%%%%%%%%%%%%%%%%%%%%%%%%%%%%
The Eq. (\ref{IDEA1}) for IDEA has been obtained from (\ref{Feqp})
by multiplying at left by $D^*_\Lm(\Om)$ and integrating over all 
angular coordinates $\Om_{N-1}$ associated with the Jacobi 
coordinates $\vec \xi_i$ for $i<N$, $ (\vec \xi_N=\vec r_{12})$ 
and then by dividing by the weight function $W_\Lm(z)$.
Eq. (\ref{IDEA1}) thus obtained is valid  for Wigner-type
 potentials, {\em i.e}, without taking into account the
 exchange operators. In general, however, nuclear potentials 
include exchange operators that give rise to pseudo-weight functions as well.
Thus, the right hand side of Eq. (\ref{Feqp}),
 after integration over $\rd\Om_{N-1}$ gives
$
     \sum_{\epsilon}W_\Lm^{(\epsilon)}(z,\om) V^{(\epsilon)}(\xi_N)
$ 
with $\xi_N=r_{12}=r\sqrt{(1+z)/2}$. For spherical nuclei the weight 
function does not contain the angular coordinates $\om$
of $\vec \xi_N$ and we may define
 \begin{equation}
    W_\Lm^{(\epsilon)}(z)=\int W_\Lm^{(\epsilon)}(z,\om) \dom
\label{welmzi}
 \end{equation}
and the effective potential in the  right hand side of 
 Eq. (\ref{IDEA1}) reads
\begin{equation}
 V(\xi_N)\equiv    V_{\rm eff}(\xi_N)=\hspace*{-2mm}
      \sum_{\epsilon= 0,\sigma,\tau,\sigma\tau}
      V^{(\epsilon)}(\xi_N)  W_\Lm^{(\epsilon)}(z) /W_\Lm^{(0)}(z) 
\label{effV}
 \end{equation}
while the effective hypercentral potential in Eq. (\ref{IDEA1}) is
\begin{equation}
   V_\Lm(r)= \int_{-1}^{+1}W_\Lm^{(0)}(z) V_{\rm eff}(\xi_N)\dz
   = \int_{-1}^{+1}\sum_{(\epsilon)}W_\Lm^{(\epsilon)}(z) 
         V^{(\epsilon)}(\xi_N)\dz
\label{hcaeff}
\end{equation}
The  nuclear potentials are given in terms of the triplet
or singlet for even or odd states.
The following relations hold between the 
$V^{(\epsilon)}$, where $\epsilon=0,\ \sigma,\ \tau, \ \sigma\tau$,
and the $V^{3+}$, $V^{1+}$, $V^{3-}$, and the $V^{1-}$
potentials
\begin{eqnarray}
\label{nuclV1}
%\nonumber
     V^0&=&\frac{1}{4}\left[ V^{1+}+V^{3+}+ V^{1-}+V^{3-}\right]\\
\label{nuclV2}
%\nonumber
     V^\sigma&=&\frac{1}{4}\left[-V^{1+}+V^{3+}- V^{1-}+V^{3-}\right]\\
\label{nuclV3}
%\nonumber
     V^\tau&=&\frac{1}{4}\left[V^{1+}-V^{3+}- V^{1-}+V^{3-}\right]\\
%\nonumber
     V^{\sigma\tau}&=&
        \frac{1}{4}\left[-V^{1+}-V^{3+}+V^{1-}+V^{3-}\right]
\label{nuclV4}
\end{eqnarray}
In terms of  Wigner, Bartlett, Heisenberg and Majorana 
potentials, the following relations hold
 \begin{equation}
        V^W=V^0,\quad V^B=V^\sigma,\quad V^H=-V^\tau,\quad V^M=-V^{\sigma\tau}
 \end{equation}
For the $N=Z$ nuclei constructed from spin and isospin
saturated HO states, like the $\alpha$-particle, the relation 
(\ref{lmn1}) still is valid and the effective potentials 
associated with the direct  and  exchange terms $I_0$ and
$I_{\alpha\beta}$ are 
 \begin{equation}
     V_{I_0}=\frac{3}{2}\left(V^{1+}+V^{3+}\right)
          + \frac{1}{2}V^{1-}
        + \frac{9}{2}V^{3-}
\label{vio}
 \end{equation}
for  $I_0$ and
 \begin{equation}
     V_{I_{\alpha\beta}}=-\frac{3}{2}\left(V^{1+}+V^{3+}\right)
          + \frac{1}{2}V^{1-}
        + \frac{9}{2}V^{3-}
\label{viab}
 \end{equation}
for  $I_{\alpha\beta}$.\par
The effective potential can be alternatively defined in terms 
of the polynomials
 \begin{equation}
      \rho^{(D)}_\Lm(z)=\sum_n^{2\ell_m} 2^n n! I_0(n)(z-1)^{2\ell_m-n}
      P^{\alpha-n, 1/2}_n(z)
\label{rhod}
\end{equation}
with $\alpha=L_m+(D-5)/2$ and
 \begin{equation}
      \rho^{(E)}_\Lm(z)=\sum_n^{2\ell_m} 2^n n! I_{\alpha\beta}(n)
          (z-1)^{2\ell_m-n} P^{\alpha-n, 1/2}_n(z)
\label{rhoab}
\end{equation}
the result being
\begin{equation}
    V_{\rm eff}(r_{ij})=\frac{1}{4}\frac{ \rho^{(D)}_\Lm(z)V_D(r_{ij})
                        -\rho^{(E)}_\Lm(z)V_E(r_{ij})}
                        {4\rho^{(D)}_\Lm(z)-\rho^{(E)}_\Lm(z)}
\label{vefde}
\end{equation}
where $V_D=2V_{I_0}$ and $V_E=2 V_{I_{\alpha\beta}}$
  and where the polynomial occurring in the weight function 
is
 \begin{equation}
         \rho^0_\Lm(z)=2\left[ 4\rho^{(D)}_\Lm(z)-\rho^{(E)}_\Lm(z)\right]
\end{equation}
It is interesting to note that the one pion exchange potential (OPEP) defined 
by
 \begin{equation}
         V_{\rm OPEP}(r_{ij})=(\vec \sigma_i\cdot\vec \sigma_j)
     (\vec \tau_i\cdot\vec \tau_j)Y(r_{ij})
\label{VOPEP}
\end{equation}
where 
$$
       Y(r_{ij})=V_0\frac{\re^{-\mu r_{ij}}}{\mu r_{ij}}
$$
for $3V_0\sim 10$\,MeV and $\mu \sim 0.7$\,fm$^{-1}$ does not
 contribute to the  direct term. Indeed, since
\begin{eqnarray}
\nonumber
         V_{ \rm OPEP}^{1+}&=& V_{ \rm OPEP}^{3+}=-3Y(r_{ij})
\\\nonumber         
        V_{ \rm OPEP}^{1-}&=&9 Y(r_{ij})
\\\nonumber         
         V_{ \rm OPEP}^{3-}&=& Y(r_{ij})
\end{eqnarray}
the contribution in the direct term disappears and amounts to 
$18 Y(r_{ij})$ in the exchange term, Eq. (\ref{viab}).\par
Since the contribution to the effective potential of
the exchange term decreases rapidly for increasing $A$,
$I_0(0)=(A/4)^2$, while  $I_{\alpha\beta}(0)=A/4$ 
%(see Tables I and II in Appendix)) 
the contribution of the OPEP, which is the
dominant term in the description of the long range part of the
 nucleon-nucleon potential, fades away for large nuclei.
\subsection{Spurious component}
%--------------------------
In the Schr\"odinger equation,  Eq. (\ref{Schr2}),
the sum over all pairs  of the two-body potential contributes
while only one component $V(\rij)$ associated with the reference 
pair $(i,j)$ appears in the amplitude of Eq. (\ref{Feq}).
The potential can be expanded on the complete potential basis 
associated with the weight function. The polynomials of degree one
$$
   P_1(z)=\left( \frac{1+z}{2}
       -\frac{1}{A-1}\right)/N=\left( \frac{\rij^2}{r^2}
              -\frac{1}{A-1}\right)/N
$$
where $N$ is the normalization constant
given by
$$
    N^{-2}=\int_{-1}^{+1}\left( \frac{1+z}{2}
       -\frac{1}{A-1}\right)^2 W_\Lm(z)\dz\,,
$$
is independent of $\Lm$ and disappears when the sum is taken over all
pairs, $\sum_{i,j>i}\rij^2/r^2=A/2$. Therefore one term occurring 
in the expansion of the potential disappears in the Schr\"odinger
equation.\par
The occurrence of this spurious component in the potential  has been 
discussed in Ref. \cite{vide1,FL92}. This component contributes in the SIDE, 
{\em i.e}  when  $V_\Lm(r)$ in Eq.~(\ref{IDEA1}) is set to zero since in 
the right hand side $P(z,r)$ is an amplitude where the reference pair
$(i,j)$ is in $S$-state while $V(r_{ij}){\cal P}^0$  operates as an
 $S$-state projected potential that vanishes when in the other amplitudes 
expanded in terms of the Jacobi coordinates, Eq. (\ref{Jacobi}),
the reference pair is not in $S$-state \cite{F90}.\par
For three-bodies this equation is known as the Faddeev equation 
for $S$-states projected potentials. In the IDEA we assume that 
the potential is local and we isolate the hypercentral part of the 
potential, $V_\Lm(r)$, leaving only the residual potential to operate, 
in the r.h.s of Eq.~(\ref{IDEA1}), on pairs in $S$-states. When the sum in 
Eq. (\ref{Feq}) is performed over all pairs $(i,j)$, one obtains
the Schr\"odinger equation Eq. (\ref{Schr2}).\par
 Since the component $P_1(z)$ in the expansion of $V(\rij)$
disappears in the sum  $\sum_{i,j>i}V(\rij)$ occurring in 
the Schr\"odinger equation, it must be canceled in $V(\rij)$
for local potentials.
Therefore, in order to avoid taking spurious part of the potential
operating in IDEA,  one should consider instead
 \begin{equation}
   \tilde V(\rij)=V(\rij)-V_1(r)P_1(z)
\label{vtilde}
\end{equation}
where 
 \begin{equation}
     V_1(r)=\int_{-1}^{+1} V(r\sqrt{(1+z)/2})P_1(z)
           W_\Lm(z)\dz
\label{v1}
\end{equation}
Since 
$$
    \sum_{\ell,k>\ell}\left( \frac{r^2_{k\ell}}{r^2}
       -\frac{1}{A-1}\right)=0
$$ 
and for the reference pair
$$
   {\cal P}^0P_1(z)=P_1(z)\,,
$$
 the projection on this pair  is
$$
   {\cal P}^0\sum_{\ell,k>\ell\atop{\ne i,j}} P_1(z_{k,\ell})=-P_1(z)
$$
which is in agreement with the coefficient
$$ 
    (A-2)\left[2P_1(-1/2)+\frac{A-3}{2} P_1(-1)\right]/P_1(1)=-1
$$
occurring  in the projection function, Eq.~(\ref{fex}).
%

 %ccccccccccccccccccccccccccccccccccccccccccc
%              secVIII.tex
%ccccccccccccccccccccccccccccccccccccccccccc%
\section{Spin-isospin Exchange Generated Elements of the State}
%%%%%%%%%%%%%%%%%%%%%%%%%%%%%%%%%%%%%%%%%%%%%%%%%%%%%%%%%%%%%5
Once the problem for the closed shell nuclei in ground state
%(where only one determinant of minimal degree exists)
has been solved, with the inclusion of two-body correlations, 
one may argue that at the level of  accuracy where many-body 
correlations are neglected, the solution obtained is complete. 
However, the exchange operators occurring in the
potential generate new states of grand orbital $L_m+2$ which 
can not be reached by solving the single IDEA equation,
 Eq.  (\ref{IDEA1}).\par
To see how to include these states, let us begin with the 
$\alpha$-particle where the problem is well known (the same 
situation holds for three nucleons in ground state). For three- 
and four-body systems, besides the space symmetric
 state associated with the spin-isospin antisymmetric state, 
the spin and isospin exchanged operators can generate also
mixed symmetry states which must be taken into consideration
in constructing the fully antisymmetric state. In such a  case
we have a sum of  space and spin-isospin states products each 
state of the product having a definite symmetry 
in the exchange of two selected particles.
To understand how such states are generated let
 \begin{equation}
      D_\alpha=|| \alpha p \ \ \beta p \ \ \alpha n \ \ \beta n ||
\label{si1}
\end{equation}
be the spin and isospin antisymmetric Slater determinant associated with 
 the fully-symmetric space state of $^4$He in ground state, 
where $\alpha(\beta)$ denotes the spin-up(down) while $p$ denotes 
the proton and $n$ the neutron. The ground polynomial, from which 
the expansion of the $^4$He in ground state starts, is $D_\alpha Y_\Lz$ 
where $Y_\Lz$ is a constant, in fact, a HH of order zero, {\em i.e}, 
with $L_m=0$. Applying, for example,   the Bartlett spin-exchange 
operator $ P^\sigma_{ij}$ on $D_\alpha$ one obtains either 1 when 
the two spins are the same or  -1 when the  exchanged spins 
between two identical particles (protons  or neutrons) are opposite,
or 0 when the exchange is between two different spins belonging
to two different particles. Indeed, in the last case the determinant has
two pairs of identical columns and thus  disappears, {\em i.e},  one
gets that $\sum_{ij>i} P^\sigma_{ij}D_\alpha=0$. Similarly,
if the isospin is exchange between 
either the first and the last column or between the second and 
third column for which $\zeta_\alpha +\zeta_\beta=0$, one gets a 
determinant with two pairs of identical columns and thus  also
disappears. \par
If now one considers the scalar operator  $\sum_{ij>i} r^2_{ij}
  P^\sigma_{ij}$ the exchange between the same columns generates 
new HPs, namely,
 \begin{equation}
     || \beta_1 p_1\vec x_1 \ \ \beta_1 p_1 \ \ \alpha_1 n_1 
\ \ \alpha_1 n_1\vec x_1 ||
\label{si2a}
\end{equation}
and
 \begin{equation}
     || \alpha_1 p_1 \ \ \alpha_1 p_1\vec x_1 \ \ \beta_1 n_1\vec x_1 
\ \ \beta_1 n_1 ||
\label{si2b}
\end{equation}
where in the expansion of the determinant with respect to the first two
rows the  scalar product $\vec x_1 \cdot \vec x_2 $ appears.
These two determinants are  HP of degree two which cannot be generated
by the Wigner  or Majorana potentials and should be included in the 
description of $^4{}$He. These are the so-called mixed symmetry states
which are coupled to the fully symmetric $S$-state through the 
spin-exchange  (Bartlett) potential. \par
Apart from  the scalar spin-exchange operator $r^2_{ij} P^\sigma_{ij}$
one can apply also the scalar isospin-exchange operator
$r^2_{ij} P^\tau_{ij}$. The generated determinants
 \begin{equation}
      || \alpha n\vec x \ \ \beta p \ \ \alpha n \ \ \beta p \vec x ||
\label{si3a}
\end{equation}
and
 \begin{equation}
      || \alpha p \ \ \beta n\vec x \ \ \alpha p\vec x \ \ \beta n ||
\label{si3b}
\end{equation}
are HPs  of opposite sign to those generated by the spin-exchange operator,
{\em i.e}
\begin{equation}
     r^2_{ij} P^\tau_{ij}D_\alpha=-r^2_{ij} P^\sigma_{ij}D_\alpha\,,
\label{si4}
\end{equation}
 and thus one can consider only HPs generated by the spin-exchange operator
and coupled to the space symmetric state $D_\alpha$ through the 
potential
\begin{equation}
  \left(   V^{\sigma}(r_{ij}) P^\sigma_{ij}
      +V^{\tau}(r_{ij}) P^\tau_{ij} \right)D_\alpha
=\frac{1}{2}\left( V^{3+}(r_{ij})-V^{1+}(r_{ij})
     \right)P^\sigma_{ij} D_\alpha
\label{si5}
\end{equation}
where $V^{(3+)}$ and  $V^{(1+)}$ are the triplet even
and singlet  even potentials respectively.
It is clear that the HPs generated by the operator 
$r^2_{ij} P^\sigma_{ij}D_\alpha$ cannot be neglected.
They constitute the mixed symmetry component of the $^4$He wave function
coupled, through the spin- and isospin-exchange
operators, to the space symmetric  component of the 
wave function.\par
Similar components are generated by the exchange operators in
nuclei. They should be taken into account  explicitly  as they bring  a 
non-negligible contribution to the binding energy. The potentials
associated with the spin-  or the isospin-exchange operators depend on 
the difference   between  the triplet $V^{3+}$  and singlet   $V^{1+}$
even potentials which, for realistic potentials, is rather weak. 
Indeed, the contribution to the binding
energy of the mixed symmetry state in three- and four-body bound
states does not exceed a few MeV. In contrast to the Bartlett potentials,
%Eq. (\ref{si5}), 
however, the Majorana potential depends on the sum of these
potentials which is large. It is responsible,
together with the Wigner potential, to nearly all the binding 
energy in few-body systems. The spin-isospin exchange operator 
does not generate mixed symmetry states in the tri-nucleon or the 
alpha-particle  system but in nuclei it does and therefore it can not 
be neglected.\par
Let us note by $D_\alpha |n,\ell,m>$ the part of a system of particles
where four nucleons are in the same space
 HO state $|n,\ell,m>$ and the spin and isospin states are
saturated   as in $D_\alpha$ (see Eq. (\ref{si1})). For instance,
 for $^{16}$O the HO wave function would be written as
\begin{equation}
    D^{HO}_\Lm=D_\alpha|0,0,0\rangle D_\alpha|0,1,-1\rangle
\,D_\alpha|0,1, 0\rangle\,D_\alpha|0,1,+1\rangle
\label{dho}
\end{equation}
where $D_\alpha |n,\ell,m>$  means that each of the four spin-isospin
states in $D_\alpha$  is in the same space $ |n,\ell,m>$ state.
Let us  further consider  the operator 
\begin{equation}
    {\cal O}=\sum_{i,j}(1-\delta_{\zeta_i \zeta_j})P^{\sigma\tau}_{ij}
\label{calo}
\end{equation}
which exchanges the spin-isospin states $\zeta_i$ and  
$\zeta_j\ne \zeta_i$. When
the exchange is between spin-isospin states in different states 
$|n,\ell,m>$ and $|n',\ell',m'>$ it generates a determinant with two 
identical pairs of columns and then disappears. Proceeding, as for 
for spin and isospin,  the scalar operator 
 \begin{equation}
   \sum_{i,j} \rij^2 (1-\delta_{\zeta_i \zeta_j})(1-\delta_{\alpha_i\alpha_j})
       P^{\sigma\tau}_{ij}\,,
\label{scop}
\end{equation}
where 
$$
       \delta_{\alpha\alpha'}=
\cases{
            1 & {\rm when}\ 
          $ n_\alpha=  n_\alpha',\la=\la', m_\alpha=m_\alpha'$\cr
           0 & {\rm otherwise}\cr
}
$$
applied to different space states $|\alpha>$ and $|\alpha'>$
generates HPs of degree $L_{m}+2$ which should be 
included in the description of  the  wave function. They should much 
contribute to the binding energy of the nucleus since they are coupled to
the main component of the wave function  through the Majorana potential
which contains the sum of the triplet and singlet even potentials.\par
It has already been shown, by solving the $^6$Li problem in  
the hypercentral $L_m$-approximation \cite{Li6}, that the contribution 
to the binding energy of the mixed symmetry state generated by the 
Majorana exchange operator is of the order of a few MeV.
For a HO Serber force which is fitted in order to give the experimental size 
and binding energy   of $^4$He, the eigen-energy calculated for $^6$Li
 without the inclusion of the mixed symmetry state is above the one
of  $^4$He \cite{Li6}. An exact calculation could not 
provide, for potentials vanishing at infinity,  an eigen-energy above 
the one of $^4$He since one should at least obtain the $^4$He binding energy
for $^6$Li when the two nucleons in the $p$-shell are in zero energy 
scattering state.\par
When the operator (\ref{calo})  operates on closed shell it generates 
two pairs of  identical columns. Therefore, the operator (\ref{scop}) in 
order to remove the cancellation of the determinant must change the space 
state of the  incriminated columns. It means that only the scalar product
$\vec x_i\cdot \vec x_j$ in $\rij^2$ must be taken into account.
But the scalar product written as 
 \begin{equation}
    \vec x_i\cdot\vec x_j=x_ix_j P_1(\cos\varphi)=x_ix_j\frac{4\pi}{3}
         \sum_{m=-1}^{+1} Y_1^{m*}(\om_i)Y_1^{m}(\om_j)\,,          
\label{xidxj}
\end{equation}
where $\varphi$ is the angle between the directions
$\om_i$ and $\om_j$ and $P_1(\cos\varphi)$ is a Legendre polynomial,
is a harmonic polynomial of degree two as the product of two
 harmonic polynomials of degree one
 in the variable $\vec x_i$ and $\vec x_j$.\par
When the operator 
 \begin{equation}
   Q^{\sigma\tau}=\sum_{i,j>i}
    \vec x_i\cdot\vec x_j\,[1-\delta_{\zeta_i \zeta_j}]
      (1-\delta_{\alpha_i\alpha_j}) P^{\sigma\tau}_{ij}
\label{qop}
\end{equation}
is applied to the state   $D^{HO}_\Lm(\vec x)$ it generates
a sum of new HO states $D^{HO}_{[L_m+2]}(\vec x)$ which are HP  of 
degree $L_m+2$. The new polynomials are indeed 
constructed according to the procedure described in Sect. II for HPs.
Let us call 
$$
        D^{\sigma\tau}_{[L_m+2]}(\vec x)=Q^{\sigma\tau}
        D^{HO}_\Lm(\vec x)
$$ 
the sum of the HO Slater determinant  generated by the 
$Q^{\sigma\tau}$ operator. Each one is constituted by the 
original  $ D^{HO}_{[L_m]}(\vec x)$ Slater determinant where
two HO individual states in the last occupied shell have been raised
to the next shell. The $ D^{HO}_{[L_m+2]}(\vec x)$ must  be 
normalized and the weight and pseudo-weight functions and projection
function must be calculated. Then the coupling between the 
$ D^{HO}_\Lm(\vec x)$ and  $D^{HO}_{[L_m+2]}(\vec x)$ states 
generated  by the spin-isospin exchange operator can be introduced
 through the exchange pseudo-weight function
$$
  \int
        D^{\sigma\tau}_{[L_m+2]}(\vec x) P^{\sigma\tau}_{ij} 
           D^{HO}_\Lm(\vec x){\rm d}\Om_{N-1}\,.
$$
%ccccccccccccccccccccccccccccccccccccccccccccccccccccccccccc  
%               secIX.tex 
%cccccccccccccccccccccccccccccccccccccccccccccccccccccccccc  
%%%%%%%%%%%%%%%%%%%%%%
\section{Applications}
%%%%%%%%%%%%%%%%%%%%%%
To solve the basic equation (\ref{IDEA1}) for a specific 
nucleus and a specific filling of the HO shells, one has first 
to evaluate the coefficients  $(\Lm|n,\ell)^{(\epsilon)}$
from which the weight functions (\ref{wlmrho})  and the  density
functions $\rho_\Lm^{(\epsilon)}(z)$, Eq. (\ref{rholma}), 
or equivalently their roots (\ref{roots}) can be obtained.
 These coefficients are evaluated from Eqs. (\ref{LnlW}), (\ref{LnlB}), 
and (\ref{LnlM}) as well as from Eq. (\ref{LnlC}) when Coulomb 
forces are included.
 The results for the closed shell nuclei 
$A=16$ and $A=40$ and for the  closed shell neutron systems $N=8$ 
and $N=20$ are given in table \ref{coeft} while the corresponding
roots for $\rho_\Lm^{(\epsilon)}(z)$ are given in table \ref{r1640}
for the $^{16}$O and $^{40}$Ca nuclei and  in table \ref{r820}
for the closed shell 8-neutron  and 20-neutron systems.
It is noted here that only the relative normalization constants 
$C^{(\epsilon)}/C^{(0)}$ are required to construct the effective 
potential defined by
\begin{eqnarray} 
\nonumber
 V_{\rm eff}(\rij)&=&
      \sum_{\epsilon= 0,\sigma,\tau,\sigma\tau}
      V^{(\epsilon)}(\rij)  W_\Lm^{(\epsilon)}(z) /W_\Lm^{(0)}(z)
\\&\equiv&
      \sum_{\epsilon= 0,\sigma,\tau,\sigma\tau}
        V^{(\epsilon)}(\rij)  \rho_\Lm^{(\epsilon)}(z) 
      /\rho_\Lm^{(0)}(z) 
\label{effVr}
\end{eqnarray}
where the nuclear potentials $V^{(\epsilon)}$ are given by
Eqs. (\ref{nuclV1}--\ref{nuclV4}) in terms of the singlet or triplet,
even or odd potentials.
The extracted ratios $C^{(\epsilon)}/C^{(0)}$ are also  given in tables 
\ref{r1640} and \ref{r820}. It is further noted that the 
 weight function for the $^4$He system, is given by 
(\ref{wlmrho}) with $\rho^{(\epsilon)}(z)=1$ for $L_m=0$.\par
The next task is the choice of the potential. Since we are 
testing here the suitability of our method in nuclear
structure calculations, we employed three interactions
with different characteristics and for which results by other 
competing methods are available. 
The first potential employed is the widely used in nuclear structure 
calculations Brink and Boeker B1 effective soft core potential \cite{B1}
which is  of a rather long range and of soft core.
Although this potential is not realistic, since it does not fit
the scattering N-N phase shifts, there exist in the literature
a lot of results obtained with it and therefore its use is warranted 
for comparison purposes. For convenience, we recall here this potential
which  has only Wigner and Majorana components  
\begin{eqnarray}
\nonumber
     V_W&=&595.55\exp(-2.041r^2)-72.212\exp(-0.512r^2)\,,\\
     V_M&=&-206.04\exp(-2.041r^2)-68.388\exp(-0.512r^2)\,.
\nonumber
\end{eqnarray}
 The second potential used is the Afnan and Tang  \cite{AT} S3  
potential. The original potential was adjusted to the static 
properties of $^4$He nucleus. However, in order to extend
its applicability to heavier nuclei, Guardiola and collaborators 
\cite{Guard} added  a repulsive part  in the singlet- and triplet-odd
components, the modified  potential thus obtained (known as 
MS3 potential) being
\begin{eqnarray}     
\nonumber  
    V_W &=&- 5.75\exp(-0.4r^2) -10.75\exp(-0.6r^2)\\
\nonumber  
      &-&41.5\exp(-0.8r^2)-81.675\exp(-1.05r^2)  +1000\exp(-3r^2)\\
\nonumber
    V_M &=& 5.75\exp(-0.4r^2) +10.75\exp(-0.6r^2)
      +41.5\exp(-0.8r^2)\\
\nonumber  
     V_B&=&-V_H= 5.75\exp(-0.4r^2)-10.75\exp(-0.6r^2)\\
\nonumber
     &+&41.5\exp(-0.8r^2)  -81.675\exp(-1.05r^2)
\end{eqnarray}  
This potential has a very strong repulsive core which give rise to strong
two-body correlations. As a third potential we employed the more
realistic soft core  interaction of Gogny, Pires, and de Tourreil (GPDT)
 \cite{GPDT} which was also used in the past in nuclear structure 
calculations.\par
Using the aforementioned potentials we firstly investigated
whether  8 and 20 neutrons, forming a closed shell system, 
can sustain  a bound state. It was found that, at least with the
potentials employed, no bound state exists. Even with the soft B1
potential a bound state can only be generated by modifying the 
potential by more than 20\%. \par
The ground state energies $E_g$   and root mean square radii $r_{\rm rms}$ for
the closed shell nuclei  $^4$He, $^{16}$O, and $^{40}$Ca
are given in tables  \ref{He}, \ref{O16}, and \ref{Ca40}.
The results were obtained in the extreme and uncoupled adiabatic
approximations and are compared with other results obtained
with cluster expansion  method (FAHT) \cite{Guard}, 
Bruekner-Hartree-Fock type (BHF) \cite{Guard}, Fermi-Hyper-Netted-Chains 
 (FNHC) \cite{FHNC},  variational Monte Carlo (VMC) \cite{VMC1,VMC2}
as well with results obtained with hyperspherical harmonics expansion
methods (HHE) \cite{HHE}. We refer also
to the work of Guardiola {\em et al.} \cite{Guard96} where more relevant 
results are compiled  using  various techniques.
The  importance of the correlations,
stemming mainly from the short range repulsion of the two-body force, 
is also inferred from these tables by comparing the IDEA results with 
those of hypercentral approximation. \par
 It should be noted that the results obtained by our method do not involve 
any adjustable parameter unlike, for example, the FHNC method whose 
results depend on the Jastrow ansatz, and in general on the  model 
wave function employed. This can be seen, for instance, in 
table \ref{O16} where the use of different ansatzs in the FHNC 
resulted in $\sim$13\,MeV difference in the ground state energy 
with B1 potential.\par
The inclusion of the Coulomb potential is of utmost  importance
when one is dealing with nuclei involved in reactions of astrophysical 
interest. Its inclusion, however, with the exact integral, differential,
or integro-differential  methods   employed in Few-Body calculations
is non-trivial and thus it is usually omitted. However, the 
incorporation of Coulomb forces  in our formalism is
straightforward. The extra repulsion  generated by the Coulomb 
potential is also given in tables \ref{He}, \ref{O16}, and \ref{Ca40}. 
As expected, the smaller the rms radius, the higher
the eigen-energy is.\par
We calculated the EAA and UAA giving respectively  a lower- and
upper-bound to the eigen energy. The  difference decreases from 
$\sim0.9$\,MeV to $\sim0.1$\,MeV for A growing from A=4 to A=40.
By taking the average for the eigen energy, the difference with 
respect to the exact value is less than 0.5\,MeV  for $^4$He and
becomes negligible for increasing A.  The values obtained are in 
agreement with the spectrum of those obtained by other methods
except for $^{40}$Ca with the MS3 potential where we got a 
significant lower binding but unfortunately variational values are 
not available for comparison. We notice that the strongest the 
core of the potential, the larger is the increase of binding 
energy brought by the  correlations. Since most realistic 
potentials have strong repulsive core, we expect that the binding 
in nuclei  originates from the correlations and that the effect 
of the  hypercentral potential is to balance the kinetic energy 
only.

As an example for the applicability of our formalism to
open shell nuclei we consider the $^{10}$B nucleus.
The ground state of this nucleus is known to be $J^\pi=3^+$, $T=0$.
Two likely ground configurations  with this state are the
$$
           \psi_1=  (4s_{1/2})(4p_{1/2})(2p_{3/2})
$$ 
configuration where the $p_{1/2}$ nucleons have the quantum numbers 
$J=0$ and $T=0$ and the
 $$
        \psi_2=(4s_{1/2})(6p_{3/2})
$$
configuration where the two holes in the, otherwise, full $p_{3/2}$
subshell are also coupled to $J^\pi=3^+$, $T=0$. 
The form of the density function $\rho^{(\epsilon)}_\Lm(z)$ 
in this case is given by \cite{FFW82,FN79a,Adam},
\begin{equation}
\label{rhoB10}
    \rho^{(\epsilon)}_\Lm(z)=\sum_{n=0}^{2\ell_m}
                \, \langle\Lm|n,\ell\rangle^{(\epsilon)}
        (1-z)^{L_m-n}(1+z)^n
\end{equation}
where the expansion coefficients are given by
\begin{equation}
\label{ketA}
      \langle\Lm|n,\ell\rangle^{(\epsilon)}=\frac{
        \Gamma(L_m+D/2)}{\Gamma(L_m+D/2-n-3/2)
         \Gamma(n-3/2)}A_n^{\epsilon}\,.
\end{equation}
The $A_n^{\epsilon}$ are calculated from the shell
structure of the nuclear state under discussion by means of
the Talmi-Moshinsky or Gogny coefficients \cite{Athesis,Brody}.
The explicit values for these coefficients are given in table 
\ref{Acoef} while the roots and the relative normalization 
constants for the polynomial $\rho^{(\epsilon)}_\Lm$  are given in 
table \ref{boronr}. The corresponding binding energy results for 
the two configurations are given in table \ref{B10}.\par
%ccccccccccccccccccccccccccccccccccccccccccccccccccccccccccc  
%               secX.tex 
%cccccccccccccccccccccccccccccccccccccccccccccccccccccccccc  
%%%%%%%%%%%%%%%%%%%%%%
%%%%%%%%%%%%%%%%%%%%%%
 \section{Conclusions}
 %%%%%%%%%%%%%%%%%%%%%%
When the structure of nuclei has been identified as associated
to the HO quantum numbers corrected by the effect of a strong 
spin-orbit force, it was taken for granted that nuclei can  
be described as an IPM. In the present work we have shown
that the HO is both an IPM and a collective model and that
the HO potential is a hypercentral potential invariant by rotation 
in the D-dimensional
space spanned by the particle coordinates in the center of mass frame.
This leads to another interpretation where the most significant
part of the interaction is hypercentral in such a way that in 
the ground state  the repulsive centrifugal barrier must be minimum.
This  interpretation also leads to the appearance of the so-called
magic numbers in nuclei whatever the nuclear potential is.\par
We have further shown that  the next improvement after the HCA, where
the wave function is described by the product  of an antisymmetric
harmonic polynomial and a function of the hyperradius,  is obtained by 
introducing the two-body correlations generated by the nuclear 
two-body potential. This results to an integro-differential equation
(IDEA) for the A-body bound system. In order to test the quality of 
this approximation where only one  harmonic polynomial  of minimal 
degree and two-body correlations are taken into account in the 
wave function,   we compared the binding energies obtained with the 
IDEA to those computed for central forces by other methods. 
We found values in agreement with the spectrum of those
available in the literature, the largest differences being 
in the case of $^{40}$Ca  with S3 potential which has
a strong repulsive  core. Unfortunately, no
variational results are available in the latter case. \par
Our formalism is based of course on the assumption that 
the NN force is the dominant one inside nuclei. 
It is well known, however,  that three-body forces are also 
essential in the description of nuclear dynamics
and their introduction might explain at least a part of the
underbinding of nucleons in nuclei interacting by two-body forces 
only. Their employment in our basic equation 
(\ref{IDEA1}) is straightforward
when they are symmetrical and separable,
$$ 
     W(\vec x)=\sum_{i<j<k\le A}V^{(3)}(r_{ij})V^{(3)}(r_{jk})\,,
$$
as described in Ref. \cite{FFS88,v3p}.\par
Up until now it was possible to solve three- and four-body problems 
with accuracy by using the Faddeev and Faddeev-Yakubovsky 
equations respectively. For more nucleons it is customary to assume 
that nucleons in  nuclei form clusters, and then to construct
effective inter-cluster interactions for the sake of solving
the problem as a three- or four-body.
The construction of the effective interactions, however,
is a formidable task with all sort of ambiguities naturally 
creeping in due mainly to insufficient scattering data 
required in the construction of the nuclear forces involved.\par
The existence or not of bound and of resonance states
 in three- and four-neutron has 
been the subject of numerous theoretical and experimental 
 works \cite{Vermark,Timo}.
 Going beyond the four-neutron system \cite{Gent} is  
also important as nuclear matter exhibits a much richer phase 
structure than light nuclei because of stronger
correlations  between nucleons. However, our results for the 8- 
and 20-neutron systems indicate that
realistic nucleon-nucleon central forces alone  are not sufficient
to generate closed shell bound systems.\par
The huge differences between the results obtained with 
the three different two-body forces used, indicates 
that one has to employ a realistic two-nucleon potential
supplemented with a three-body force that accounts for the missing 
binding energy.  The use of effective forces, either at the two-body 
level, or  cluster-cluster potentials, or optical
potentials in processes where the details of the interior part
of the wave function plays an important role, could provide
misleading results. Such  processes are, for example,
the electromagnetic reactions, in which  the transitions 
are guided by strict selection rules  and  depend on the details  
of the wave functions in each channel  and their
characteristics in the interior region.  
Therefore, the inclusion of the  underlying correlations in an unambiguous 
way is of crucial importance.

%&&&&&&&&&&&&&&&&&&&&&&&&&&&&&&&&&&&&&&&&&&&&&&&&&&&&&&&&&&&&&&&&&
%
%ccccccccccccccccccccccccccccccccccccccccccccccccc
%              appA.tex
%ccccccccccccccccccccccccccccccccccccccccccccccccc
\begin{appendix}
%%%%%%%%%%%%%%%%%%%%%%%%%%%%%%%%%%%%
\section{Jacobi Coordinates}
\label{gcoord}
 Let us define the Jacobi coordinates for equal mass particles
 %s
 \begin{eqnarray}
\nonumber
	 &&\vec \xi_N=(\vec x_2-\vec x_1)\\
\nonumber
 	 &&\vec \xi_{N-1}=\sqrt{3}(\vec x_3-\vec X_3)\\
\nonumber
         &&\vdots\\
\nonumber
	 &&\vec \xi_{N-i+1}=\sqrt{\frac{2i}{i+1}}(\vec x_{i+1}-\vec X_i)
\\\nonumber&&\hspace{1.7cm}
             =\sqrt{\frac{2(i+1)}{i}}(\vec x_{i+1}-\vec X_{i+1})\\
 \nonumber&&
        \vdots\\
&&	 \vec \xi_1=\sqrt{\frac{2A}{A-1}}(\vec x_A-\vec X)
 \label{Jacobi}
 \end{eqnarray}
 with $N=A-1$, $
 	 \vec X_i=\frac{1}{i}\sum_{p=1}^i \vec x_p
 $, and
 \begin{equation}
	 r^2=\sum_{i=1}^N\xi^2_i=2\sum_{i=1}^A(\vec x_i-\vec X)^2
	   =\frac{2}{A}\sum_{i,j>i}(\vec x_i-\vec x_j)^2\,,
 \label{Jacobir}
 \end{equation}
 and  where $\vec X\equiv \vec X_A$ corresponds to the center of mass.
 The normalization is such that for $i=N$ one gets the 
 vector  $\vec r_{12}$.\par
%
%%%%%%%%%%%%%%%%%%%%%%%%%
For describing the coordinates one  introduces the kinematical 
rotation vector \cite{F83,krv} 
\begin{equation} 
   \vec r(\delta,\varphi)=\cos\delta
          \vec \xi_N+\sin\delta\cos\varphi \vec \xi_{N-1}
    +({\rm terms \ in \ } \vec\xi_j,\ \ j<N-1)
\label{krov}
\end{equation}
where $\delta$ and $\varphi$ are angular parameters, 
$\vec \xi_N=\vec x_i-\vec x_j$, and 
$\vec \xi_{N-1}=\sqrt3(\vec x_k-1/3(\vec x_i+\vec xj+\vec x_k))$.
For $\delta=0$, $\vec r(0)=\vec x_i-\vec x_j$ is the reference pair, for
$\delta=2\pi/3$, $\varphi=0$  $\vec r(2\pi/3)=\vec x_k-\vec x_i$
is a connected pair, while for 
$\delta=\pi/2$,    $\vec r(\pi/2)$ is a disconnected pair.\par
%
%%%%%%%%%%%%%%%%%%%%%%%%%
 One may introduce also the  hyperspherical coordinates 
of Zernike and Brinkman  \cite{ZB} 
 \begin{equation}
	 \xi\equiv\xi_N=r\cos\phi_N\,,
		 \qquad \rho=r\sin\phi_N
 \label{Jacobiphi}
 \end{equation}
 with
 \begin{equation}
	 \xi^2+\rho^2=r^2\,,\qquad \rho^2=\sum_{j=2}^{A-1}\xi_j^2\,.
 \label{xirho}
 \end{equation}
 For the other  Jacobi coordinates we have
\begin{equation}
\nonumber
 \begin{array}{l}
     \xi_{N-1}=r\sin\phi_N\cos\phi_{N-1}\\
     \vdots\\
     \xi_j=r\sin\phi_N\cdots \sin\phi_{j+1}\cos\phi_j\,,\\
     \vdots\\
     \xi_2=r\sin\phi_N\cdots \sin\phi_3\cos\phi_2\,,\\
     \xi_1=r\sin\phi_N\cdots \sin\phi_2
\end{array}
 \label{Jacrest}
 \end{equation}
 where $\phi_1=0$. It is clear that in the Zernike-Brinkman
system of coordinates one has, apart from the hyperradius $r$,  $N-1$ 
angles $\phi_i$ and $2N$ angles $\om_i$, the volume element being
\begin{equation}
     \Dom_{i+1}=\Dom_i\,\sin^{(I-4)}\phi_{i+1}\,\cos^2\phi_{i+1}\rd\phi_{i+1}
       \dom_{i+1}
\label{domi}
\end{equation}
where $I=3(i+1)$. 
 The $\Om$ coordinates are separated into two parts, first  the
 $z=\cos2\phi$, ($\phi=\phi_N$), and $\om=\om_N$, the angular 
coordinates of $\vec \xi_N$,
  and second  the $\Om_{N-1}$ for the other hyperspherical coordinates
 $(\phi_i,\om_i)$, $i<N$ where $\om_i$ are the angular 
 coordinates of $\vec \xi_i$.
 Let $\rd\Om_{N-1}$ be the surface element of the  unit hypersphere
 $\rho=1$ in the $D-3$ dimensional space spanned by the Jacobi 
 coordinates $\vec \xi_i$, $i<N$. The volume element in the 
 $D$-dimensional space is  then given by
 \begin{eqnarray}
 \nonumber	
	 \rd^{3N}\xi&=& r^{D-1}\rd r\rd \Om\\
 \nonumber	
	 {\rm d}\Om&=&(\sin\phi)^{D-4}
		 \cos^2\phi\,\dph \,\dom\,\rd\Om_{N-1}\\
 \nonumber
		 &=&\frac{1}{2^{D/2}}(1-z)^{(D-5)/2}(1+z)^{1/2}
		 \,\dz\,\dom\,\rd \Om_{N-1}\\
	       &=&W_0(z)\,\dz\,\dom\,\rd\Om_{N-1}
 \label{dOm}
 \end{eqnarray}
 where
 $$
	 \cos\phi=\frac{r_{12}}{r}=\frac{\xi}{r}
 \qquad       z=\cos2\phi=2\frac{r_{12}^2}{r^2}-1\,.
 $$

In general the surface element ${\rm d}\Omega_i$ is defined to be the
part of the surface ${\rm d}\Omega$ which contains the coordinates $\omega_j$
and $z_j$ for $j\le i$, {\em i.e}
\begin{equation}
    {\rm d}\Omega_i = {\rm d}\omega_1\prod_{j=2}^i  2^{-3j/2}
	  w_j(z_j) {\rm d}z_j{\rm d}\omega_j
\label{dOj}
\end{equation}
where  $ w_j(z_j)$ is defined by 
\begin{equation}
    w_j(z_j)= (1-z_j)^{(3j-5)/2}(1+z_j)^{1/2}
\end{equation}
Knowing the element ${\rm d}\Omega_j$ we may construct the element of
 ${\rm d}\Omega_{j+1}$ via
\begin{equation}
    {\rm d}\Omega_{j+1} = {\rm d}\Omega_j (\sin\phi_{j+1})^{D-4}
		\cos\phi_{j+1}^2 {\rm d}\phi_{j+1} {\rm d}\omega_{j+1}
\label{dO0j}
\end{equation}
where $D=3(j+1)$.

The kinetic energy operator $T$, for equal mass particles, is given by
\begin{equation}
  T=-\frac{\hbar^2}{m}
	 \sum_{j=1}^{A-1}\nabla_{\xi_j}^2
	 -\frac{\hbar^2}{2M}\nabla^2_X\,.
\label{lapla}
\end{equation}
We are interested to express the translationally invariant part of this
equation in terms of the $\{ \omega_i, z_i\}$. Due to the product 
structure of d$\Omega$ it can be written as
\begin{equation}
         T=-\frac{\hbar^2}{m}\left[ \frac{\partial^2}{\partial r^2}
           +\frac{3A-4}{r}\frac{\partial}
        {\partial r}+  \frac{L^2(\Omega)}{r^2}\right]
\label{Lapl}
\end{equation}
where  $L^2(\Omega)$ is the grand orbital operator. Let $L_i^2(\Omega_1)$
be the operator associated with the first vectors $\{ {\bf \xi}_1,
\cdots, {\bf \xi}_i \}$. It can be written in terms of the one for the
 vectors  $\{ {\bf \xi}_1,\cdots, {\bf \xi}_{i-1}  \}$ as follows
\begin{equation}
   L_i^2(\Omega_i)= D(z_i)+2\frac{\ell^2(\omega_i)}{1+z_i}
	+\frac{2}{1-z_i} \,L_{i-1}^2(\Omega_{i-1})
\end{equation}
where
\begin{equation}
     D(z_i)=\frac{4}{w_i(z_i)}\frac{\partial}{\partial z_i}(1-z_i^2)
	 w_i(z_i)\frac{\partial}{\partial z_i}
 \end{equation}
 and
 $$ 
       L^2(\Omega)\equiv L^2_N(\Omega)\,.
$$
 The $ l^2(\omega_i)$ is the orbital operator normalized according to 
 \begin{equation}
     [\ell^2(\omega)+\ell(\ell+1)]Y_{\ell m}(\omega_i)=0\,.
 \end{equation}
%

%
%ccccccccccccccccccccccccccccccccccccccccccccccc
%               appB.tex
%ccccccccccccccccccccccccccccccccccccccccccccccc
%
\section{Various Coefficients}
%%%%%%%%%%%%%%%%%%%%%%%%%%%%%%%%
\subsection{The $3L$-Coefficients}
%%%%%%%%%%%%%%%%%%%%%%%%%%%%%%%%%%
The $3L$-coefficients are defined by the expansion
of a product of two HO eigenfunctions \cite{Kumar}
\begin{eqnarray}
\label{L3d}
        \psi_\alpha^*(\vec x)\psi_\beta(\vec x)&=&\frac{2}{\pi^{1/4}}
        \sum_{n,\ell,m}
        \left [ \matrix {
          n_\alpha     & n_\beta    & n    \cr
         \ell_\alpha   & \ell_\beta & \ell   \cr
        }\right] \\
&\times&
     \langle Y_{\ell_\alpha}^{m_\alpha}|Y_{\ell}^{ m\, *}| 
           Y_{\ell_\beta}^{ m_\beta}  \rangle 
       \re^{-x^2/2}\psi_{n,\ell,m}(\vec x)
\nonumber
\end{eqnarray}
where $\alpha$ and $\beta$ stand for the HO quantum numbers 
$n_\alpha,\ \ell_\alpha,\  m_\alpha$ and 
$n_\beta,\ \ell_\beta,\  m_\beta$.\par
By integrating over the $\vec x$ space for the HO quantum numbers 
$\alpha$, $\beta$ with $n=0,\ \ell=0,\  m=0$, and 
\begin{equation}
       \psi_{0,0,0}(\vec x)=\frac{2}{\pi^{1/4}}
        \re^{-x^2/2}Y_0^0\,,
\end{equation}
one finds the normalization
\begin{equation}
 \left [ \matrix {
          n_\alpha     & n_\beta    & 0    \cr
         \ell_\alpha   & \ell_\beta & 0   \cr
   }\right]
          = \delta_{\alpha\beta}\,.
\label{norm3L}
\end{equation}
Multiplying Eq. (\ref{L3d}) by
$$
     \left[ \frac{n! \Gamma(3/2)}{\Gamma(n+\ell+3/2)}\right]^{1/2} 
           Y_{\ell}^{m\,*}(\om) x^\ell L_n^{\ell+1/2}(x^2)
$$ 
and integrating over $x$ one gets \cite{MF75}
%
%\begin{widetext}
\begin{eqnarray}
\nonumber 
\hspace{-7mm} \left[ \begin{array}{ccc}
                    n_\alpha &n_\beta &n \cr
                   \ell_\alpha &\ell_\beta &\ell \cr
                 \end{array}\right]
         &=& \sqrt{2\pi^{1/2}} 
%\nonumber &&  
\left[\frac{n_\alpha!}{\Gamma(n_\alpha+\la+3/2)}
                \frac{n_\beta!}{\Gamma(n_\beta+\lb+3/2)}
                \frac{n!}{\Gamma(n+\ell+3/2)}\right]^{1/2}\\
&\times&
     \int_0^\infty x^{\la+\lb+\ell} L_{n_\alpha}^{\la+1/2}(x^2)        
        L_{n_\beta}^{\lb+1/2}(x^2)        
        L_{n}^{\ell+1/2}(x^2)\,\re^{-x^2}\,   x^2\dx\,.
\label{intL3}
\end{eqnarray}
%\end{widetext}
%
For $ n=\ell=0$ one finds  the normalization (\ref{norm3L}). Using the 
expansion for the Laguerre polynomials for $\alpha$ and  $\beta$ gives
for the integral in Eq. (\ref{intL3})
\begin{eqnarray}
\nonumber
 I&=&
    \sum_{m,m'}\frac{(-1)^{m+m'}}{m!{m'}!}
             {n_\alpha+\la+1/2 \choose n_\alpha-m}
              {n_\beta+\lb+1/2 \choose n_\beta-m'}\\
\nonumber&\times&
     \int_0^\infty x^{2(m+m'+(\la+\lb-\ell)/2)} x^{2\ell}
        L_{n}^{\ell+1/2}(x^2)   x^2\dx\,.
\end{eqnarray}
This last integral is given, according to Eq. (\ref{ddnn}), by
\begin{equation}
       \int_0^\infty x^{2(\nu+\ell)} 
        L_{n}^{\ell+1/2}(x^2)   x^2\dx
         =  \frac{(-1)^n}{2}{\nu\choose n} \Gamma(\nu+\ell+3/2)
\end{equation}
where here $\nu=  m+m'+(\la+\lb-\ell)/2$.\\

The $3L$-coefficients are thus  given by
%\begin{widetext}
\begin{eqnarray}
\nonumber
\hspace{-7mm}  \left[ 
         \begin{array}{ccc}
                    n_\alpha &n_\beta &n \cr
                   \ell_\alpha &\ell_\beta &\ell \cr
           \end{array}
   \right]
          &=& (-1)^n\sqrt{\frac{\pi^{1/2}}{2}}
\\\nonumber&\times&
       \left [ \frac{n_\alpha!}{\Gamma(n_\alpha+\ell_\alpha+3/2)}
         \frac{ n_\beta!}{\Gamma(n_\beta+\ell_\beta+3/2)}
     \frac{ n!}{\Gamma(n+\ell+3/2)}
        \right]^{1/2}\\
\nonumber
          &\times&\sum_{mm'}\frac{(-1)^{m+m'}}{m!m'!}
            {{n_\alpha+\ell_\alpha+1/2}\choose{n_\alpha-m}}
            {{n_\beta +\ell_\beta+1/2}\choose {n_\beta-m'}}\\
          &\times&  {{m+m'+(\ell_\alpha +\ell_\beta-\ell)/2}
              \choose {n}}
         \Gamma(m+m'+\frac{\ell_\alpha +\ell_\beta+\ell+3}{2})
\end{eqnarray}
%\end{widetext}
We recall that $a\choose b$ denotes the binomial coefficients, and
that $0\le m\le n_\alpha$,  $0\le m'\le n_\beta$, and 
 $n\le m +m'+(\la+\lb-\ell)/2$.
%
%\subsection{ Closed Shell Nuclei Coefficients}
%%%%%%%%%%%%%%%%%%%%%%%%%%%%%%%%%%%%%%%%%%%%%
\subsection{The   $(\alpha|n,\ell|\beta)$ coefficients} 
%%%%%%%%%%%%%%%%%%%%%%%%%%%%%%%%%%%%%%%%%%%%%%%%%%%%%%
To obtain the  coefficients $(\alpha|n,\ell|\beta)$ 
which appear in the Fourier transform   (\ref{ftpsi}),
we expand the plane wave
\begin{equation}
       \re^{i\vec k\cdot\vec x}
       =4\pi
	 \sum_{\ell=0}^{\infty}\, \sum_{m=-\ell}^{+\ell}\, i^\ell
	 j_{\ell}(kr)\, Y^{m\,*}_\ell(\omega) Y^{m}_\ell(\omega_k)
 \label{pwe}
 \end{equation}
 where  $(k,\om_k)$ and $(r,\om)$ are the polar coordinates of $\vec k$ and 
$\vec r$ respectively and $j_\ell$ is the spherical  Bessel function,
$j_\ell(\rho)=\sqrt{\pi/2\rho}J_{\ell+1/2}(\rho)$. Then, using the integral
\begin{equation}
\hspace{-7mm}
         \int_0^\infty\left(\frac{x}{b}\right )^\ell\, L_n^{\ell+1/2}
         (x^2/b^2)\frac{J_{\ell+1/2}(kx)}{\sqrt{kx}}\re^{-(x/b)^2}x^2\dx
       =\frac{b^3}{2^{n+(\ell+3)/2}}\frac{y^{2n+\ell}}{n!}
          \re^{-y^2/2}
\end{equation}
where $ y=kb/\sqrt{2}$, one obtains 
\begin{eqnarray}
%\nonumber
   (\alpha|n,\ell|\beta)=\left[
       \frac{2\pi^{3/2}}{ 2^{2n+\ell}n! \Gamma(n+\ell+3/2)}
       \right]^{1/2}
%\\&\times&
       \langle Y^{\ma}_{\la}|Y^{m *}_\ell|  Y^{\mb}_{\lb}\rangle
            \left[ \begin{array}{ccc}
                    n_\alpha &n_\beta &n \cr
                   \ell_\alpha &\ell_\beta &\ell \cr
                 \end{array}\right]
\label{anlnb}
\end{eqnarray}
The $\langle Y^{\ma}_{\la}|Y^{m *}_\ell|  Y^{\mb}_{\lb}\rangle$
can be easily expressed in terms of the Clebsch-Gordan coefficients
\cite{Rose}.\par
For $n_\alpha=n_\beta=0$, as for instance for the first $p-$ and $d$-shells,
\begin{eqnarray}
\nonumber
  &&\hspace{-5mm} (0,\la,m_\alpha|n,\ell,m|0,\lb,m_\beta)=
       \frac{\pi(-1)^n}{ 2^{n+\ell/2} \Gamma(n+\ell+3/2)}\\
&&
 \times
        {(\la+\lb-\ell)/2\choose n}\frac{\Gamma((\la+\lb+\ell+3)/2)}
        {\left[\Gamma(\la+3/2)\Gamma(\lb+3/2)\right]^{1/2}}
       \langle Y^{\ma}_{\la}|Y^{m *}_\ell|  Y^{\mb}_{\lb}\rangle
\label{anlnb1}
\end{eqnarray}
with $m=m_\beta-m_\alpha$, and $|\la-\lb|\le2n+\ell\le\la+\lb$.
The coefficient (\ref{anlnb}) is the product of two terms where  only the first
depends on the azimuthal numbers $\ma$ and $\mb$ but not 
on $n_\alpha$ and $n_\beta$.
\subsection{The  $I_0(\ell_m,n)$  coefficient}
%%%%%%%%%%%%%%%%%%%%%%%%%%%%%%%%%%%%%%%%%%%%%%%%%%%%%
For closed--shell nuclei the coefficient $I_0(\ell_m,n)$ can be 
calculated from (\ref{I0a}).
Table \ref{I0} gives the first 11 values of $I_0$ for 
$\ell_m=0,\cdots,5$. 
The rest can be easily calculated  either 
from (\ref{I0a}) or from the general relation \cite{FFW82} 
\begin{equation}
\label{I0g}
    	I_0(\ell_m,n)=\sum_{n_1,n_2,\ell_1}
	\,\left[ \sum_\alpha (\alpha|n_1,\ell_1)\right]
        \,\left[ \sum_\alpha (\alpha|n_2,\ell_1)\right]
\end{equation}
with $n=n_1+n_2+\ell_1$.
\subsection{The $I_{\alpha\beta}(\ell_m,n)$  coefficient}
%%%%%%%%%%%%%%%%%%%%%%%%%%%%%%%%%%%%%%%%%%%%%%%%%%%%%
They are given by
\begin{equation}
  I_{\alpha\beta} (\ell_m;n,0) =\sum_{n_1,n_2,\ell_1\atop {n=n_1+n_2+\ell_1}}
	\,\sum_{\alpha\beta} (\alpha|n_1\ell_1|\beta)
        \,(\beta|n_2\ell_1|\alpha)
\label{Iab}
\end{equation}
where the sum over $\alpha$ and $\beta$ means a sum over all H.O. states
(each must be taken once only either for $\alpha$ or $\beta$).
The calculation of this coefficient is tedious. However, for closed-shell
nuclei and  for small values of $n$ it reduces to  the following
analytical expressions:
\begin{eqnarray}
\nonumber
&& I_{\alpha\beta}(\ell_m,0)
       =\frac{1}{3!}\frac{(\ell_m+3)!}{\ell_m!}=\frac{A}{4}\,,\\
&&I_{\alpha\beta}(\ell_m,1)=0\,,\\
\nonumber&&
I_{\alpha\beta}(\ell_m,2)=\frac{1}{4!}\frac{(\ell_m+3)!}{(\ell_m-1)!}
      =\frac{A}{16}\ell_m\,,\\
\nonumber&&
 I_{\alpha\beta}(\ell_m,3)=-\frac{1}{5!}\frac{(\ell_m+3)!}{(\ell_m-2)!}\,,
\\
\nonumber&&
I_{\alpha\beta}(\ell_m,4)=\frac{1}{6!}\frac{5\ell_m-1}{6}
      \frac{(\ell_m+3)!}{(\ell_m-2)!} \,,\\
\nonumber&&
I_{\alpha\beta}(\ell_m,5)=-\frac{1}{6!}\frac{\ell_m+1}{12}
      \frac{(\ell_m+3)!}{(\ell_m-3)!}\,,\\
\nonumber&&
I_{\alpha\beta}(\ell_m,6)=-\frac{9\ell_m^2-3\ell_m-32}{5!(\ell_m+1)}
       I_{\alpha\beta}(\ell_m,5)\,,\\
\nonumber&&
 I_{\alpha\beta}(\ell_m,2\ell_m)=\left( \frac{1}{\ell_m!}\right)^2 \,,\\
\nonumber&&
I_{\alpha\beta}(\ell_m,2\ell_m-1)
      =-\frac{2}{(\ell_m-2)!}\frac{1}{\ell_m!}\,,\\
\nonumber&&
I_{\alpha\beta}(\ell_m,n>2\ell_m)=0\,.
\end{eqnarray}
Values of these coefficients of up to $\ell_m=7$ are given in Table \ref{Iab1}

\subsection{Coefficients in the $jj$-coupling scheme}
%%%%%%%%%%%%%%%%%%%%%%%%%%%%%%%%%%%%%%%%%%%%%%%%%%%%% 
For describing the state of the nucleus, the Slater 
determinant (\ref{sla1}) has
been filled with individual HO states
each associated with its spin-isospin state. Instead, one can combine 
the spin $\vec s$ and angular momentum $\vec \ell$ to generate the 
total angular momentum $\vec j$ of   projection $m$ and describe the individual
states in terms of the isospin $|t>$ and the $|n,\ell,j,m>$ states. The 
total angular momentum is constructed as usual,
\begin{equation} 
   \psi_{\ell j m}(\om)=\sum_{m',m_s}\langle \ell,1/2,m',m_s|j,m\rangle
     Y^{m'}_\ell(\om)\chi_{m_s}\,,
\label{j1}
\end{equation}
where $\chi_{m_s}$ is the spin variable with $m_s=\pm 1/2$ and the bracket 
is a Clebsch-Gordan coefficient.\\

The description of the individual states in terms of angular $\ell$ 
and total angular momentum $j$ is well adapted to nuclei because the
 two-body spin-orbit operator is actually a sum over individual 
spin-orbit operators (see  Eq. (\ref{lso}). In the $jj$-coupling
scheme we have to calculate the new coefficient $(\alpha|n,\ell|\beta)$
 where this time  $|\alpha>$ stands for
 $|n_\alpha,\la,j_\alpha, m_\alpha>$ and $|\beta>$ for
 $|n_\beta,\lb,j_\beta, m_\beta>$. The HO Slater determinant is now
\begin{equation}
      D^{HO}_\Lm(\vec x)=||t^j_i\psi_{n_j,\ell_j,j_j,m_j}(\vec x_i)||
\label{j2}
\end{equation}
where
\begin{equation}
   \psi_{n,\ell,j,m}(\vec x)=\left[\frac{2n!}{b^3
      \Gamma(n+\ell+3/2)}\right]^{1/2}
  \psi_{\ell,j,m}(\om) y^\ell L^{\ell+1/2}_n ( y^2) \re^{-y^2/2}
\label{j2a}
\end{equation}
where $y=x/b$.\par
The  $(\alpha|n,\ell|\beta)$ coefficients in $jj$-coupling must be 
constructed from the HO individual functions (\ref{j2a}). One starts
from the matrix elements where the integral is 
performed over $\dom$
%
%\begin{widetext}
\begin{eqnarray}
\nonumber
&&   \langle \la,\ja,\ma|Y^{m\,*}_\ell|\lb,\jb,\mb\rangle=\frac{1}
  {\sqrt{4\pi}} (-1)^{\mb-1/2}
    \left[\hat \la \hat \lb\hat \ell \hat \ja\hat \jb\right]^{1/2}
\\&&\ \ \ \ \ \times
\left(\begin{array}{ccc}
       \la& \lb &\ell\cr 0& 0& 0\cr
\end{array}\right)
\left(\begin{array}{ccc}
       \jb& \ja &\ell\cr -\mb& \ma& m\cr
\end{array}\right)
\left\{\begin{array}{ccc}
       \la& \lb &\ell\cr \jb& \ja& 1/2\cr
\end{array}\right\}
\label{j3}
\end{eqnarray}
in terms of the $3j$ and $6j$ coefficients. Here $\hat x=2x+1$,  
The one-body coefficient
$  (\alpha|n,\ell|\beta)$ in $jj$-coupling is obtained by substitution of 
(\ref{j3}) for  $\langle Y^{\ma}_{\la}|Y^{m *}_\ell|  Y^{\mb}_{\lb}\rangle$
in Eq. (\ref{anlnb})
\begin{eqnarray}
\nonumber
   (\alpha|n,\ell|\beta)&=&\left[
       \frac{2\pi^{3/2}}{ 2^{2n+\ell}n! \Gamma(n+\ell+3/2)}
       \right]^{1/2}\\
&\times&
   \langle \la,\ja,\ma|Y^{m\,*}_\ell|\lb,\jb,\mb\rangle
            \left[ \begin{array}{ccc}
                    n_\alpha &n_\beta &n \cr
                   \ell_\alpha &\ell_\beta &\ell \cr
                 \end{array}\right]
\label{j4}
\end{eqnarray}
When $\ell=0$ then
\begin{equation} 
        \langle \la,\ja,\ma|Y^{0}_0|\lb,\jb,\mb\rangle=\frac{1}{\sqrt{4\pi}}
          \delta_{\alpha\beta}
\label{j5}
\end{equation}
and
\begin{equation}
 \hspace{-5mm}(\alpha|n,0|\beta)=\left(\frac{\pi}{4}\right)^{1/4}
     \left[ 2^{2n} n! \Gamma(n+3/2)\right]^{-1/2}
\left[\begin{array}{ccc}
       n_\alpha& n_\beta &n\cr \la& \lb& 0\cr
\end{array}\right]
\delta_{\la\lb}\delta_{\ja\jb}\delta_{\ma\mb}
\end{equation}
In the direct term the sum over $-\ja\le\ma\le\ja$ for a complete 
subshell $n_\alpha,\la,\ja$ gives
\begin{equation}
    \sum_{\ma=-ja}^\ja(a|n,0)=(2\ja+1)\left(\frac{\pi}{4}\right)^{1/4}
    \left[ 2^{2n}n!\Gamma(n+3/2)\right]^{-1/2}
     \left[\begin{array}{ccc}
       n_\alpha& n_\alpha &n\cr \la& \la& 0\cr
\end{array}\right]
\label{j6}
\end{equation}
In the $jj$-coupling scheme the Slater determinant is constructed
according to (\ref{j2}).
Instead of the four spin-isospin cases occurring in Eq. (\ref{faben}) 
we have to consider only two cases for $d_{ij}(\vec x_1,\vec x_2)$ in 
Eq. (\ref{dhoex}): Either the two nucleons are the same or we
have to deal with the neutron-proton pair. In the first case
$t^i-t^j=0$. For the Wigner  $(\epsilon=0)$ and isospin exchange
$\epsilon=\tau$ weight and pseudo-weight functions,
the matrix element  (\ref{j4}) must be used in Eqs. (\ref{dcoe1})
and  (\ref{ecoe1}) to obtain the direct and exchange coefficients.
The coefficients  $(\Ll|\nu,\lambda)^{(\epsilon)}$ are given by  Eq. 
(\ref{LnlW}) for the weight function $(\epsilon=0)$
and by Eq. (\ref{LnlM}) for the isospin exchange $(\epsilon=\tau)$ 
pseudo-weight function  where $t$ is substituted for $\zeta$.\par
 The spin exchange operator generates new HO states
 $\psi_{\alpha'}(\vec x_i)\psi_{\beta'}(\vec x_j)$ with
\begin{equation}
      P^\sigma_{ij}\psi_\alpha(\vec x_i)\psi_\beta(\vec x_j)=
     \sum_{{\ma',\mb',M\atop\ja',\jb'\, J}}
     \langle \alpha',\beta'|P^\sigma_{ij}|\alpha,\beta\rangle
     \psi_{\alpha'}(\vec x_i)\psi_{\beta'}(\vec x_j)
\end{equation}
where
\begin{eqnarray}
\nonumber
\hspace{-7mm}&&
     \langle \alpha',\beta'|P^\sigma_{ij}|\alpha,\beta\rangle=
     -\left[\hat j_\alpha \hat j_\beta \hat j_{\alpha'}
       \hat j_{\beta'}\right] (-1)^{j_\beta+j_{\beta'}}
\\ &&\hspace*{10mm} \times
     \langle j_\alpha j_\beta m_\alpha m_\beta |JM\rangle 
   \langle j_{\alpha'} j_{\beta'} m_{\alpha'} m_{\beta'} |JM\rangle 
\left\{\begin{array}{ccc}
       \la& 1/2 &\ja\cr 
       1/2& \lb& \jb\cr
       \ja'& \jb'& J\cr
\end{array}\right\}
\label{j7}    
\end{eqnarray}
where $\hat j=2j+1$, 
$\alpha'=|n_{\alpha'}\ell_{\alpha'}j'_{\alpha'}m_{\alpha'}\rangle$, 
$\beta'=|n_{\beta'}\ell_{\beta'}j'_{\beta'}m_{\beta'}\rangle$, 
the brackets are  Clebsch-Gordan coefficients, while the brace is a $9j$
coefficient.\par
When the states $|\alpha>$ and $|\beta>$ are coupled to give 
a definite total angular momentum $J$ of projection $M$, like 
for two nucleons  outside a closed shell (e.g $^6$Li), the effect of 
the spin-isospin exchange operator is
%\begin{widetext}
\begin{eqnarray}
\nonumber&&
 P^\sigma_{ij}\left[\psi_\alpha(\vec x_i)\otimes\psi_\beta(\vec x_j)\right]^M_J
              =-\left[\hat \ja\hat \jb\right]^{1/2}
\\\nonumber&&\ \ \ \
\times\sum_{\ja',\jb'}(-1)^{\jb+\jb'}\left[\hat\ja'\hat\jb'\right]^{1/2}
\left\{\begin{array}{ccc}
       \la& 1/2 &\ja\cr 
       1/2& \lb& \jb\cr
       \ja'& \jb'& J\cr
\end{array}\right\}
\big[\psi_{\alpha'}(\vec x_i)\otimes \psi_{\beta'}(\vec x_j)\big]^M_J
\label{j8}    
\end{eqnarray}
When the spin exchange occurs, the direct coefficient is given by 
Eq. (\ref{dcoe1}) where
\begin{eqnarray}
\nonumber
  (\alpha,\beta|n_1,\ell_1,n_2,\ell_2|\alpha\beta)^{(0)}&=&\\
&&\hspace*{-2cm}\sum_{{\ma',\mb',M\atop\ja',\jb'\, J}}
     \langle \alpha',\beta'|P^\sigma_{ij}|\alpha,\beta\rangle
(\alpha|n_1,\ell_1|\alpha')(\beta|n_2,\ell_2|\beta')
\end{eqnarray}
where $(\alpha|n,\ell |\alpha')$ etc,  given by  (\ref{j4}), is 
substituted  for $(\alpha|n_1,\ell_1)^*(\beta|n_2,\ell_2)$.\par
The exchange coefficient is obtained by using 
$(\beta\alpha|n_1\ell_1 n_2\ell_2|\alpha\beta)^{(\sigma)}$ instead 
of $(\alpha|n_1\ell_1|\beta)^*(\beta| n_2\ell_2|\alpha)$ in Eq.
(\ref{ecoe1}). The spin $(\epsilon=\sigma)$ and isospin 
$(\epsilon=\sigma\tau)$ exchange coefficient $(\Ll|\nu,\lambda)^{(\sigma)}$
and $(\Ll|\nu,\lambda)^{(\sigma\tau)}$ are given respectively by 
(\ref{LnlW}) and (\ref{LnlM})  where $t$ is  substituted for $\zeta$.
\end{appendix}
%%%%%%%%%%%%%%%%%%%%%%%%%%%%%%%%%%%

%%%%%%%%%%%%%%%%%%%%%%%%%%%%%%%%%%%%%%%%%%%%%%%%%%%%%%
%                REFERENCES
%%%%%%%%%%%%%%%%%%%%%%%%%%%%%%%%%%%%%%%%%%%%%%%%%%%%%%%

%\end{references}

%&&&&&&&&&&&&&&&&&&&&&&&&&&&&&&&&&&&&&&&&&&&&&&&&&&&&&&&&&&&&&&&&&
%%%%%%%%%%%%%%%%%%%%%%%%%%%%%%%%%%%%
%              TABLES
%%%%%%%%%%%%%%%%%%%%%%%%%%%%%%%%%%%%
\newpage
%%%%%%%%%%%%%%%%%%%%%%%%%%%%%%%%%%%%%%%%%%%%%%%%%%%%%
\begin{table}[hb]
\caption{\label{coeft}
The  coefficients $(L_m | n,\ell)^{(\epsilon)}$
for the closed shell nuclei A= 16 and A=40 and for the
 A=8 and A=20 closed shell neutron systems.  }

\begin{tabular}{|c| c c c c|}
\hline
\multicolumn{1}{|c} {$\epsilon$ } &
\multicolumn{1}{|c} {\ \ $^{16}$O \ \ }&
\multicolumn{1}{c} {\ \ $^{40}$Ca \ \ }&
\multicolumn{1}{c} {\ \ $8$n \ \ }&
\multicolumn{1}{c|} {\ \ $20$n \ \ }\cr
\hline

\hline
   WIGNER &   120.    & 780. &   28.     & 190.0 \cr
          &   -64.    &-800. &   -16.    & -200.0\cr    
          &     6.    & 270. &    1.     & 65.0  \cr       
          &           & -38. &           & -9.0  \cr
          &           & 1.5  &           & 0.25  \cr
\hline
   BARTLETT &  48.   & 360   &   8.      & 80.0  \cr
            &   -32. &-400.  &  -8       & -100.0\cr
            &     0. & 120.  &  -1.      &25.0   \cr
            &        & -16.  &           & -3.0      \cr
            &        &   0   &           & 0.250      \cr
\hline
   HEISENBERG &   48.  & 360. &   28.    & 190.0  \cr
              &   -32. &-400. &  -16.    &-200.0  \cr
              &     0. & 120. &   1.     & 65.0  \cr     
              &        & -16. &          & -9.0 \cr
              &        &  0   &          & 0.25   \cr
\hline
   MAJORANA &  0.     & 120.  &   8.     &80.0     \cr
            &   -16 . & -200. &  -8 .    &-100.0 \cr
            &    -6.  &  30.  &  -1.     &25.0    \cr      
            &         &  -2.  &          &-3.0     \cr
            &         & -1.5  &          & 0.25     \cr
\hline
   COULOMB & 28.  & 190.  & 0    &   0      \cr
           & -16. & -200. & 0    &   0      \cr
           &  1.  &  65.  & 0    &   0      \cr
           &      &  -9.  &       &         \cr
           &      & 0.25  &       &         \cr
\hline
\end{tabular}
\end{table}
\newpage
%%%%%%%%%%%%%%%%%%%%%%%%%%%%%%%%%%%%
%               II
%%%%%%%%%%%%%%%%%%%%%%%%%%%%%%%%%%%%
\begin{table}[hb]
\caption{\label{r1640} Roots and relative normalization constants 
$C^{(\epsilon)}/C^{(0)}$
for $^{16}$O and $^{40}$Ca systems.}
\begin{tabular}{|c|c|c|c|c|}
\hline
\multicolumn{1}{|c|}{$\epsilon$} &
\multicolumn{2}{c|}{$^{16}$O}   &
\multicolumn{2}{c|}{$^{40}$Ca} \cr
\hline
\multicolumn{1}{|c|}{}&
\multicolumn{1}{c|}{Roots}&
\multicolumn{1}{c|}{$C^{(\epsilon)}/C^{(0)}$}&
\multicolumn{1}{c|}{Roots}&
\multicolumn{1}{c|}{$C^{(\epsilon)}/C^{(0)}$}\cr
\hline
 WIGNER
 & -1.2590158014&1.0 & -1.1554853491                 &1.0  \cr         
 & -1.1462382460&  &-1.0152702740                  &  \cr
 &              & &(-0.9828722545,    0.0692920115)  &  \cr
 &              & &(-0.9828722545,   -0.0692920115)  &  \cr
\hline
BARTLET
 &  -1.0         &-0.2085327360 & +1.0     & -0.1036460531 \cr
 &   1.0         &              & -1.0                         &  \cr
 &               &              & (-0.9723433872, 0.0576076251) &  \cr         
 &               &              & (-0.9723433872, -0.0576076251) & \cr      
\hline
HEISENBERG
 & -1.0          &-0.2085327360 & +1.0                & -0.1036460531 \cr
 & 1.0           &              & -1.0                          &  \cr
 &               &              & ( -0.9723433872,0.0576076251)&  \cr
 &               &              &( -0.9723433872,-0.0576076251)&  \cr
\hline
 MAJORANA
 & -0.9283077625 &-1.5213318399 &  -0.9879721459      &-1.2591151327  \cr
 &-0.6527109110  & &-0.7766080532               &  \cr        
 &               & & (-0.9604317388,     0.0448314378)&  \cr
 &               & &(-0.9604317388,    -0.0448314378)&  \cr
\hline
COULOMB  
 & -1.9906725214  &0.1319112107 &-1.4266427288&  0.1493923245 \cr
 & -1.0483085670  & &-1.0070162824 &  \cr
 &                & & (-0.9781410398,  0.0634856828)&  \cr
 &                & &(-0.9781410398, -0.0634856828)&  \cr
\hline
\end{tabular}
\end{table}
\newpage
%%%%%%%%%%%%%%%%%%%%%%%%%%%%%%%%%%%%
%               III
%%%%%%%%%%%%%%%%%%%%%%%%%%%%%%%%%%%%
\small
\begin{table}[hb]
\caption{\label{r820}
 Roots and relative normalization constants 
$C^{(\epsilon)}/C^{(0)}$ for 8n and 20n systems.   }
\begin{tabular}{|c|c|c|c|c|}
\hline
\multicolumn{1}{|c|}{$\epsilon$} &
\multicolumn{2}{c|}{8n  } &
\multicolumn{2}{c|}{20n}\cr
\hline
\multicolumn{1}{|c|}{}&
\multicolumn{1}{c|}{Roots}&
\multicolumn{1}{c|}{$C^{(\epsilon)}/C^{(0)}$}&
\multicolumn{1}{c|}{Roots}&
\multicolumn{1}{c|}{$C^{(\epsilon)}/C^{(0)}$}\cr
\hline
 WIGNER
     & -8.6346734966   &1.0 &  -2.1965713866&1.0  \cr
     &  -1.1135782516  & &  -1.0145962721&  \cr
     &                 & &  (-0.9502197308,0.1301817431)&  \cr
     &                 & &(-0.9502197308,-0.1301817431)&  \cr
\hline
BARTLET
     & -0.9146656586  &-10.3986013985 &-0.9870485512& -2.4763777976  \cr
     &  -0.0227923104 & &-0.3424370373&  \cr
     &                &  &(-0.9502197308, 0.1037682674)&  \cr
     &                & &(-0.9502197308,-0.1037682674)&  \cr
\hline
HEISENBERG
    & -8.6346734966    &1.0 & -2.1965713866& 1.0 \cr
    & -1.1135782516    & &-1.0145962721&  \cr
    &                  &&(-0.9502197308,     0.1301817431)&  \cr
    &                  &&(-0.9502197308,    -0.1301817431)&  \cr
\hline
 MAJORANA
    &-0.9146656586    &-10.3986013985&-0.9870485512& -2.4763777976  \cr
    & -0.0227923104   &&-0.3424370373&  \cr
    &                  &&(-0.9270670083,     0.1037682674)&  \cr 
    &                  &&(-0.9270670083,     -0.1037682674)&  \cr 
\hline
\end{tabular}
\end{table}
\newpage
%%%%%%%%%%%%%%%%%%%%%%%%%%%%%%%%%%%%
%               IV
%%%%%%%%%%%%%%%%%%%%%%%%%%%%%%%%%%%%

\begin{table}[htb]
\caption{\label{He}
Ground state energies $E_g$ (in MeV)  and root
mean square radius $r_{rms}$ (in fm) for $^4$He. 
The IDEA results are those of the uncoupled adiabatic 
approximation (UAA). The results $a$ in FHNC are with
Gaussian and  $b$ with Euler correlations. 
${\rm E_C}$ is the increase in the binding energy due to the Coulomb
interaction.  }

\begin{tabular}{|l|c c c c c |c|}
\hline
\multicolumn{1}{|l|}{Potential}&
\multicolumn{1}{c}{HCA}&
\multicolumn{1}{c}{HCA+Coul}&
\multicolumn{1}{c}{EAA}&
\multicolumn{1}{c}{UAA}&
\multicolumn{1}{c}{${\rm E_C}$}&
\multicolumn{1}{|c|}{Other Methods}\cr
\hline
%%%%%%%%%%%%%%%%%%%%%%%%
 B1\cite{B1}
%   POT &   HCA   &HCA+Coul& EAA      &  UAA   &  Coul  & Other   \cr
        & 29.292  & 28.473 &39.169    & 38.162  & 0.810             
&  FHNC/1$^a$ \cite{FHNC}\hfill   37.7  \cr
        & (1.497) &(1.502) & (1.420)  &(1.422)  & (1.425)   
&  FHNC/1$^b$ \cite{FHNC}\hfill \ \  37.9  \cr
&&&&&&  FAHT/III \cite{Guard}\hfill \ \   36.6  \cr
&&&&&&  BHF \cite{Guard}  \hfill    \ \  36.9  \cr
&&&&&&  VMC \cite{VMC1,VMC2}\hfill  \ \  36.4  \cr
\hline
%-------------------------------------------------------
  MS3\cite{AT,Guard}
%   POT &   HCA   &HCA+Coul& EAA      &  UAA   &  Coul  & Other\cr
        & 7.177  & 6.516   &28.070    & 26.760  & 0.805             
&  FHNC/1$^a$ \cite{FHNC}\hfill   24.7  \cr
        & (1.894)  &(1.910) & (1.443)  &(1.454)  & (1.458)   
&  FAHT/III \cite{Guard}\hfill \ \   24.2  \cr
&&&&&&  BHF \cite{Guard}  \hfill    \ \  25.0  \cr
&&&&&&  VMC \cite{VMC1,VMC2}\hfill  \ \ 23.9-26.5\cr
\hline
%-------------------------------------------------------
GPDT\cite{GPDT}
%   POT &   HCA    & HCA+Coul& EAA      &  UAA   &  Coul  & Other   \cr
        &  14.199  & 13.436 &18.848    & 18.175   & 0.757   &       \cr
        & (1.651)  & (1.661) & (1.593)  &(1.603)  & (1.608)    &  \cr
\hline
\end{tabular}
\end{table}
\newpage
%%%%%%%%%%%%%%  O-16  %%%%%%%%%%%%%%%
%%%%%%%%%%%%%%%%%%%%%%%%%%%%%%%%%%%%
%               V
%%%%%%%%%%%%%%%%%%%%%%%%%%%%%%%%%%%%
\begin{table}[hb]
\caption{\label{O16}
Same as Table \protect{\ref{He} but for $^{16}$O}}
\begin{tabular}{|l|c c c c c |l|}
\hline
\multicolumn{1}{|l|}{Potential}&
\multicolumn{1}{c}{HCA}&
\multicolumn{1}{c}{HCA+Coul}&
\multicolumn{1}{c}{EAA}&
\multicolumn{1}{c}{UAA}&
\multicolumn{1}{c}{${\rm E_C}$}&
\multicolumn{1}{|l|}{Other Methods}\cr
\hline
 B1\cite{B1}
%   POT &   HCA   &HCA+Coul& EAA      &  UAA   &  ${\rm E_C}$ & Other   \cr
        & 106.529 & 93.135 &164.777   &164.332   & 13.727            
&  FHNC/1$^a$ \cite{FHNC}\hfill \ \ 150.4\cr
        & (2.604) &(2.628) & (2.559)  &(2.566)  & (2.587)   
&  FHNC/1$^b$ \cite{FHNC}\hfill \ \ 152.4  \cr
&&&&&&  FAHT/III \cite{Guard}\hfill \ \ 163.7   \cr
&&&&&&  BHF \cite{Guard}  \hfill    \ \ 163.7  \cr
&&&&&&  VMC \cite{VMC1,VMC2}\hfill  \ 150.9$\pm$ 0.3  \cr
&&&&&&   IDEA\cite{Brizzi}\hfill\ \  165.2  \cr 
&&&&&&   HHE\cite{HHE}   \hfill    \ \  152.10   \cr
&&&&&&   ${\rm E_C}$\cite{HHE}   \hfill    \ \  13.78   \cr

\hline
%-------------------------------------------------------
  MS3\cite{AT,Guard}
%   POT &   HCA   &HCA+Coul& EAA      &  UAA   &  ${\rm E_C}$  & Other   \cr
        & 12.745 & 1.443   &103.261   &102.793  & 13.843             
&  FHNC/1$^b$ \cite{FHNC}\hfil\ \ 105.3   \cr
        & (3.079)  &(3.167) & (2.539)  &(2.535)  & (2.563)   
&  FAHT/III \cite{Guard}\hfill \ \  107.7  \cr
&&&&&&  BHF \cite{Guard}  \hfill    \ \ 118.6  \cr
&&&&&&  IDEA \cite{Brizzi}  \hfill    \ \ 103.2  \cr
\hline
%-------------------------------------------------------
GPDT\cite{GPDT}
%   POT &   HCA   &HCA+Coul& EAA      &  UAA   & ${\rm E_C}$ & Other   \cr
        &61.867  &47.740 &100.796   & 100.688 & 14.305      &
                  HHE\cite{HHE}   \hfill      94.63   \cr                      
        & (2.4791  &(2.513) & (2.484)  &(2.484)  & (2.528)  &   
              ${\rm E_C}$    \cite{HHE}\hfill 14.71    \cr
\hline
\end{tabular}
\end{table}
\newpage
%%%%%%%%%%%%%%%%%%%%%%%%%%%%%%%%%%%%%%%%%%%%%%%%%%%%%
%                      VI
%%%%%%%%%%%%%%%%%%%%%%%%%%%%%%%%%%%%%%%%%%%%%%%%%%%%%%%%%%%%
%%%%%%%%%%%%%%%%%%%%%%%%   Ca-40 %%%%%%%%%%%%%%%%%%%%%%%%%%%
\begin{table}[hb]
\caption{\label{Ca40}
Same as Table \protect\ref{He} but for $^{40}$Ca}
\begin{tabular}{|l|c c c c c |l|}
\hline
\multicolumn{1}{|l|}{Potential}&
\multicolumn{1}{c}{HCA}&
\multicolumn{1}{c}{HCA+Coul}&
\multicolumn{1}{c}{EAA}&
\multicolumn{1}{c}{UAA}&
\multicolumn{1}{c}{${\rm E_C}$}&
\multicolumn{1}{|l|}{Other Methods}\cr
\hline
 B1\cite{B1}
%   POT &   HCA   &HCA+Coul& EAA      &  UAA   & ${\rm E_C}$  & Other   \cr
        & 323.355 & 250.844&475.381    &475.345   & 75.461            
&  FHNC/1$^a$ \cite{FHNC}\hfill \ \ 471.0\cr
        & (3.341) &(3.397) & (3.251)  &(3.251)  & (3.300)   
&  FHNC/1$^b$ \cite{FHNC}\hfill \ \ 482.0  \cr
&&&&&&  FAHT/III \cite{Guard}\hfill \ \ 478.0   \cr
&&&&&&  BHF \cite{Guard}  \hfill    \ \ 507.2  \cr
&&&&&&  VMC \cite{VMC1,VMC2}\hfill  \ 483.0$\pm$ 0.4  \cr
 &&&&&&  $\tilde {\rm WFA}$ \cite{Brizzi}\hfill\ \  447.8  \cr 
 &&&&&&  IDEA \cite{Brizzi}\hfill\ \  483.0  \cr 
 &&&&&& HHE \cite{HHE}\hfill  468.14          \cr 
 &&&&&& ${\rm E_C}$ \cite{HHE}\hfill  76.0          \cr 
\hline
%------------------------------------------------------------
  MS3\cite{AT,Guard}
%   POT &   HCA   &HCA+Coul& EAA      &  UAA   &  ${\rm E_C}$ & Other   \cr
        & 48.419 &no bound    &259.447   &259.328  & 72.149   
&  FHNC/1$^b$ \cite{FHNC}\hfill \ \ 350.0  \cr
        & (3.868)  &          & (3.375)  &(3.372)  & (3.457)   
&  FAHT/III \cite{Guard}\hfill \ \  335.6  \cr
&&&&&&  BHF \cite{Guard}  \hfill    \ \ 354.0  \cr
 &&&&&& IDEA  \cite{Brizzi}\hfill\ \ 272.54   \cr 
\hline
%-------------------------------------------------------------
GPDT\cite{GPDT}
%   POT &   HCA    & HCA+Coul & EAA      &  UAA    &  Coul  & Other\cr
        & 255.952  & 171.852  &  376.331 & 376.110 & 86.31    &
                     HHE \cite{HHE} \hfill  363.53          \cr
        & (2.868)  & (2.951)  & (2.842)  & (2.842) & (2.912)     &
                     ${\rm E_C}$ \cite{HHE} \hfill  86.0           \cr
\hline
\end{tabular}
\end{table}
%\end{widetext}
\newpage
%%%%%%%%%%%%%%  B-10  %%%%%%%%%%%%%%%
%%%%%%%%%%%%%%%%%%%%%%%%%%%%%%%%%%%%
%               VII
%%%%%%%%%%%%%%%%%%%%%%%%%%%%%%%%%%%%

\begin{table}[hb]
\caption{\label{Acoef}
 The coefficients $A_n^{\epsilon}$, Eq. (\ref{ketA}) for the 
 two configurations considered for the $^{10}$B nucleus.}
\begin{tabular}{|c|ccc|ccc|}
\hline
\multicolumn{1}{|c|}{$\epsilon$} &
\multicolumn{3}{c|}{ $(4s_{1/2})(4p_{1/2})(2p_{3/2})$ }
  &
\multicolumn{3}{c|}{$(4s_{1/2})(6p_{3/2})$}\cr
\hline
\multicolumn{1}{|c|}{}&
\multicolumn{1}{c}{$A_0^\epsilon$}&
\multicolumn{1}{c}{$A_1^\epsilon$}&
\multicolumn{1}{c|}{$A_2^\epsilon$}&
\multicolumn{1}{c}{$A_0^\epsilon$}&
\multicolumn{1}{c}{$A_1^\epsilon$}&
\multicolumn{1}{c|}{$A_2^\epsilon$}\cr
\hline
 WIGNER  & 41/2 &19 &11/2     &41/2 & 19 & 11/2\cr
%\hline
BARTLET &1/2&46/3 & 1/2       &1/2 &40/3 & 1/2\cr
%\hline
HEISENBERG
         &-1/2 & 12 & -1/2    &-1/2 & 12 & -1/2 \cr
%\hline
 MAJORANA
         &-171/8 &  323/12 &  -51/8 &   -156/8 &   284/12   & -36/8\cr
\hline
\end{tabular}
\end{table}
\newpage

%%%%%%%%%%%%%%%%%%%%%%%%%%%%%%%%%%%%
%               VIII
%%%%%%%%%%%%%%%%%%%%%%%%%%%%%%%%%%%%

\begin{table}[hb]
\caption{\label{boronr}
 Roots and relative normalization constants 
$C^{(\epsilon)}/C^{(0)}$ for the $^{10}$B nucleus.}
\begin{tabular}{|c|c|c|c|c|}
\hline
\multicolumn{1}{|c|}{$\epsilon$} &
\multicolumn{2}{c|}{ $(4s_{1/2})(4p_{1/2})(2p_{3/2})$ }
  &
\multicolumn{2}{c|}{$(4s_{1/2})(6p_{3/2})$ }\cr
\hline
\multicolumn{1}{|c|}{}&
\multicolumn{1}{c|}{Roots}&
\multicolumn{1}{c|}{$C^{(\epsilon)}/C^{(0)}$}&
\multicolumn{1}{c|}{Roots}&
\multicolumn{1}{c|}{$C^{(\epsilon)}/C^{(0)}$}\cr
\hline
 WIGNER
     &  -2.4267116231  &   1.0         &   -2.4267116231   &  1.0        \cr
     &  -1.2816012954  &               &   -1.2816012954   &             \cr
\hline
BARTLET
     & -1.0057745600  & -0.6712940280 & -1.0066449150    & -0.5602373564 \cr
     & 1.5278733354   &               & 1.6322400952     &               \cr
\hline
HEISENBERG
    & -0.9926668226   &-0.8464804834 & -0.9926668226     & -0.8464804834 \cr
    &  0.5786209032   &              &  0.5786209032     &               \cr
\hline
 MAJORANA
    &-0.8743416859    &-0.5924519571 & -0.8590657779     & -2.1204747128  \cr
    &  4.6270236827   &              & 0.2787982623      &                \cr
\hline
\end{tabular}
\end{table}
\newpage

%%%%%%%%%%%%%%%%%%%%%%%%%%%%%%%%%%%%
%               IX
%%%%%%%%%%%%%%%%%%%%%%%%%%%%%%%%%%%%
\begin{table}[hb]
\caption{\label{B10}
{Binding energies and rms radii for the  $^{10}$B} nucleus}
\begin{tabular}{|l|c c | c c |}
\hline
\multicolumn{1}{|l|}{Potential}&
\multicolumn{2}{c|}{ $(4s_{1/2})(4p_{1/2})(2p_{3/2})$ }
  &
\multicolumn{2}{c|}{$(4s_{1/2})(6p_{3/2})$ }\cr
\hline
\multicolumn{1}{|c|}{}&
\multicolumn{1}{c}{EAA}&
\multicolumn{1}{c|}{UAA}&
\multicolumn{1}{c}{EAA}&
\multicolumn{1}{c|}{UAA}\cr
\hline
\hline
%   POT       &  EAA   & UAA      &  EAA     &  UAA       \cr
 B1\cite{B1}  & 54.787 &  54.542  &  41.955  &   41.615   \cr
              &(2.699) & (2.703)  &  (2.753) & (2.758)    \cr
\hline
%-------------------------------------------------------
  MS3\cite{AT,Guard}  & 33.475  & 32.941    &  22.504 & 22.252 \cr
                      &(2.648)  &(2.665)    & ( 2.756) & (2.764) \cr
\hline
%-------------------------------------------------------
GPDT\cite{GPDT}  &17.928  & 17.827  &9.350        &  9.269         \cr
                 &(3.139) &(3.144)  &(3.361)        &(3.369)         \cr

\hline
\end{tabular}
\end{table}
\newpage
%ttttttttttttttttttttttttttttttttttttttttttttttttttttttttt
%%%%%%%%%%%%%   

\begin{center}
\begin{table}[phtb]
	\caption{\label{I0} The $I_{0}(\ell_m,n)$ coefficient} 
	\begin{tabular}{ccccccc}
		\hline
		        & \multicolumn{6}{c} {$\ell_{m}$}	\\
		\cline{2-7}
$n$
	& 0   & 1    & 2  & 3  & 4  &  5\\
\hline
%  |  0  |  1   | 2   | 3      | 4     |  5  
%%-----------------------------------------------------------
0  & 1   & 16   & 100 & 400    &1225    &3136  \\
1  &     & -8   &-100 & -600   & -2450  &-7840 \\
2  &    &  1   &  35 & 345    & 1960   & 8036 \\
3  &    &      & -5  & -290/3 &-2450/3 &-13328/3\\
4  &    &      & 1/4 & 14     &1169/6  &4424/3 \\
5  &    &      &     & -1   & -329/12  & -4634/15    \\
6  &    &      &     & 1/36   &161/72  & 749/18 \\
7  &    &      &     &        &-7/72   &-161/45   \\
8  &    &      &     &        &1/576   &17/90\\
9  &    &      &     &        &     &-1/180 \\
10 &    &      &     &        &     &1/14400 \\
\hline
	\end{tabular}
\end{table}
\end{center}
\newpage
%%%%%%%%%%%%%%%%%%%%%%%%%%%%%%%%%%%%%%%%%%%%%%%%%%%%%%%%%
%%%%%%%%%% 
\begin{center}
\begin{table}[phtb]
	\caption{\label{Iab1}
The $I_{\alpha\beta}(\ell_{m},n)$ coefficient. The symbol $[-n]$
denotes the exponential $10^{-n}$.}  
 	\centering
	\begin{tabular}{c|cccccccc}
		\hline
		        & \multicolumn{8}{c} {$\ell_{m}$}	\\
		\cline{2-9}
$n$
	& 0   & 1    & 2  & 3  & 4  &  5  & 6  & 7\\
\hline
%  |  0  |  1   | 2   | 3      | 4     |  5       |  6   |  7     
%%-----------------------------------------------------------
0  & 1   &  4   & 10  & 20     & 35     & 56        & 84      & 120 \\
1  & 0   &  0   &  0  &  0     & 0      &  0        &  0      &  0  \\
2  &     &  1   &  5  & 15     & 35     & 70        & 126     & 210 \\  
3  &     &      & -1  & -6     &-21     & -56       &-126     & -252\\  
4  &     &      & 1/4 & 7/3    &133/12  &112/3      & 203/2   &  238\\  
5  &     &      &     & -1/3   & -35/12 & -14       & -49     & -140 \\
6  &     &      &     & 1/36   & 35/72  & 623/180   & 15.9833 &56.583\\
7  &     &      &     &        &-1/24   &-8/15      & -3.4833 &-15.833\\
8  &     &      &     &        &1/576   &5.1389[-2] &0.51458  & 3.1268  \\
9  &     &      &     &        &     &-2.777[-3] &-5.0463[-2] &-0.43638 \\
10 &     &      &     &        &     &6.9444[-5] &3.1713[-3]&4.2774[-2]\\
11 &     &      &     &        &     &       &-1.1574[-4] &-2.8770[-3] \\
12 &     &      &     &        &     &       & 1.9290[-6] &1.2676[-4] \\
13 &     &      &     &        &     &           &    &-3.3069[-6]  \\
14 &     &      &     &        &     &           &    &3.9367[-8]  \\
\hline
	\end{tabular}
\end{table}
\end{center}

\end{document}